\documentclass{article}

\usepackage{arxiv}

\usepackage[utf8]{inputenc} 
\usepackage[T1]{fontenc}    
\usepackage{hyperref}       
\usepackage{url}            
\usepackage{booktabs}       
\usepackage{amsfonts}       
\usepackage{nicefrac}       
\usepackage{microtype}      
\usepackage{lipsum}
\usepackage{graphicx}

\usepackage{xcolor}
\usepackage{tikz}
\usepackage{amsmath}
\usepackage{amssymb}

\usepackage{lineno,hyperref}
\usepackage{algorithm2e}
\usepackage{textcomp} 
\usepackage{subcaption}
\usepackage{adjustbox}

\modulolinenumbers[5]
\usetikzlibrary{calc}
\usetikzlibrary{intersections}
\usetikzlibrary{through}

\title{Data Acquisition and Image Processing for Solar Irradiance Forecasting}

\author{
 Guillermo Terr\'en-Serrano \\
  Department of Electrical and Computer Engineering \\
  The University of New Mexico \\
  Albuquerque, NM 87131, United States\\
  \texttt{guillermoterren@unm.edu} \\
 \And
  Manel Mart\'inez-Ram\'on \\
  Department of Electrical and Computer Engineering \\
  The University of New Mexico \\
  Albuquerque, NM 87131, United States\\
  \texttt{manel@unm.edu} \\
}

\begin{document}

\maketitle

\begin{abstract}
    The energy available in Micro Grid (MG) that is powered by solar energy is tightly related to the weather conditions in the moment of generation. Very short-term forecast of solar irradiance provides the MG with the capability of automatically controlling the dispatch of energy. To achieve this, we propose a method for statistical quantification of cloud features extracted from long-ware infrared (IR) images to forecast the Clear Sky Index (CSI). The images are obtained using a data acquisition system (DAQ) mounted on a solar tracker. We explain how to remove cyclostationary bias in the data caused by the devices in the own DAQ. We investigate a method to obtain the CSI, after the detrending of Global Horizontal Irradiance (GHI) measurements. We propose a method to fusion multiple exposures of circumsolar visible (VI) light images. We implement a method for extracting physical features using radiometric measurements of the IR camera. We introduce a model to remove from IR images both the effect of the atmosphere scatter radiation, and the effect of the Sun direct radiation. We explain how to model of diffuse radiation of the IR camera window, which is produce by water spots and dust particles stack to the germanium lens of the DAQ enclosure. The frames, that were used to model the camera window, are selected using an atmospheric condition model. This model classifies the sky four different categories: clear, cumulus, stratus, and nimbus. We introduce a geometric transformation of the size of the pixels to their actual dimension in a plane of the atmosphere which is at a given height. This transformation is performed according to the elevation angle of the Sun and field of view (FOV) of the camera. We compare the error between the transformation and an approximation of transformation.
\end{abstract}

\keywords{Sky Imaging \and Machine Learning \and Long-wave Infrared Camera \and Solar Forecasting}

\tableofcontents

\section{Technical Specification}

We built a system composed of Visible (VI) and Infrared (IR) solar radiation cameras, a tracking system, and a pyranometer, to collect the data. A weatherproof enclosure contains the two USB cameras, and it is mounted on top of two servomotors. The cameras are connected to a motherboard that runs a solar tracking algorithm. This motherboard is placed inside of a different watherproof enclosure together with a router, a servomotors control unit, a data storage system, plus their respective power supplies. The enclosure’s degree of protection is IP66. This is an international standard utilized for electronic equipment that provides protection against dust and water.

\subsection{Infrared Sensor}

The IR sensor used to capture the images is a FLIR$\copyright$ Lepton 2.5 camera\footnote{http://www.flir.com} which is mounted on Pure Thermal 1 board\footnote{https://groupgets.com} manufactured and distributed by Group Gets. The Lepton 2.5 sensor produces thermal imaging by measuring long-wave infrared. It captures IR radiation with a nominal response wavelength band from 8 to 14 microns. 
	
The dimensions of a Lepton 2.5 are 8.5 x 11.7 x 5.6 mm. It has $51^{\circ}$ HFOV and $63.5^{\circ}$ diagonal with lens type f1.1 silicon doublet. The resolution is 80 horizontal x 60 vertical active pixels. The thermal sensitivity is $<50$ mK. This module model integrates digital thermal image processing functions that includes automatic thermal environment compensation, noise filter, non-uniformity correction and gain control. This sensor can produce on stream an image in $<0.5$ seconds. It operates at 150 mW nominally, and has a low power standby mode. The Pure Thermal 1 board adds functionalities to the IR sensor such as thermal video transmission over USB which works with USB Video Class (UVC) library on Windows, Linux, Mac and Android. It also includes on-board image processing without needing for an external system. The microprocessor is a STM32F411CEU6 ARM. It also includes a thermopile contactless temperature sensor to manually calibrate a Lepton module. The operational temperatures for Flat Filed Correction (FFC) function range from $-10^{\circ}$C to $65^{\circ}$C. The captured images are digitalized in Y16 format. It is a single channel format that only quantifies intensity levels.

\begin{figure}[!htbp]
    \begin{subfigure}{0.3275\textwidth}
        \centering
        \includegraphics[scale = 0.15]{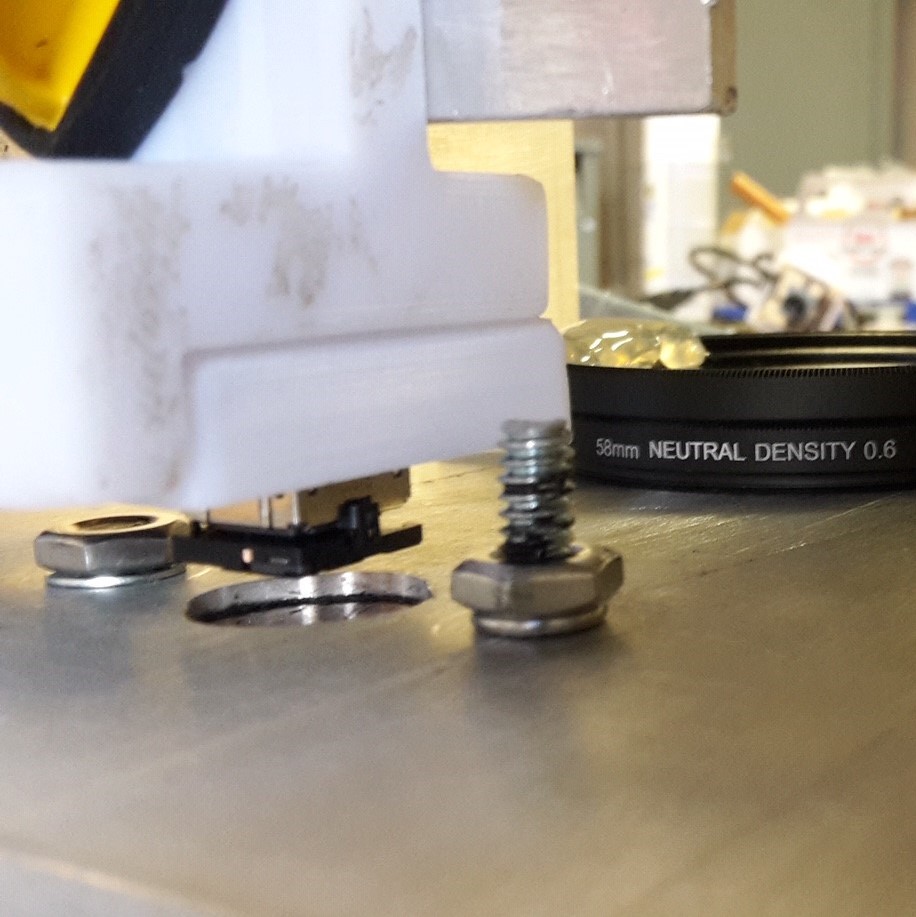}
    \end{subfigure}
    \begin{subfigure}{0.3275\textwidth}
        \centering
        \includegraphics[scale = 0.15]{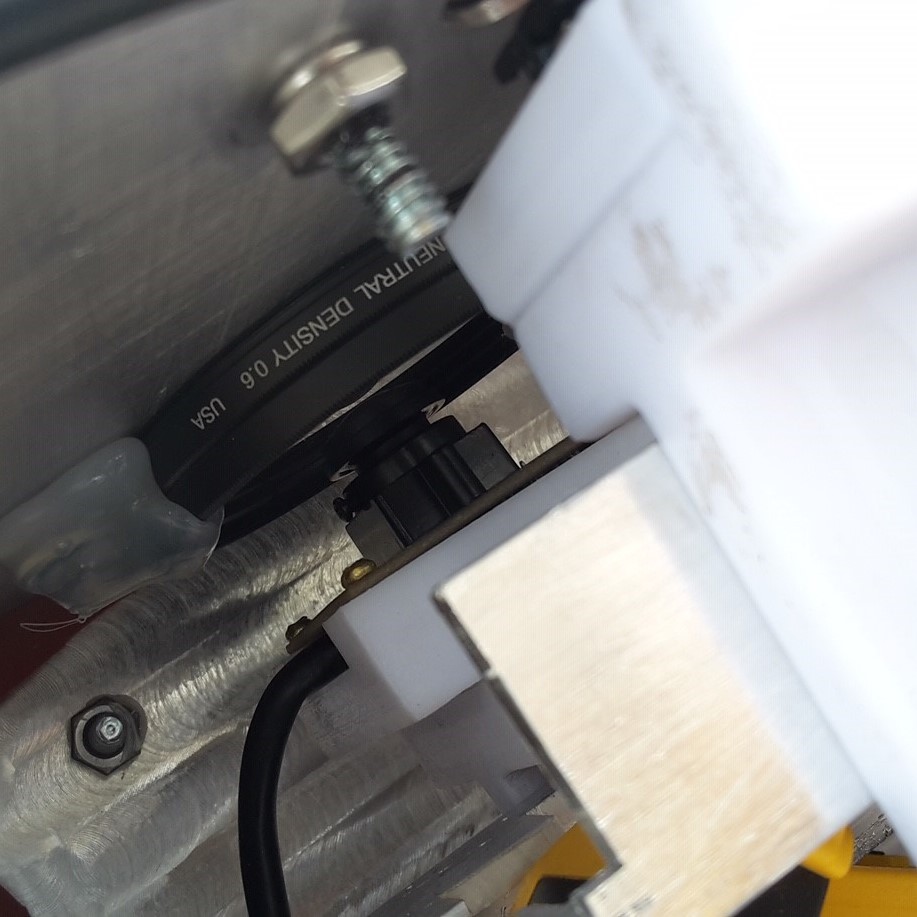}
    \end{subfigure}
    \begin{subfigure}{0.3275\textwidth}
        \centering
        \includegraphics[scale = 0.15]{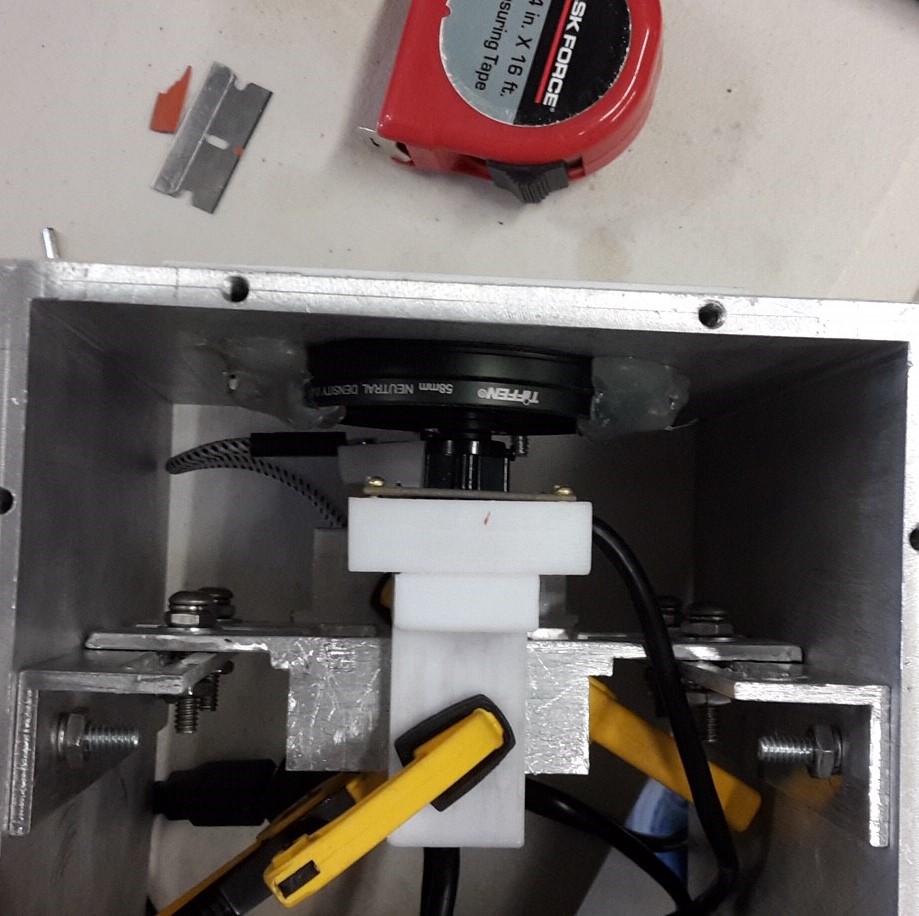}
    \end{subfigure}
    \caption{Details of IR and VI cameras screwed on their individual 3D printed support inside the custom made enclosure, and placed in front of their respective apertures. The supports are fixed to the enclosure's structured by means of adjustable clamps.}
\label{fig:interior}
\end{figure}

\subsection{Visible Sensor}

The VI camera is a 5 megapixels USB 2.0 color sensor manufacture by ELP$\copyright$. The sensor is an OV5640 with maximum resolution of 2592 horizontal x 1944 vertical pixels and $170^{\circ}$ FOV with fisheye type lens that is adjustable within the range of 2.1 to 6 mm. The pixel’s size is 1.4 x 1.4 microns, and the image area is 3673.6 x 2738.4 microns. The image stream rate is 30 frames per seconds at 640 x 480 pixels’ resolution and using MJPEG compression format. The communication protocol is UVC and its interface is USB 2.0 high speed. The dynamic range is 68 dB, and it has a mount-in shutter that can control frame’s exposure time. The camera-board has built-in functions that can be enable for Automatic Gain Control (AGC), Automatic Exposure Control (AEC), Automatic White Balance (AWB) and Automatic Black Focus (ABF). The sensor's brightness, contrast, hue, saturation, sharpness, gamma, white balance, exposure and focus can be adjusted by software. Power consumption in VGA resolution is 150mW. The dimensions of camera board are 38 x 38 mm. The recommended operational temperatures for stable images range from $0^{\circ}$C to $60^{\circ}$C. The camera's output format is YUYV for images and MJPEG for video. YUYV is 3 channels format in the color space of YUV format which represent luminance intensity, blue color and red color, respectively. MJPEG is lossy compression format. UVC drivers can operate in Windows, Linux, MAC and Android.

\begin{figure}[!htbp]
    \begin{subfigure}{0.3275\textwidth}
        \centering
        \includegraphics[scale = 0.1625]{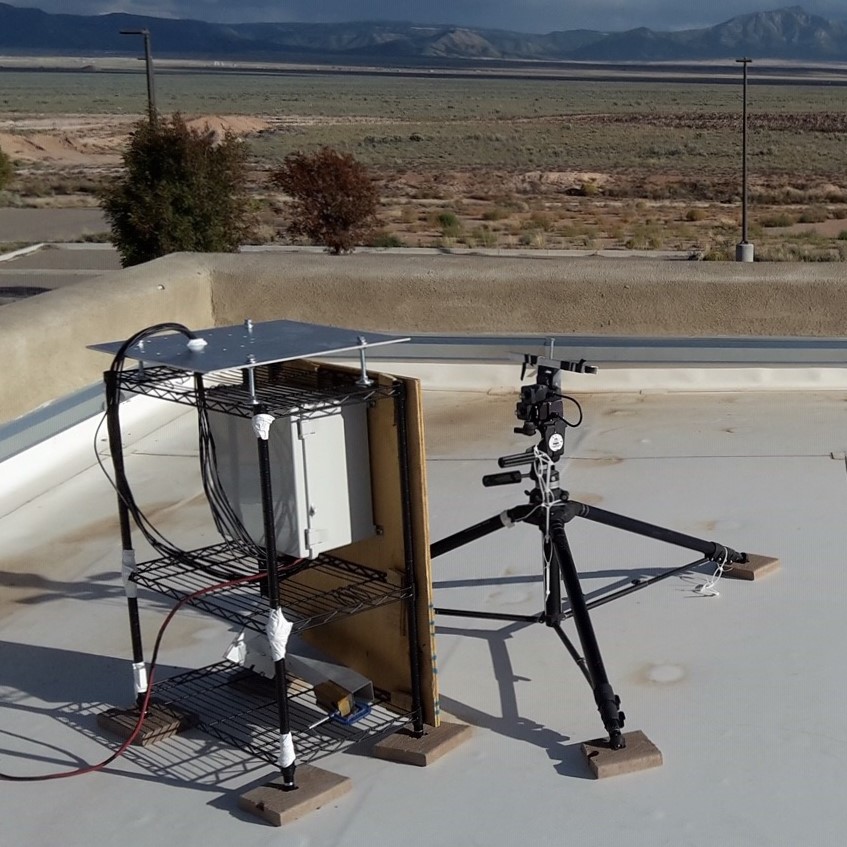}
    \end{subfigure}
    \begin{subfigure}{0.3275\textwidth}
        \centering
        \includegraphics[scale = 0.178]{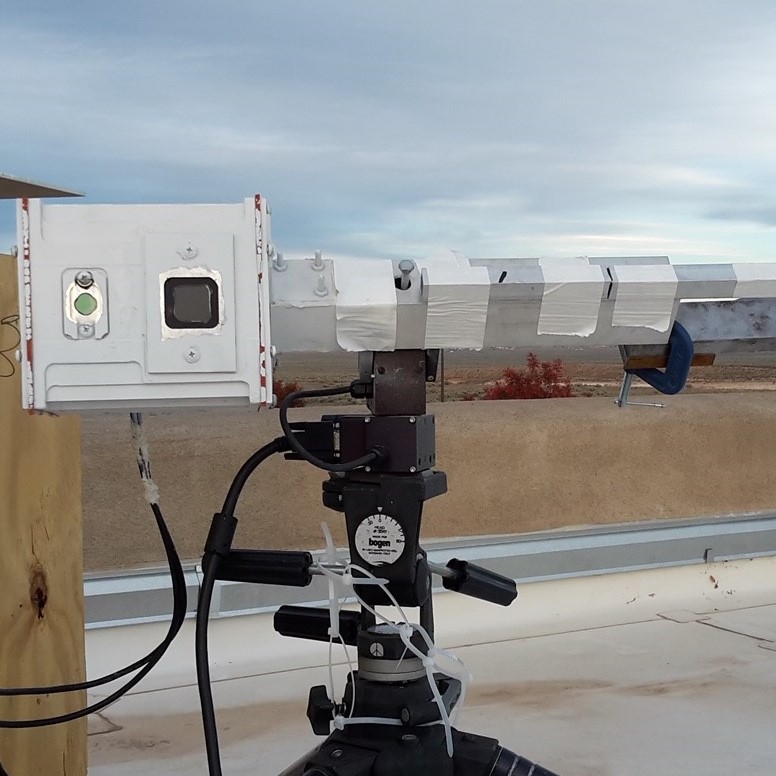}
    \end{subfigure}
    \begin{subfigure}{0.3275\textwidth}
        \centering
        \includegraphics[scale = 0.15]{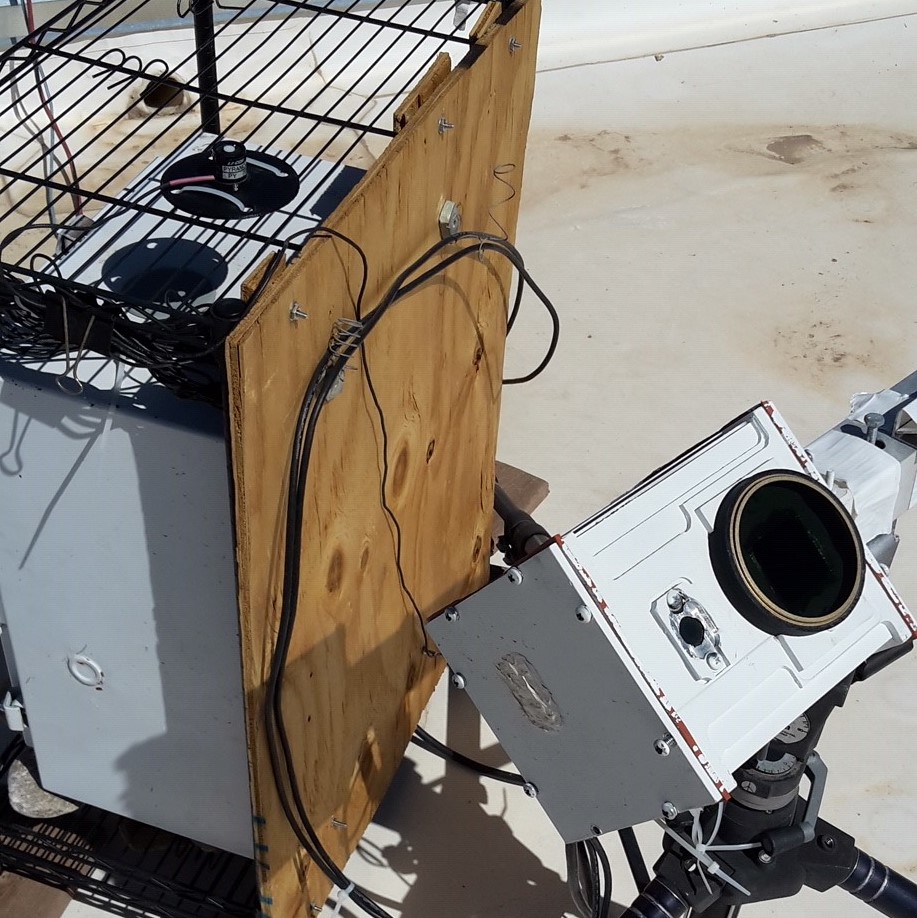}
    \end{subfigure}
    \caption{Tripod and junction box installed on the roof area of an UNM's facility. Solar tracker's servo motors mounted on the tripod, and screwed to a metallic structure which counterbalances the own enclosure's weight. The pyranometer is adjusted to a 3D printed support placed on top of the junction box's rack. Germanium and neutral density lenses for the IR and VI camera respectively, to filter the light beams and protect the cameras from weather hazards.}
\label{fig:exterior}
\end{figure}

\subsection{Solar Tracker}
The device used for the tracking is a commercial model manufacture by FLIR$\copyright$ and distributed by moviTHERM. The model is PTU-E46-70. The tracking system is driven by two servo motors mounted to allow rotation on tilt and pan axes. The tracker is rated for a payload of 4.08 kg. Hence, the entire system must not weight more than this rate. Rotational speed of the tracker is $60^{\circ}$ per second, and it has a resolution of $0.003^{\circ}$.
	
The degrees of freedom of the tilt servomotor allows vertical maneuvers of $78^{\circ}$ with vertical limit from the horizontal of $-47^{\circ}$ to $+31^{\circ}$. The pan servomotor has more degrees of freedom. This allows tracking of the sun throughout a day. Its range is $\pm 159^{\circ}$. The motors’ velocity is adjustable up to $0.003^{\circ}$ per second. The tracker requires an input voltage range of 12-30 VDC. It consumes 13 W in full power mode, 6 W low-power mode and 1 W in holding power off mode. The system is controlled via Ethernet from a host computer. The control unit is connected to the servomotors via a DB-9 female connector.
	
Regarding to the mechanical description of the tracker. The system weight is 1.36 kg, its dimensions are 7.6 cm height x 13 cm width x 10.17 cm depth. The control unit weight 0.227 kg, and its dimensions are 3.18 cm height x 8.26 cm width x 11.43 cm depth.
	
\subsection{Pyranometer}

As part of the development, a pyranometer sensor has been designed and tested to provide a device with high portability and accuracy. Here it is presented the circuit design for a signal conditioning of the pyranometer sensor model LI-COR LI-200, and it is summarized the architecture, components and software. The prototype was designed to send data to a laptop, Raspberry Pi or any other computer through USB communications, to which a simple python script needs to be added and executed. For portability purposes, the designed board is provided with a 5 V power supply, so that Raspberry Pi can be also connected to it. 
	
The design objectives were to condition and measure the signal of a pyranometer sensor by converting it to a 12-bits digital signal and saving the data by timestamps in a csv file. This is done in a daily basis between 5 am to 9 pm. The system must be portable, easy to setup, and reliable. 
	
The pyranometer output voltage range from 0 to 20 mV, where 10 mV represent a radiation power density of 1,000 $W/m^2$. A 120 V power outlet is available to power the device. The system must be connected to the internet and it has logging capabilities. 
	
\begin{figure}[!htbp]
    \centering
	\includegraphics[width=0.95\linewidth]{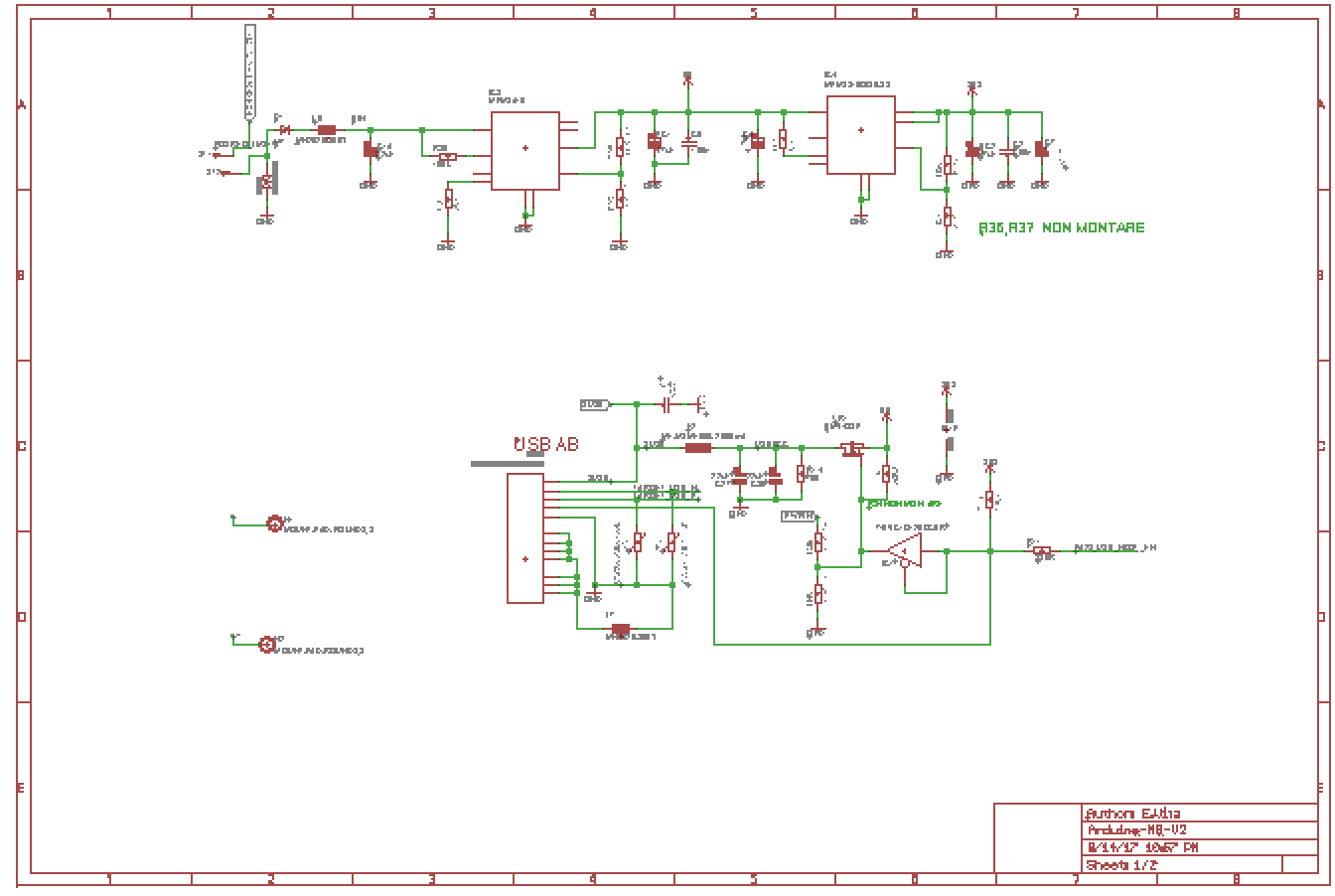}
	\includegraphics[width=0.95\linewidth]{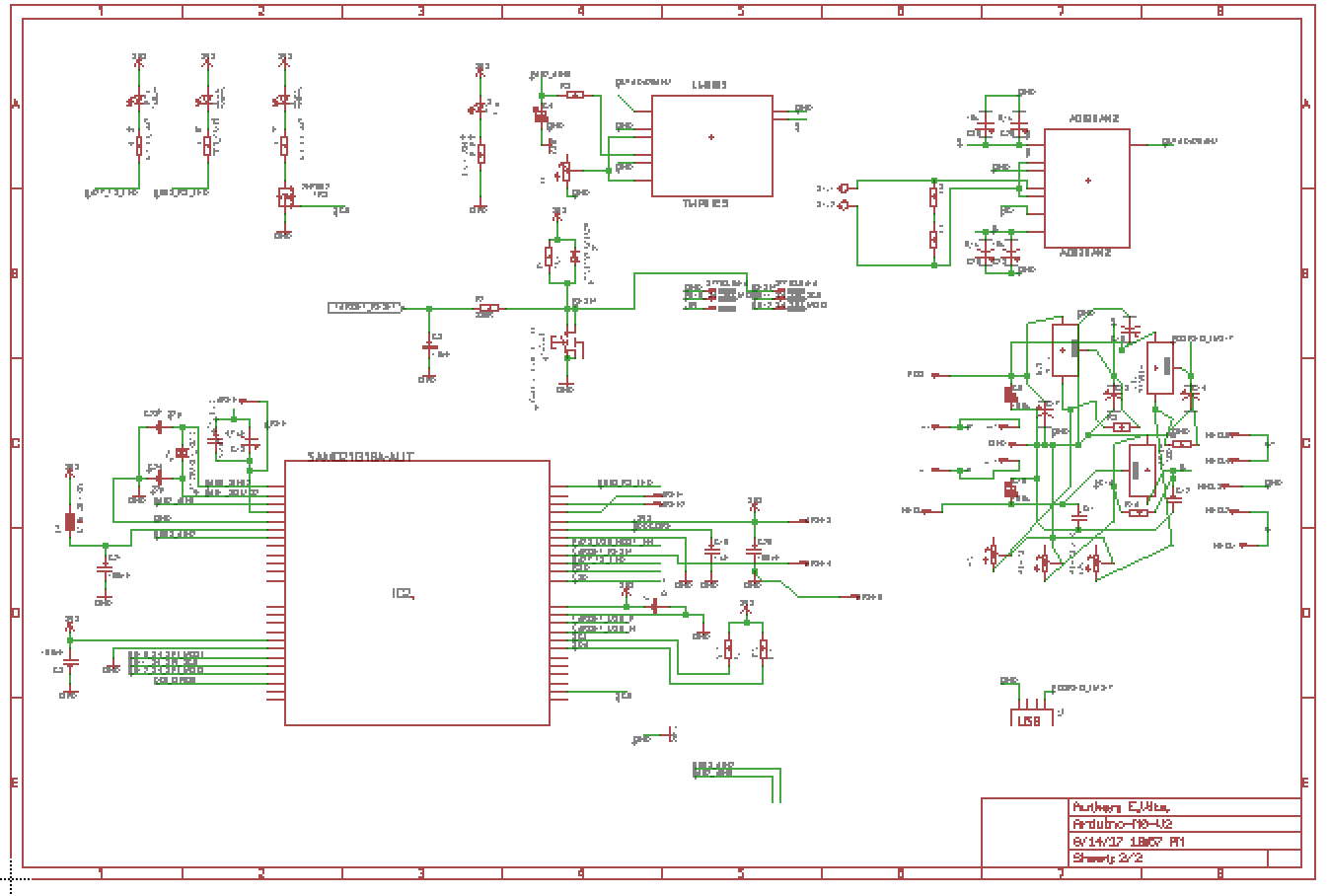}
	\caption{ADC circuit’s schematics. The drawings show the connections for the all parts in the circuit. The analog circuit which measures the voltage level of the pyranometer is plotted on the top picture. On the bottom picture details the interconnections from the microprocessor to all circuit parts.}
	\label{fig:circuit_schematics}
\end{figure}

The designed solution uses a differential amplifier with 13 dB gain that conditions the signal to a DAC. A microprocessor is used to measure the signal output from the ADC, and transmits it by USB port. The board is connected to a Raspberry Pi to save the signal's data in files. This Raspberry Pi is connected to the internet, so it is also able to transmit the data. The circuit schematics are shown in Figure \ref{fig:circuit_schematics}.
	
\section{Data Acquisition}

The tracking system and DAQ software is currently operative. The software was programmed in single Python 2.7 script. The system is placed on the top roof area of the Mechanical Engineering building at UNM central campus (35.0821, -106.6259). The DAQ sessions can be visually monitored through a webpage. This webpage contained real-time data acquired from IR and VI sensors, the Sun's position algorithm and measures from the pyranometer. All devices were interconnected via a LAN network built by DHCP server.

\subsection{Sun Position}
The sun position is calculated to update the servomotors’ position every second. The following set of equations was used to calculate azimuth and elevation angles\footnote{http://www.pveducation.org/}. These angles correspond to the servomotors’ pan and tilt axes. The tracking system must be aligned to the true North or South depending local, geographic coordinates define as $\left( \lambda, \phi \right)$, which are longitude $\lambda$, and latitude $\phi$. It also must be leveled with respect to the horizontal and vertical axis. The first step is to calculate the Local Standard Time Meridian (LSTM) which is the reference sun time between the Greenwich meridian and another time zone. LSTM is calculated per the following equation:	
\begin{equation} 
	\mathbf{LSTM} = 15^\circ \cdot \Delta T_{GTM},
\end{equation}	
where $\Delta T_{GMT}$ is the difference of the Local Time (LT) from Greenwich Mean Time (GMT) and it is measured in hours. Equation of Time (EoT) is an empirical equation that considers the eccentricity of earth's orbit and its axial tilt. It is expressed in minutes and is described by,
\begin{equation} 
	\mathbf{EoT} = 9.87  \cdot  \sin(2\mathbf{B}) - 7.53 \cdot  \cos(\mathbf{G}) - 1.5  \cdot \sin(\mathbf{B}),
\end{equation}
where,
\begin{equation} 
	\mathbf{B} = \frac{360}{365} \cdot \left( d - 81 \right).
\end{equation}
	
$\mathbf{B}$ units are degrees and $d$ corresponds to the days since the beginning of the year. The variation of the LST in a local effect due to a different longitude is account by the Time Correction Factor (TC), is which quantified in minutes, and its equation is
\begin{equation} 
	\mathbf{TC} = 4 \cdot \left( \lambda - \mathbf{LSTM} \right) + \mathbf{EoT},
\end{equation}
the constant 4 is related to the rotation of the earth, which rotates $1^{\circ}$ per 4 minute, and $\lambda$ is the longitude. Local Solar Time (LST) is calculated by using the previous correction and Local Time (LT). Following equation is expressed in hours
\begin{equation} 
	\mathbf{LST} = \mathbf{LT} + \frac{\mathbf{TC}}{60}.
\end{equation}
	
The Earth rotates $15^{\circ}$ per hour, so each hour away from the solar noon corresponds to an angular position. The Hour Angle (HRA) converts LST to the degrees that the sun moves on the sky. Before noon, angles are negative and after, the angles obtained are positive. The expression for this is:
\begin{equation} 
	\mathbf{HRA} = 15^\circ \cdot \left( \mathbf{LST} - 12 \right).
\end{equation}
	
The Sun's declination angle $\delta$ varies seasonally, and it accounts for the angular difference between the equator and the Earth's center. $\delta$ is obtained from this formula,
\begin{equation} 
	\delta = 23.45^\circ \cdot \sin \left[ \frac{360}{365} \cdot \left( d - 8 \right) \right],
\end{equation}
where $d$ represents the day since the beginning of the year. The rotations on the tracker’s pan and tilt axes are given by Elevation $\epsilon$ and Azimuth $\alpha$ angles. The Elevation angle quantifies the Sun's angular height, and the azimuth $\alpha$ is the Sun's angular position on the horizon. The angles are calculate from these equation,
\begin{align} 
	\epsilon &= \sin^{-1} \left[\sin\delta  \cdot \sin\phi + \delta \cdot \cos\phi \cdot \cos(\mathbf{HRA}) \right], \\
	\alpha &= \cos^{-1} \left[ \frac{\sin\delta \cdot \cos\varphi - \cos\delta \cdot \sin\varphi \cdot \cos(\mathbf{HRA})}{\cos \xi } \right],
\end{align}
where $\mathbf{HRA}$ is the hour angle, $\phi$ is latitude, and 
\begin{align} 
\xi = \sin^{-1} \left[ \sin\delta  \cdot \sin\phi + \delta \cdot \cos\phi \cdot \cos(\mathbf{HRA}) \right].
\end{align}
Alternatively, $\xi$ can be also calculated such as $\xi = 90^\circ + \phi - \delta$. The Zenith angle $\zeta$ is the difference between the Elevation angle $\epsilon$ and the vertical, so that
\begin{equation} 
	\zeta = 90^\circ - \epsilon.
\end{equation}
	
The tracking system initializes an hour after sunrise and stops an hour before sunset. This considers that sun’s elevation is too low to generate energy by PV systems. Inverters require a minimum consumption of power which is reached when it is above and below these time limits. The sunrise $\rho$ and sunset $\sigma$ are expressed in decimal hours and they are given by following equations,
\begin{align}
	\rho &= 12 - \frac{1}{15^\circ}  \cdot\cos^{-1} \left(\frac{-\sin\phi \cdot \sin\delta}{\cos\phi \cdot \cos\delta}\right) - \frac{ \mathbf{TC} }{60}, \\
	\sigma &= 12 + \frac{1}{15^\circ} \cdot \cos^{-1} \left(\frac{-\sin\phi \cdot \sin\delta}{\cos\phi  \cdot \cos\delta} \right) - \frac{ \mathbf{TC} }{60},
\end{align}
where $\mathbf{TC}$ is the time correction.

\subsection{Noise Attenuation}

The captures from the IR cameras are noisy. So, we propose to attenuate the noise by averaging $N$ consecutive frames taken at capture $k$. Our cameras fps are lower than the computer socket reading speed, so we need to assure that the frame $\mathbf{x}^f$ read from the buffer is new, or it exits in our set. We define a capture as a set of frames $\mathcal{X} = \emptyset$. When we retrieve the first $\mathbf{x}^f$, we add it to our set $\mathcal{X} = \{\mathbf{x}^f\}$. We now read new $\mathbf{x}^f$, and to find out if our frame $\exists  \mathbf{x}^f \in \mathcal{X}$, we calculate the normalized cross-correlation $\rho_{\mathbf{x}_i, \mathbf{x}^f}$ between each frames $\mathbf{x}_i \in \mathcal{X}$, and $\mathbf{x}^f$. Then, if $\{\rho_{\mathbf{x}_i, \mathbf{x}^f}\}_{i = 1}^f \neq 1, \ \mathcal{X} = \mathbf{x}_i \cup \mathbf{x}^f, \ \forall i = \{1, \dots, f\}$. We retrieve frames $\mathbf{x}^f$ until $f = N$. The formula for the normalize cross-correlation $\rho_{\mathbf{x}_i, \mathbf{x}^f}$ is
\begin{equation}
\{\rho_{\mathbf{x}_i, \mathbf{x}^f}\}_{i = 1}^f  = \left\{ \frac{  \left(\mathbf{x}_i - \bar{\mathbf{x}}_i \right) \cdot \left(\mathbf{x}^f -  \bar{\mathbf{x}}^f \right)}{ \sqrt{\left(\mathbf{\mathbf{x}}_i - \bar{\mathbf{x}}_i\right)^2} \cdot \sqrt{\left(\mathbf{x}^f - \bar{\mathbf{x}}^f\right)^2}} \right\}_{i = 1}^f,
\end{equation}
where $\bar{\mathbf{x}} = \frac{1}{ND} \cdot \sum_{i = 1}^N \sum_{j = 1}^D x_{ij}$, which is the variable sample mean. Now, we define $\mathbf{I}^k_i$ as the set of different frames from capture $k$ such as
\begin{equation}
\mathbf{I}^k_i \in \mathbb{N}^{[1, 2^{16}]},\quad \forall i = 1, \dots, N.
\end{equation}

We average them together to attenuate the noise such as $ \bar{\mathbf{I}}^k = \frac{1}{N} \cdot \sum_{i = 1}^N \mathbf{I}^k_i $. So now, we have a single frame that is defined as $\bar{\mathbf{I}}^k\in \mathbb{N}^{[1, 2^{16}]}$. 
\begin{figure}[!htbp]
    \begin{subfigure}{0.245\textwidth}
        \centering
        \includegraphics[scale = 0.135]{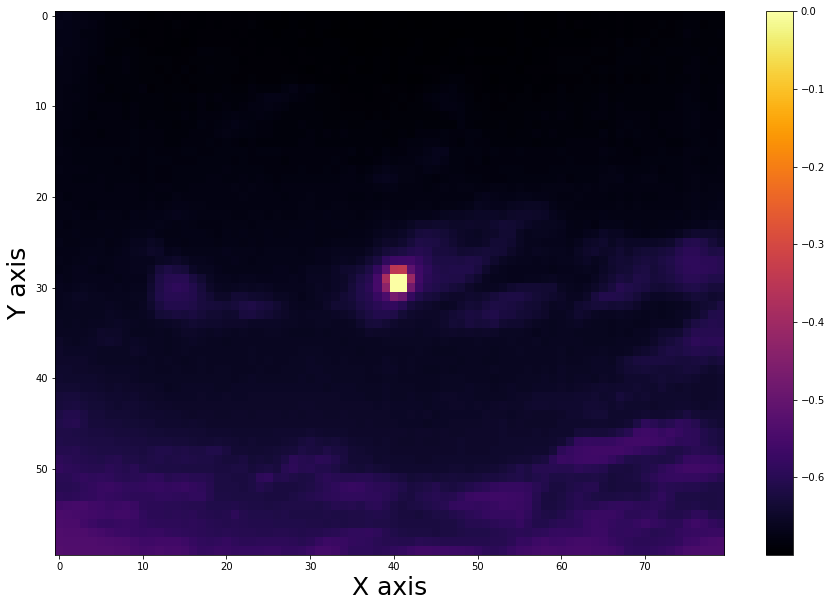}
    \end{subfigure}
    \begin{subfigure}{0.245\textwidth}
        \centering
        \includegraphics[scale = 0.135]{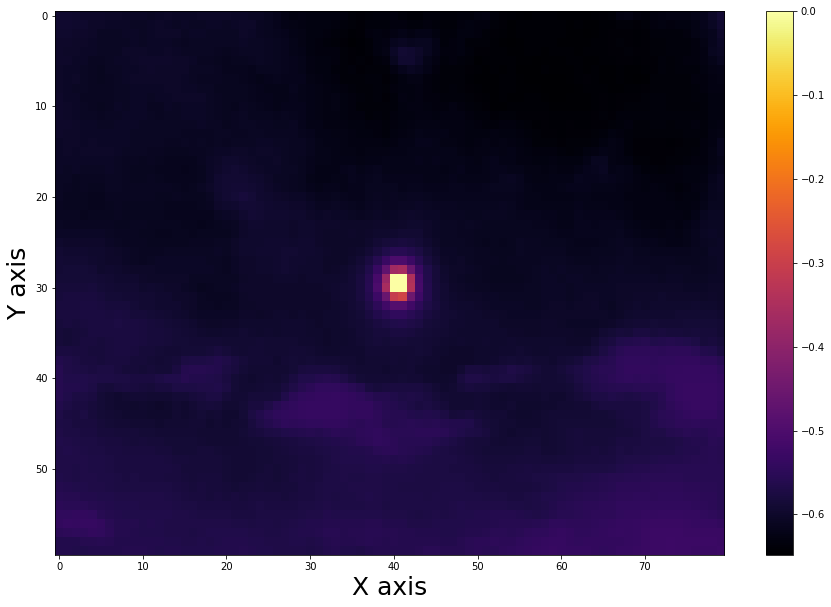}
    \end{subfigure}
    \begin{subfigure}{0.245\textwidth}
        \centering
        \includegraphics[scale = 0.135]{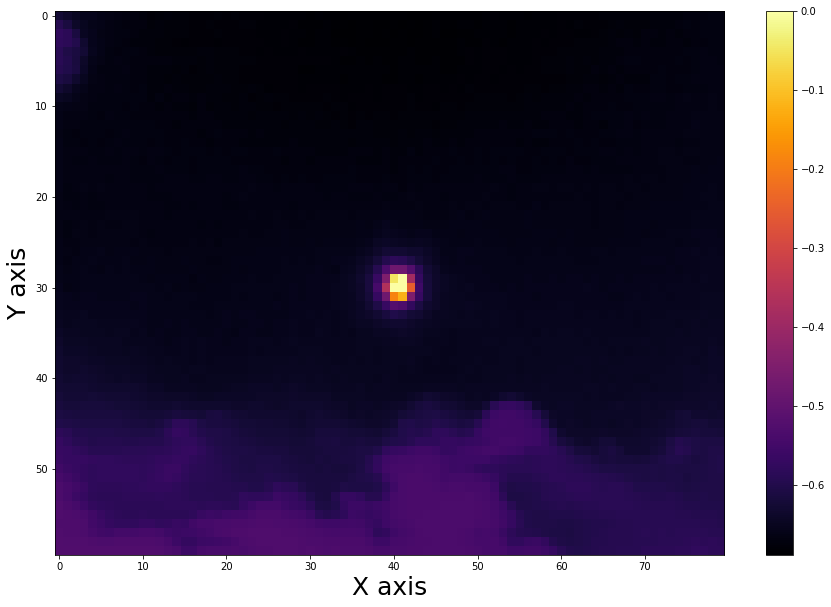}
    \end{subfigure}
    \begin{subfigure}{0.245\textwidth}
        \centering
        \includegraphics[scale = 0.135]{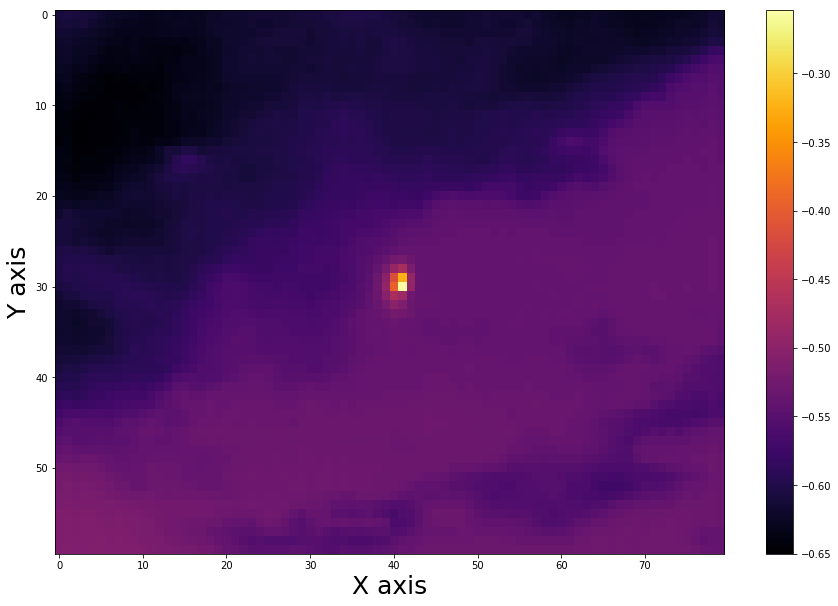}
    \end{subfigure}
    \caption{This figure shows the infrared images after applying the noise attenuation algorithm. The intensity of the pixels in the images is displayed in log-scale. The resolution of the images is $60 \times 80$.}
\label{fig:infrared}
\end{figure}

\subsection{Exposure Switching}

The scattering effect produced by the Sun radiation in the VI images avoids to detect the clouds in the circumsolar region, when the exposition time in capture is large. The light in the outer region of the image is dim when the exposition time is short. In order to sort this problem, we propose to switch the exposition time in each VI capture $\mathbf{x}^f$ at the same time that we attenuate the noise as we did with the IR captures. So, we have a set of $N$ frames with $E$ exposure times for each VI capture $k$ such as
\begin{equation}
\mathbf{I}^k_{i,j} \in \mathbb{N}^{[1, 2^8]},\quad \forall i = 1, \dots, N,\quad \forall j = 1, \dots, E.
\end{equation}

We average the $N$ frames for each $E$ exposition time to attenuate the noise
$ \bar{\mathbf{I}}^k_j = \frac{1}{N} \cdot \sum_{i = 1}^N \mathbf{I}^k_{i,e} $. And, we obtain $E$ frames $\bar{\mathbf{I}}^k_j\in \mathbb{N}^{[1, 2^8]}$ per VI capture. 

\section{Data Processing}

The pyranometer's measurements has two error that we name as amplitude and shifting errors. The amplitude error is due to the calibration of the device, the pyranometer measures GSI at a higher $W/m^2$ than the real GSI. The shifting error is due to inclination of the pyranometer's support platform. The pyranometer Li-200 is calibrated with incidence angle correction, so that functionality we were capable of recover the days that has a shifting in the measurements because of the artificial increment in real incidence angle that the support inclination produces.

\subsection{Physical Model of Global Solar Irradiance}

The GSI on a flat surface normal to the ground has 3 components \cite{MASTERS2004},
\begin{align}
    \mathcal{I}_{GSI}  = \mathcal{I}_{Direct} + \mathcal{I}_{Diffuse} + \mathcal{I}_{Reflected}.
\end{align}

The Direct component, when the body of reference is normal to the ground, is a function of the Sun's elevation angle $\epsilon$,
\begin{align}
    \mathcal{I}_{Direct} = \mathcal{I}_{DN} \cdot \sin \epsilon,
\end{align}
and the Direct Normal (DN) component, that is
\begin{align}
    \mathcal{I}_{DN} = \theta_1 \cdot \exp \left( - \frac{\theta_2}{\sin \epsilon} \right),
\end{align}
$\theta_1$ is the surface altitude with respect to the see level, and $\theta_2$ is the air mass coefficients.

The Diffuse component is the solar irradiance scattered by hitting particles in the atmosphere, as the water droplets forming clouds. The expression for a flat surface normal to the ground is
\begin{align}
    \mathcal{I}_{Diffuse} = \theta_3 \cdot \frac{\mathcal{I}_{DN}}{2}.
\end{align}

The coefficient $\theta_3$ defines the proportion of irradiance that is scattered for an atmospheric condition.

The Reflected component depends on the surfaces tilt angle. For our case, as the tilt angle is $0^{\circ}$, that means that is a flat surface normal to the ground, the expression is
\begin{align}
    \mathcal{I}_{Reflected} = \theta_4 \cdot \mathcal{I}_{DN} \cdot \left( \theta_3\ + \sin \epsilon \right),
\end{align}
$\theta_4$ is the reflective coefficient of the ground material where the surface is placed normal to it.

Finally, we can put together this three equations and obtained the solution for $\mathcal{I}_{GSI}$ as function of the coefficients set $\boldsymbol{\theta}^{(1)} = \{ \theta_1, \dots, \theta_4\} \in \mathbb{R}$ and the Sun elevation angle $\epsilon$, as
\begin{align}
    \mathcal{I}_{GSI} \left( \boldsymbol{\theta}^{(1)}, \epsilon \right) = \mathcal{I}_{DN} \cdot \sin \epsilon + \theta_3 \cdot \frac{\mathcal{I}_{DN}}{2} + \theta_4 \cdot \mathcal{I}_{DN} \cdot \left( \theta_3\ + \sin \epsilon \right),
\end{align}
where $\mathcal{I}_{DN} = \theta_1 \cdot \exp \left( - \frac{\theta_2}{\sin \epsilon} \right)$.

Now, we obtain these coefficients $\boldsymbol{\theta}^{(1)}$ from a set of theoretical formulae,  
\begin{align}
    \theta_1 &= 1160 + 75 \cdot \sin \left( \frac{360}{N} \cdot \left( d - 275 \right) \right), \\
    \theta_2 &= 0.174 + 0.035 \cdot \sin \left( \frac{360}{N} \cdot \left( d - 27 \right) \right), \\
    \theta_3 &= 0.095 + 0.4 \cdot \sin \left( \frac{360}{N} \cdot \left( d - 27 \right) \right), 
\end{align}
that are in function of the year day $d$, and the days on a year $N$ \cite{MASTERS2004}. The coefficient $\theta_4$ amounts for the reflected radiation. As our pyranometer is mounted on a black surface, we assume that $\mathcal{I}_{reflected} = 0$, for that we set $\theta_4 = 0$.

\begin{figure}[!htbp]
    \begin{subfigure}{0.49\textwidth}
        \centering
        \includegraphics[scale = 0.225]{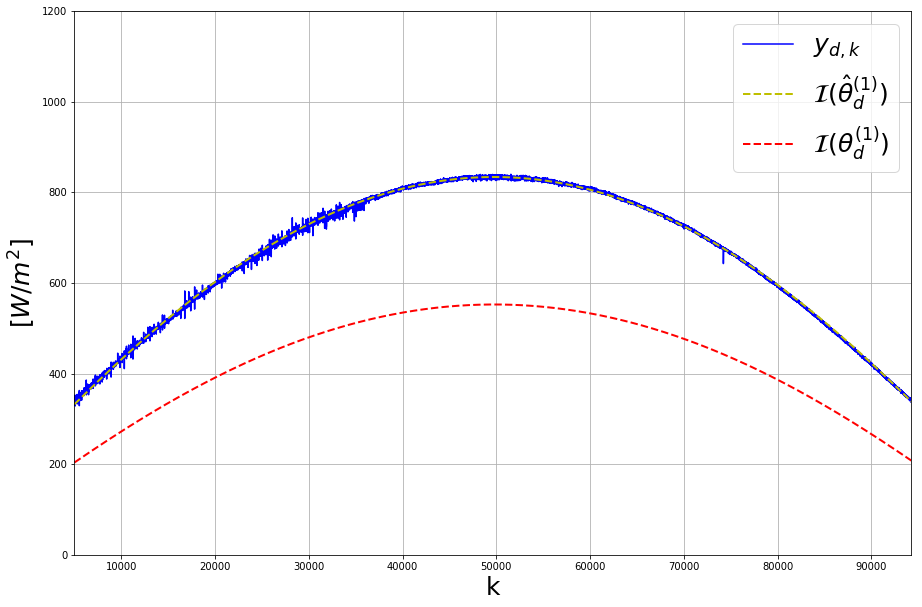}
        \includegraphics[scale = 0.225]{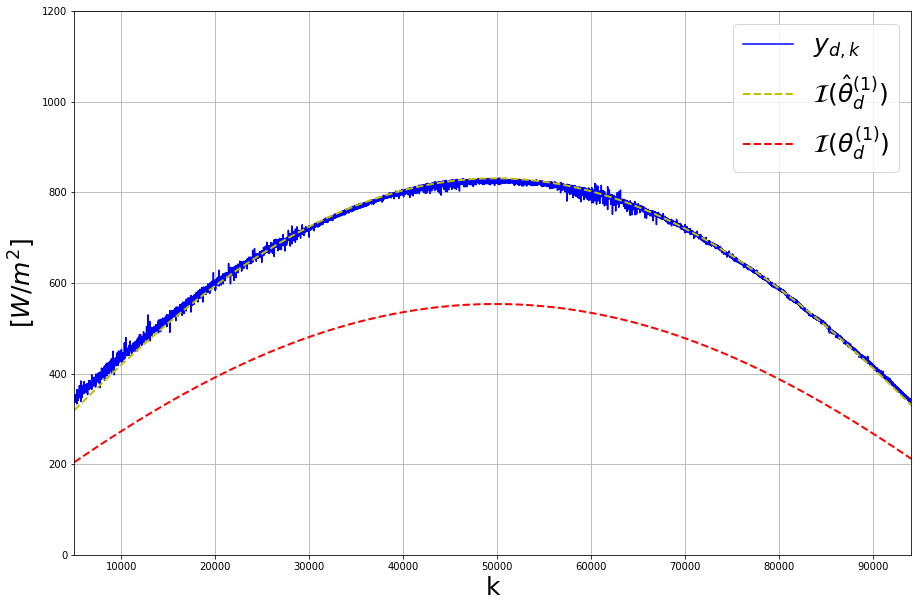}
        \includegraphics[scale = 0.225]{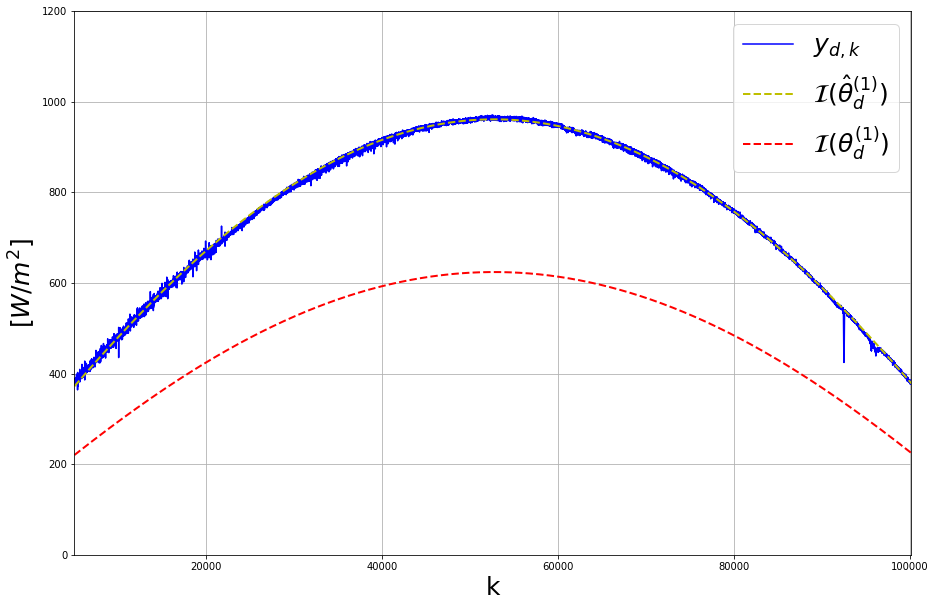}
        \includegraphics[scale = 0.225]{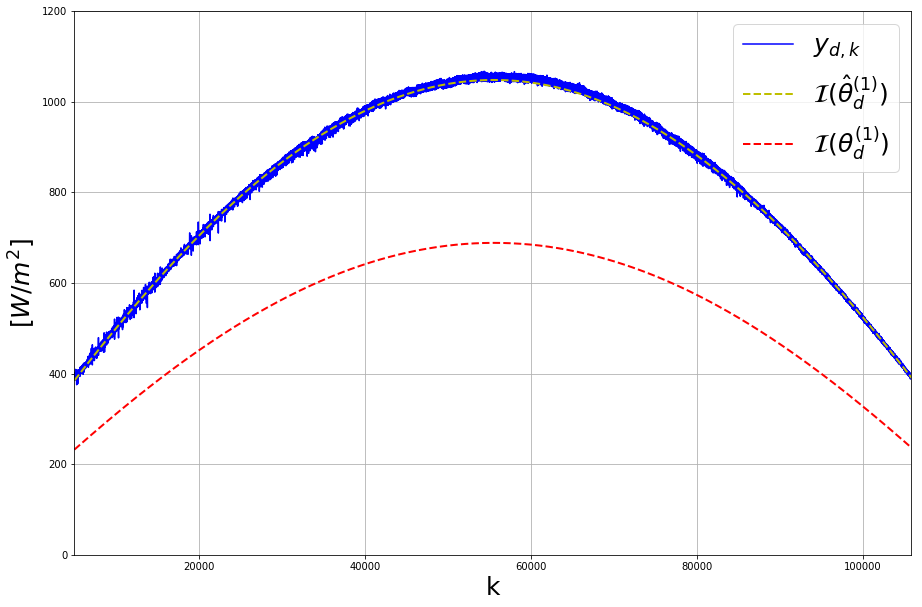}
    \end{subfigure}
    \begin{subfigure}{0.49\textwidth}
        \centering
        \includegraphics[scale = 0.225]{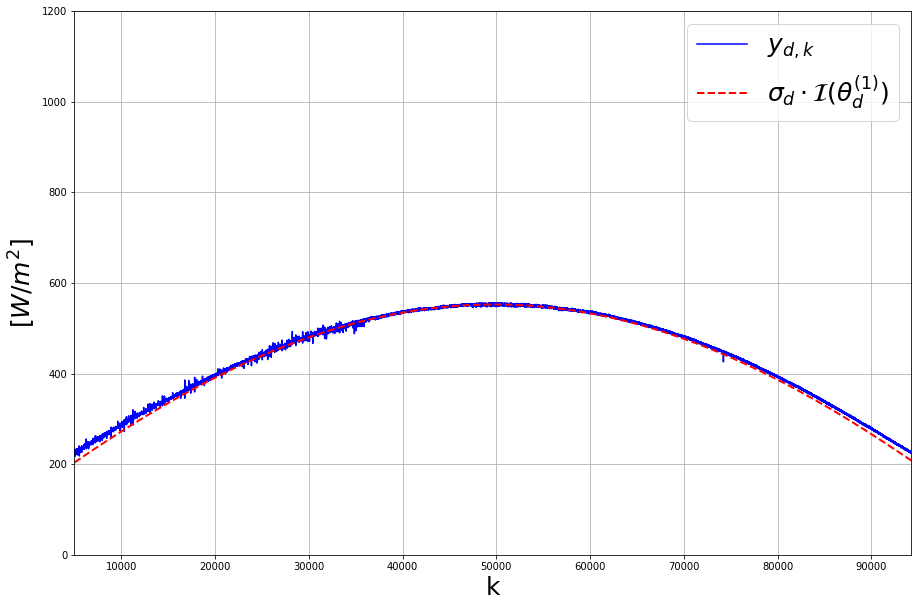}
        \includegraphics[scale = 0.225]{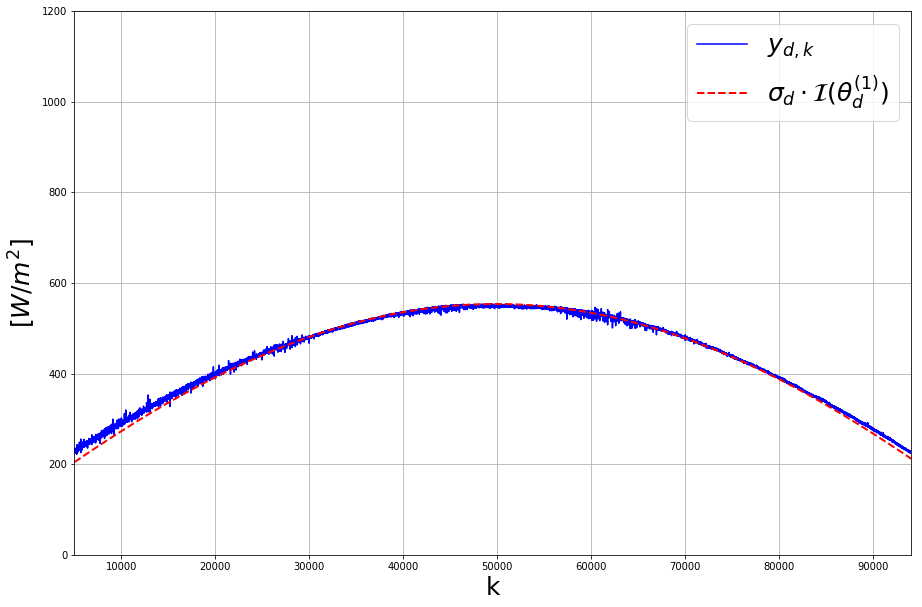}
        \includegraphics[scale = 0.225]{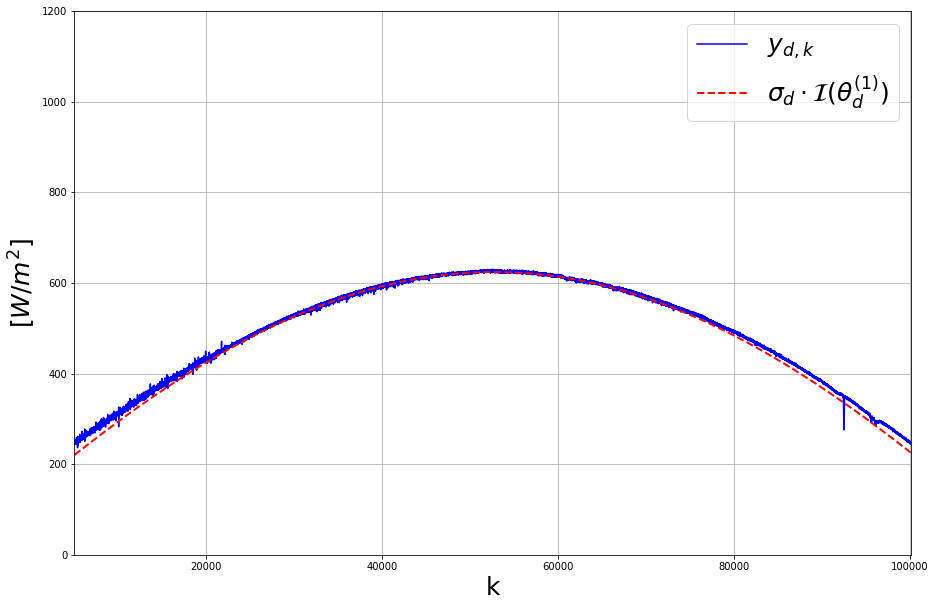}
        \includegraphics[scale = 0.225]{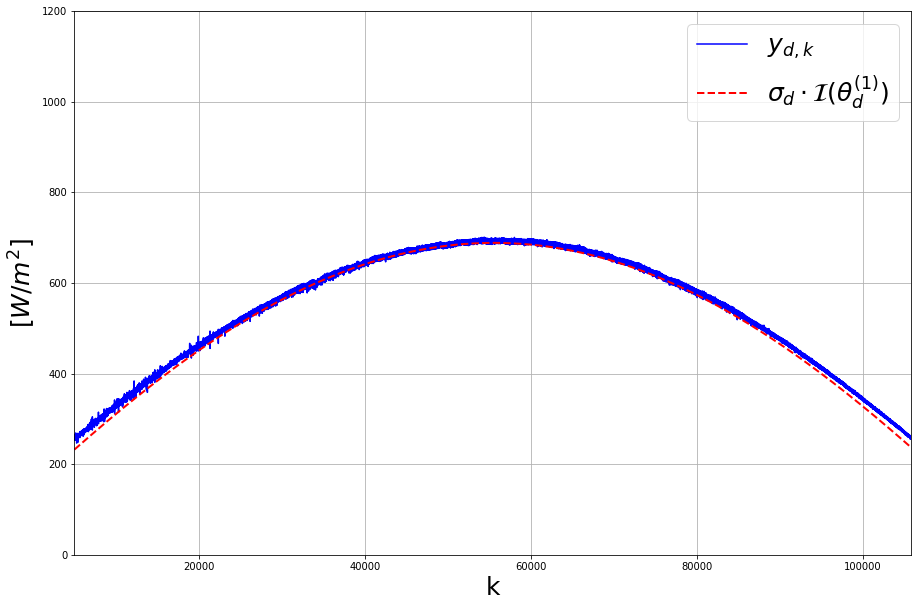}
    \end{subfigure}
    \caption{The graphs on the left column shows the GSI measurements from the pyranometer, the empirical coefficients in yellow, and the theoretical coefficients in red. The column on the right shows the GSI measurements, and theoretical model amplified by a $\sigma_d$.}
\label{fig:amplitude_samples}
\end{figure}

\subsection{Pyranometer Bias Correction}

\begin{itemize}

\item Pyranometer Amplitude Adjustment 

The amplitude of the pyranometer's measurements does not match the theoretical solar radiation and the measures are are noisy at the maximum amplitude point, see Figure \ref{fig:amplitude_samples}. In order to overcome to the problem of identifying this point, we propose to optimize the theoretical model to fit the measurements instead, this optimization problem is described in Physical Model Parameters' Adjustment. In this way, we will obtain two smooth irradiance functions but with different amplitudes. We proceed to calculate the attenuation in the amplitude between the function evaluated with the theoretical coefficients $\boldsymbol{\theta}^{(1)}_d$, and the function evaluated with the optimized model coefficients $\hat{\boldsymbol{\theta}}^{(1)}_d$,
\begin{align}
     \sigma_d = \frac{ \max \left[ \mathcal{I}_{GSI} \left( \hat{\boldsymbol{\theta}}^{(1)}_d, \boldsymbol{\epsilon}_d \right) \right]}{\max \left[ \mathcal{I}_{GSI} \left( \boldsymbol{\theta}^{(1)}_d, \boldsymbol{\epsilon}_d \right) \right]}, \quad \forall d =  1, \dots, D, \quad \sigma_d \in \mathbb{R}^{[1, 2]}.
\end{align}

The amplitudes samples $\sigma_d$ are the dots in blue color in the left graph of figure \ref{fig:bias}. We can observe that the amplitude bias follows a periodic function, so that we propose to model the bias by cycle-stationary sinusoidal function that is,
\begin{align}
    \mathcal{C} \left( \boldsymbol{\theta}^{(2)}, d \right) = \theta_1 \cdot \sin \left( \theta_2 + \frac{2\pi d}{N}\right) + \theta_3,
\end{align}
where $N$ is the days on a year, $d$ is the day of the year, and $\boldsymbol{\theta}^{(2)} = \{ \theta_1, \theta_2, \theta_3 \} \in \mathbb{R}$ is the parameters' set.

We propose to correct the pyranometer's measurements such as,
\begin{align}
    \mathcal{R}_{GSI} \left(\boldsymbol{\theta}^{(1)}_d, \epsilon_{d, k} \right) = \frac{y_{d, k}}{\sigma_d}, \quad y_{d, k}  \in \mathbb{R}^+,
\end{align}
and find the optimal set of parameters $\boldsymbol{\theta}^{(2)}$ that minimize the error function,
\begin{align}
    \hat{\boldsymbol{\theta}}^{(2)} = \arg\min_{ \boldsymbol{\theta}^{(2)}} \ \mathcal{E} \left( \boldsymbol{\theta}^{(2)} \right).
\end{align}

We define the total sum of residuals as the error function,
\begin{align}
  \mathcal{E} \left( \boldsymbol{\theta}^{(2)} \right) = \frac{1}{DK} \cdot \sum_{d = 1}^D  \sum_{k = 1}^K  | r_{d, k} |,
\end{align}
and we calculate the residuals of our model's fit as,
\begin{align}
    r_{d, k} = \frac{y_{d, k}}{\mathcal{P} \left( \boldsymbol{\theta}^{(2)}, d_d \right)} - \mathcal{I}_{GSI} \left( \hat{\boldsymbol{\theta}}^{(1)}_d, \epsilon_k \right), \quad r_{d, k}  \in \mathbb{R},
\end{align}
where $\mathbf{y}_d$ is the pyranometer's measure, and $\boldsymbol{\epsilon}_d$ is the Sun's elevation angle, and $d_d = \left\{ d_d \ | \ d_d \in \mathbb{N}^{\left[1, N \right]}, \ d = 1, \dots, D \right\}$ is the day of the year. We minimize the error function using the gradient w.r.t. $\boldsymbol{\theta}^{(2)}$ that is,
\begin{align}
    \frac{\partial \mathcal{E} \left( \boldsymbol{\theta}^{(2)} \right) }{\partial \theta^{(2)}_i} = - \frac{1}{DK} \sum_{d = 1}^D  \sum_{k = 1}^K \left[ \mathcal{I}_{GSI} \left( \hat{\boldsymbol{\theta}}^{(1)}_d, \epsilon_{d,k} \right) \frac{ r_{d,k } }{\left| r_{d, k} \right|} \frac{ \partial \mathcal{P} \left( \boldsymbol{ \theta}^{(2)}, d_d \right) }{\partial \theta^{(2)}_i}  \right].
\end{align}

The results of model's fit are displayed on the right graph in Figure \ref{fig:bias}.

\begin{figure}[!htbp]
    \begin{subfigure}{0.495\textwidth}
        \centering
        \includegraphics[scale = 0.225]{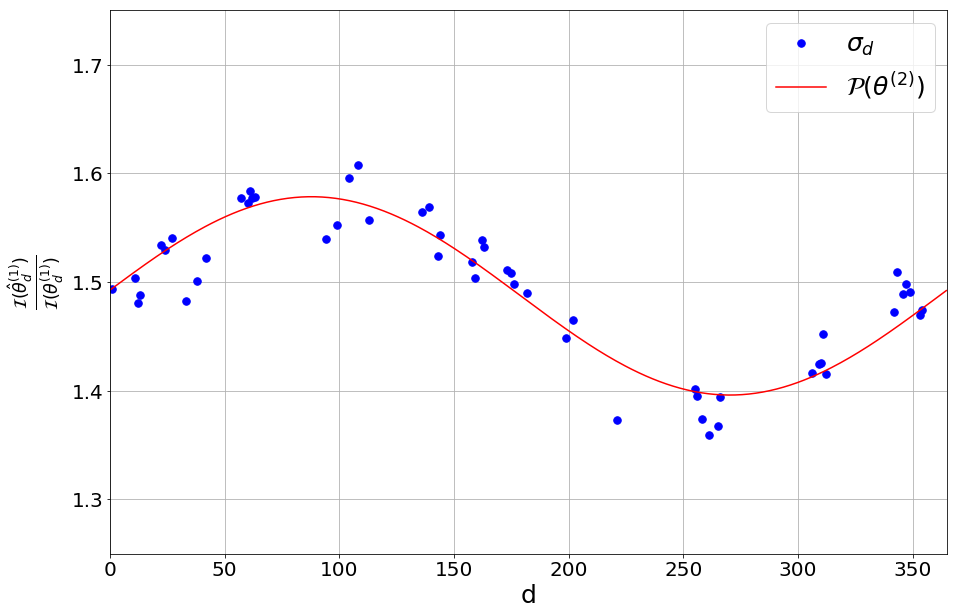}
    \end{subfigure}
    \begin{subfigure}{0.495\textwidth}
        \centering
        \includegraphics[scale = 0.225]{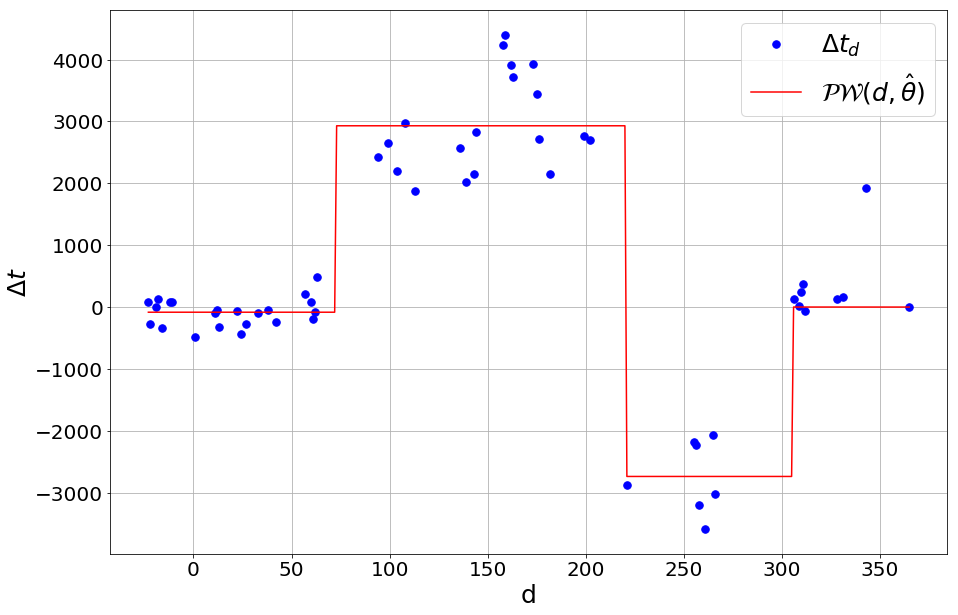}
    \end{subfigure}
    \centering
    \caption{In the left plot appears the estimated amplitude $\sigma_d$ for each clear day $d$ in blue. The proposed model $\mathcal{P} \left( \boldsymbol{\theta}^{(2)} \right)$ is in red. In the right plot appears the piecewise model fit to shifts in the time series.}
    \label{fig:bias}
\end{figure}

\item Pyranometer Shifting Adjustment 

The shifting caused by the inclination of the pyranometer's support is corrected calculating the correlation between the theoretical GSI and the real GSI, and finding where is the maximum,
\begin{align}
    t_{max} = \arg\max \ \mathrm{Corr} \left( \mathcal{I} \left[ t \right], \mathcal{R} \left[ t \right] \right),
\end{align}
so that $\Delta t = \frac{t_{N}}{2} - t_{max}$ is the displacement, and the $\mathrm{sign} \left( \Delta t \right)$ indicates the shifting direction in the time series. 

Once we know the displacement in the maxima, we shift the theoretical irradiance $\mathcal{I}$ to detrend the pyranometer's measurements $\mathcal{R}$ and obtain the CSI. If know multiply the CSI by the theoretical irradiance without shift, we have the corrected pyranometer's measurements $\mathcal{R}^{\prime}$,
\begin{align}
    \mathcal{R}^\prime \left[ t \right] = 
    \begin{cases} 
    \frac{\mathcal{R} \left[ \Delta t:t_N \right]}{\mathcal{I} \left[ 1:t_N - \Delta t \right]} \cdot \mathcal{I} \left[  \Delta t:t_N \right] & \mathrm{sign} \left( \Delta t \right) = 1 \\
    \frac{\mathcal{R} \left[ 1:t_N - \Delta t \right]}{\mathcal{I} \left[ \Delta t:t_N \right]} \cdot \mathcal{I} \left[ 1:t_N - \Delta t\right]
    & \mathrm{sign} \left( \Delta t \right) = -1 
    \end{cases} \ \forall t = 1, \ldots, t_N - \Delta t
\end{align}
Despite the fact of the correction, there is information loss. The correct time series $\mathcal{R}^\prime$ is $\Delta t$ shorter. However, this do not suppose as problem as we only use information when $ \epsilon > 15^\circ$.

We approximate the shifts in clear sky days by a piecewise model, see Figure \ref{fig:bias}. The piecewise model has 4 constant levels that are the mean of the shifts at the same level. We process the measures in the dataset to have a shift according to this model.

\begin{figure}[!htbp]
    \begin{subfigure}{0.495\textwidth}
        \centering
        \includegraphics[scale = 0.225]{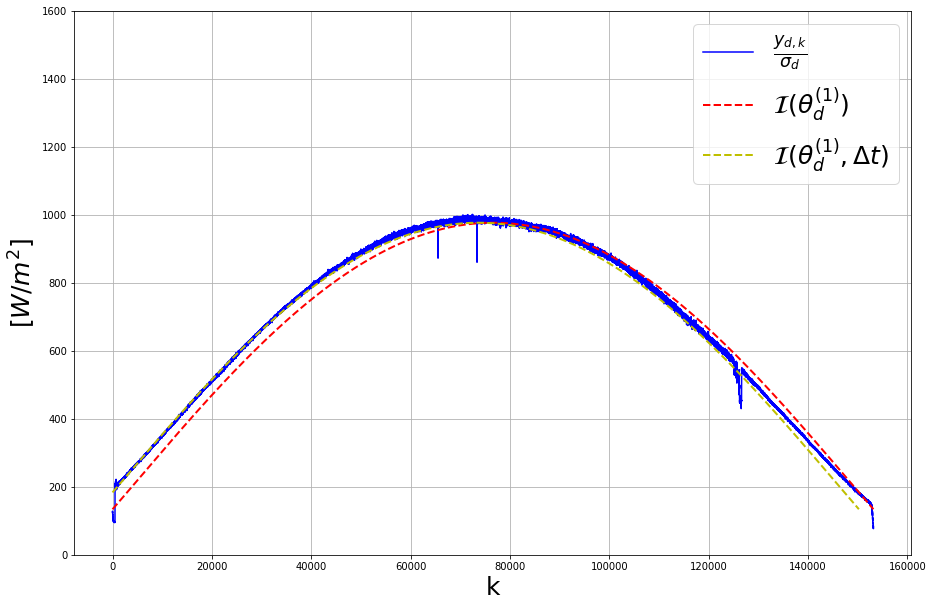}
        \includegraphics[scale = 0.225]{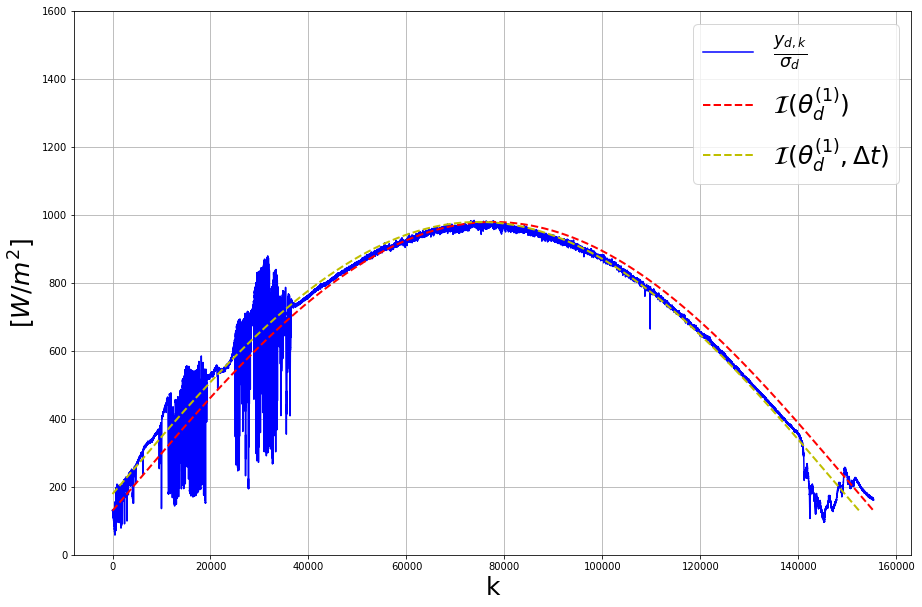}
        \includegraphics[scale = 0.225]{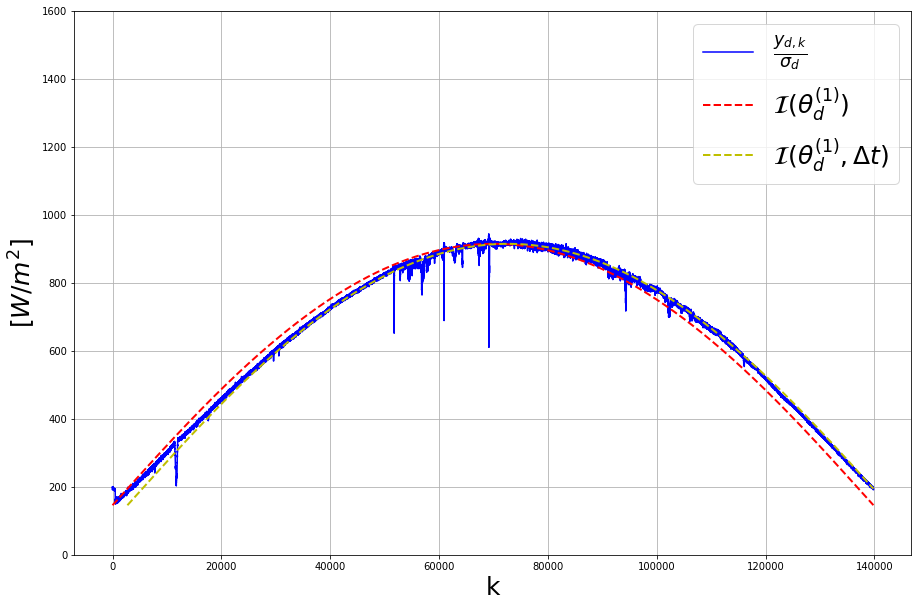}
        \includegraphics[scale = 0.225]{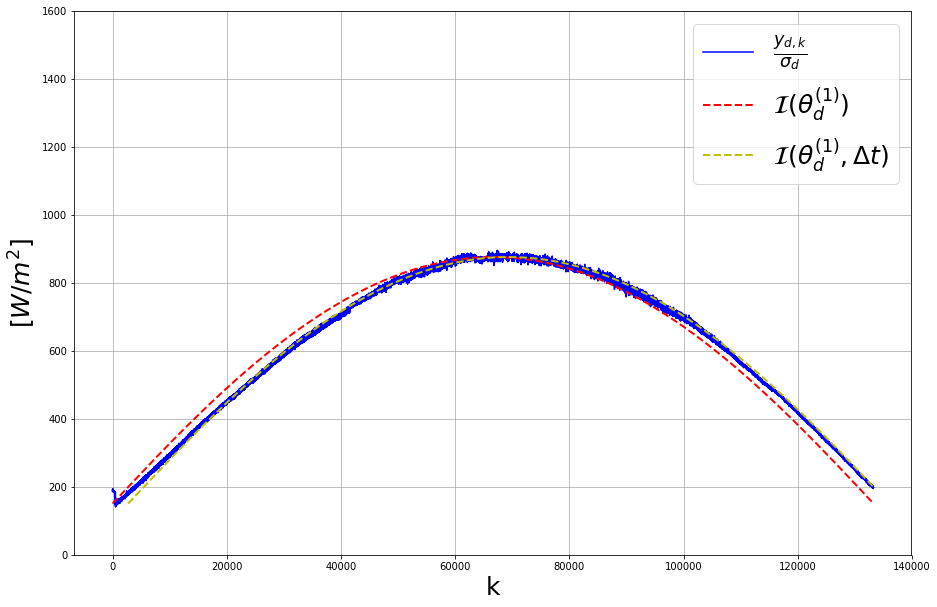}
    \end{subfigure}
    \begin{subfigure}{0.495\textwidth}
        \centering
        \includegraphics[scale = 0.225]{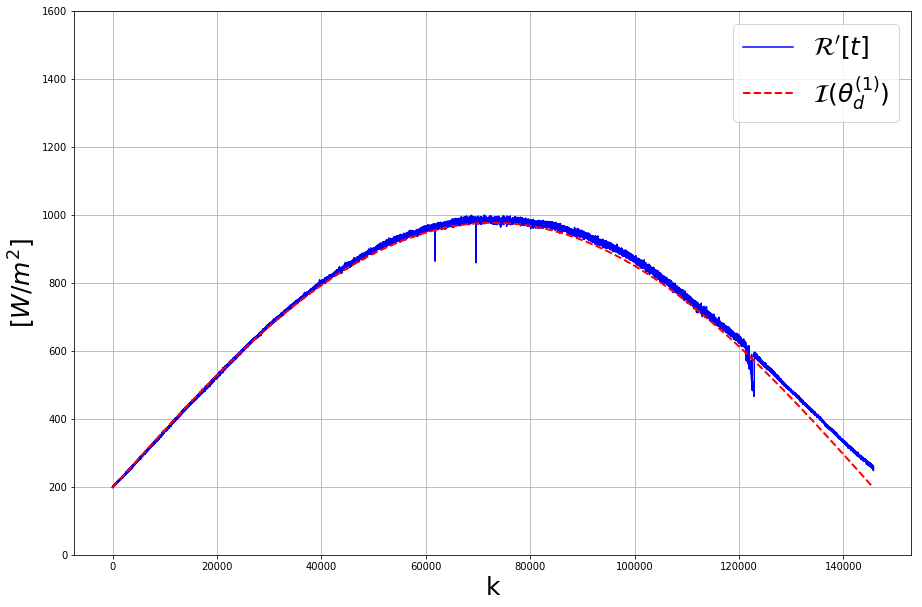}
        \includegraphics[scale = 0.225]{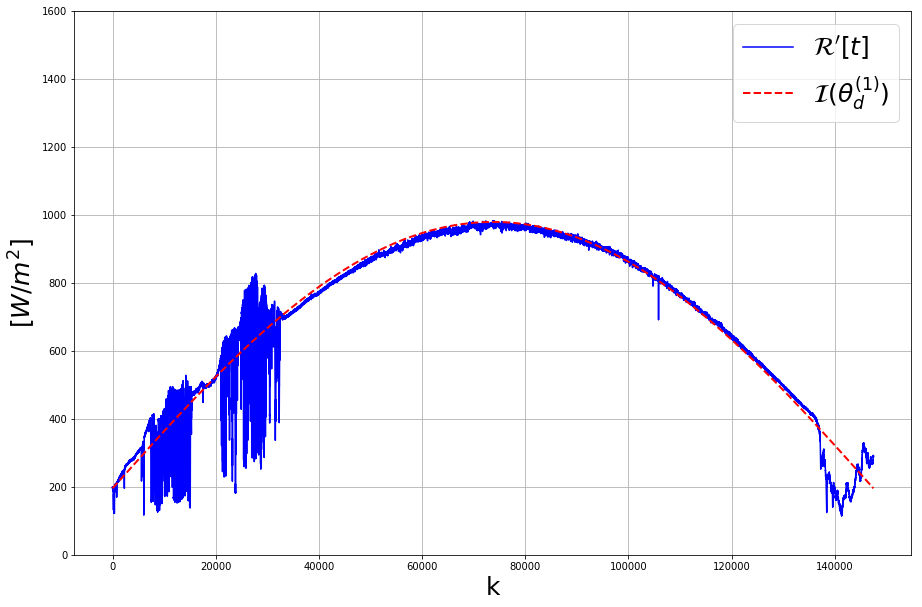}
        \includegraphics[scale = 0.225]{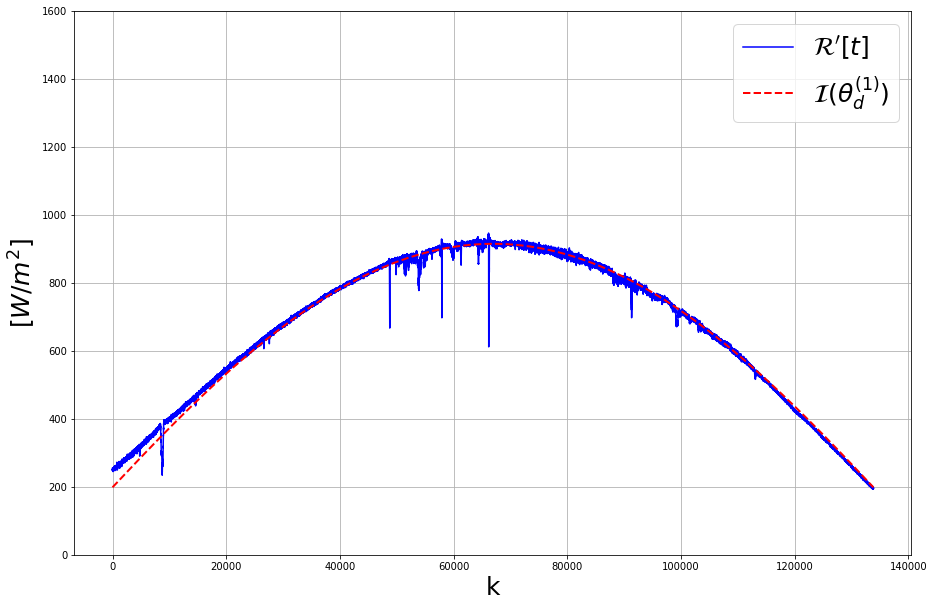}
        \includegraphics[scale = 0.225]{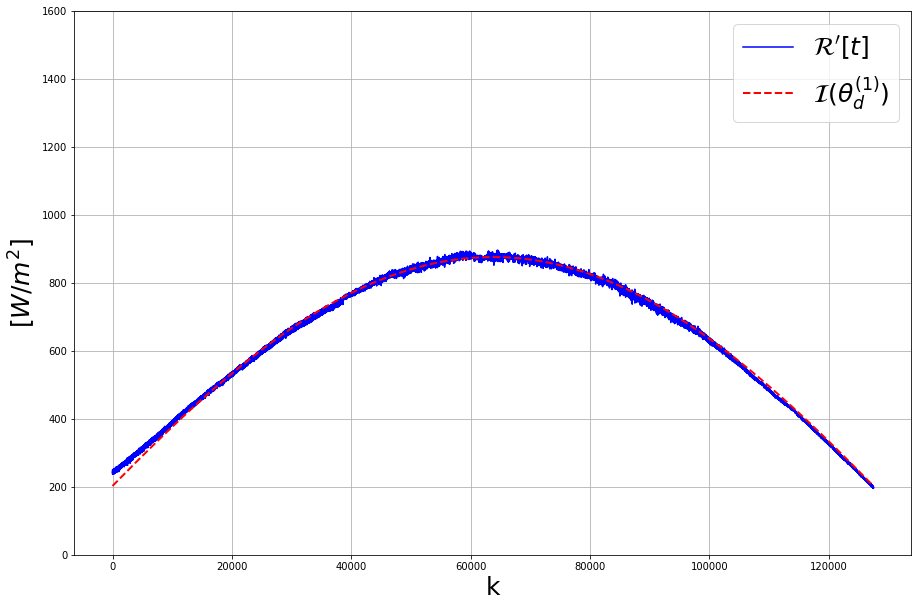}
    \end{subfigure}
    \caption{The graphs in the left show the displacement in the pyranometer's time series with respect to the theoretical GSI. The graphs in the right show the results after correcting the time series.}
\label{fig:shifting_samples}
\end{figure}

\item Physical Model Numerical Parameters Adjustment

To adjust the physical model, we have a data set of $D$ days of clear sky. In each day, the data set is composed by $K_d$ measurements of irradiance $\mathbf{y}_d = \{ y_{d,k} \mid  y_{d,k} \in \mathbb{R}^+, k = 1, \ldots, K_d \}$, and their corresponding elevation angles $\mathbf{e}_d = \{ \epsilon_{d,k} \mid  \epsilon_{d,k} \in \mathbb{R}^{ \left[ 1,\frac{\pi}{2} \right] }, \ k = 1, \ldots, K_d \}$. We solve the following optimization problem to estimate each day optimal set of a parameters $\hat{\boldsymbol{\theta}}^{(1)}_d$,
\begin{align}
    \hat{\boldsymbol{\theta}}^{(1)}_d = \arg\min_{\boldsymbol{\theta}^{(1)}_d} \quad \mathcal{E} \left( \boldsymbol{\theta}^{(1)}_d \right),\quad \forall d = 1, \dots, D.
\end{align}

The function that we propose to minimize is the Root Mean Square Error (RMSE),
\begin{align}
    \mathcal{E} \left(\boldsymbol{\theta}^{(1)}_d \right) = \frac{1}{K} \sum_{k = 1}^K \sqrt{ \left( y_{d, k} - \mathcal{I}_{GSI} \left( \boldsymbol{\theta}^{(1)}_d, \epsilon_{d, k} \right) \right)^2}.
\end{align}

The error function's gradient w.r.t $\theta_{d,i}$ is,
 \begin{align}
    \frac{\partial \mathcal{E} \left( \boldsymbol{\theta}^{(1)}_d \right)}{\partial \theta_{d,i}} = - \frac{1}{K} \sum_{k = 1}^K \left[ \frac{ y_{d, k} - \mathcal{I}_{GSI} \left( \boldsymbol{\theta}^{(1)}_d, \epsilon_{d, k} \right) }{ \left| y_{d, k} - \mathcal{I}_{GSI} \left( \boldsymbol{\theta}^{(1)}_d, \epsilon_{d, k} \right) \right| } \cdot \frac{\partial \mathcal{I}_{GSI} \left( \boldsymbol{\theta}^{(1)}_d, \epsilon_{d, k} \right) }{\partial \theta_{d, i}} \right].
\end{align}

\item Steepest Descent

We propose to estimate the each parameters' set $ \boldsymbol{\theta}^{(j)} = \{ \boldsymbol{\theta}^{(1)}, \boldsymbol{\theta}^{(2)}, \boldsymbol{\theta}^{(3)}\}$ by Steepest Descent \cite{NOCEDAL2006}. So, we start by a random initialization of the parameters such as $\boldsymbol{\theta}^{(j)}_0 \sim \mathcal{U} \left(0, 1 \right)$, and we update the parameters following this formula that uses the function gradient,
\begin{align}
    \theta_{i, n + 1}^{(j)} &= \theta_{i, n}^{(j)} - \eta \cdot \frac{ \partial \mathcal{E} \left( \boldsymbol{\theta}^{(j)}_n \right) }{ \partial \theta_{i, n}^{(j)} }, \quad \forall n = 0, \ldots, N,
\end{align}
where $\eta \in \mathbb{R}^+$ is a very small number. Therefore, we might have a series of parameters' updates that converge to a function minimum such as,
\begin{align}
    \hat{\boldsymbol{\theta}}^{(j)}_N &= \arg\min_{ \boldsymbol{\theta}^{(j)}} \ \left\{ \mathcal{E} \left( \boldsymbol{\theta}^{(j)}_n \right) \geq \mathcal{E} \left( \boldsymbol{\theta}^{(j)}_{n + 1} \right), \dots, \mathcal{E} \left( \boldsymbol{\theta}^{(j)}_{N} \right) < \mathcal{E} \left( \boldsymbol{\theta}^{(j)}_{N + 1} \right) \right\}.
\end{align}

\end{itemize}

\begin{figure}[!htbp]
    \begin{subfigure}{0.245\textwidth}
        \centering
        \includegraphics[scale = 0.20]{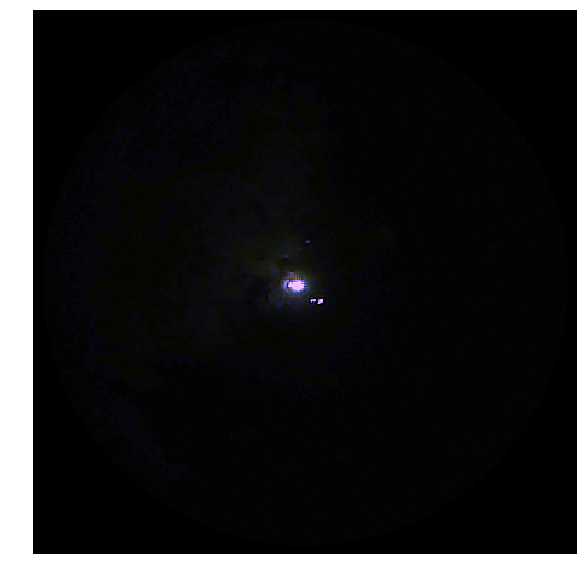}
        \includegraphics[scale = 0.20]{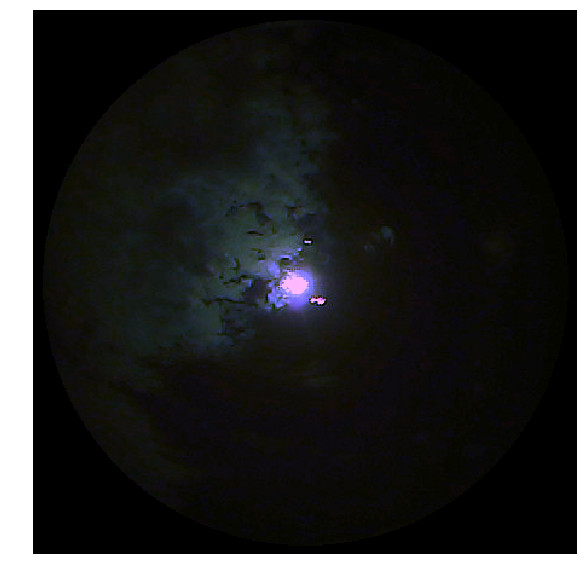}
        \includegraphics[scale = 0.20]{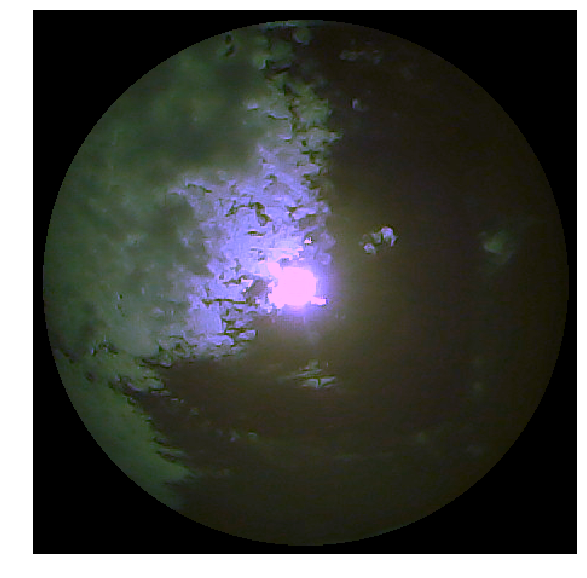}
        \includegraphics[scale = 0.20]{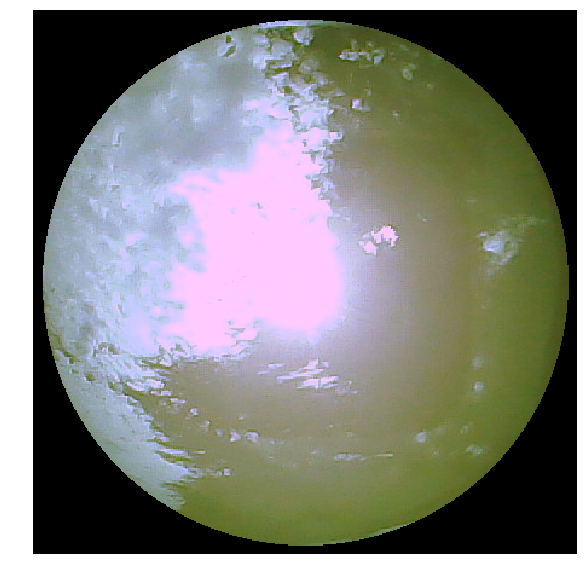}
        \includegraphics[scale = 0.20]{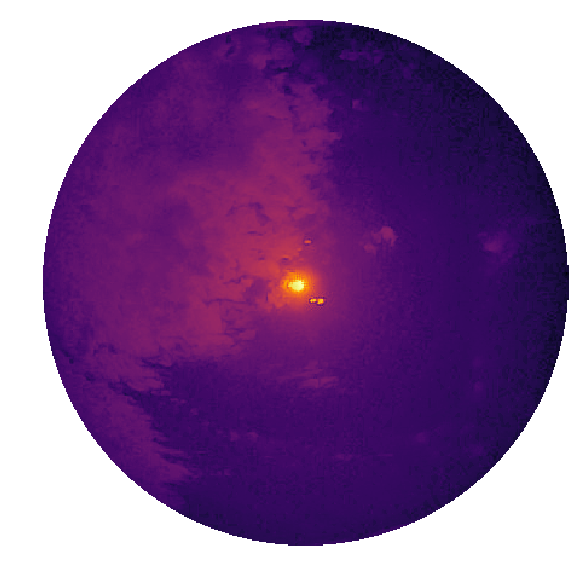}
    \end{subfigure}
    \begin{subfigure}{0.245\textwidth}
        \centering
        \includegraphics[scale = 0.20]{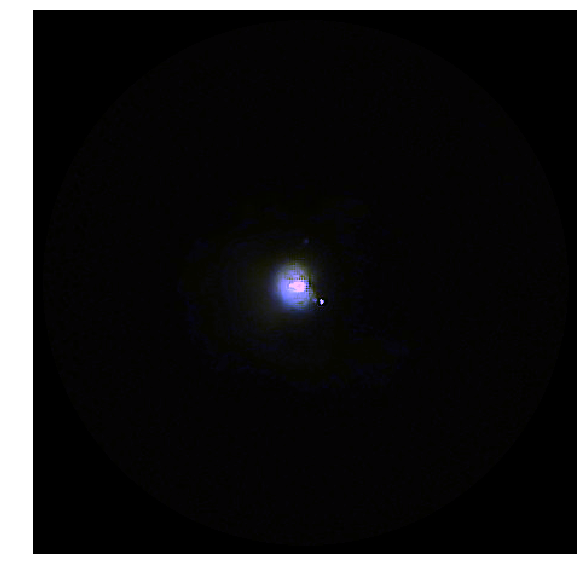}
        \includegraphics[scale = 0.20]{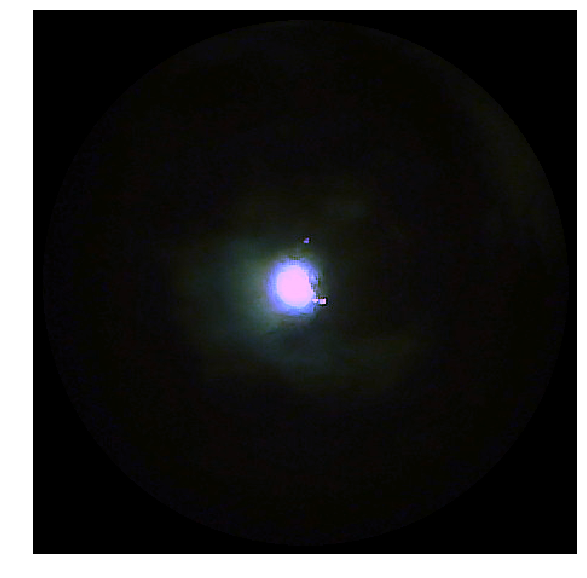}
        \includegraphics[scale = 0.20]{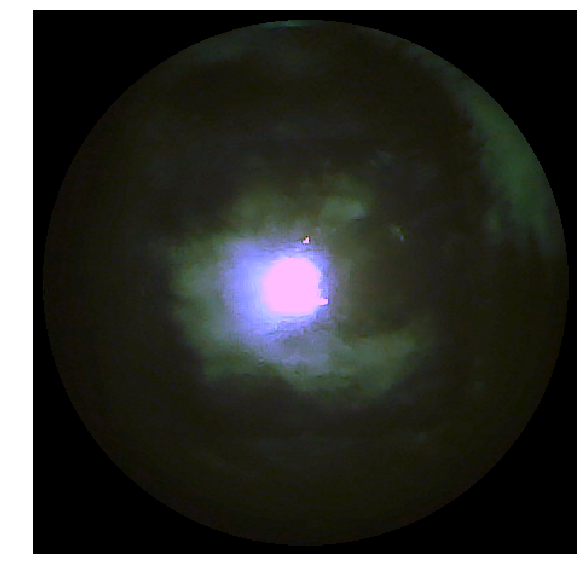}
        \includegraphics[scale = 0.20]{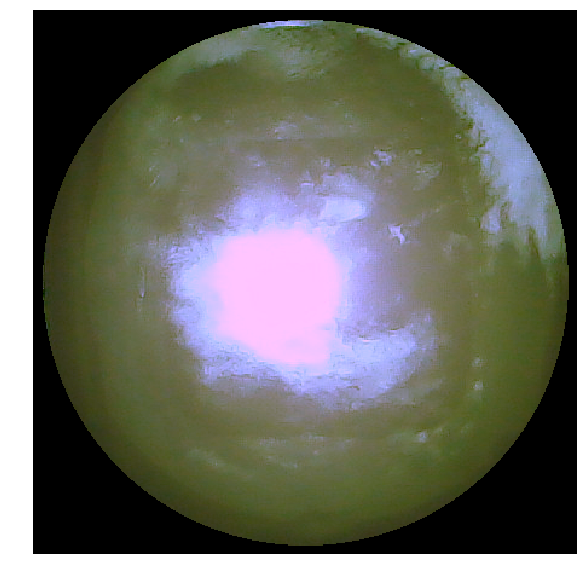}
        \includegraphics[scale = 0.20]{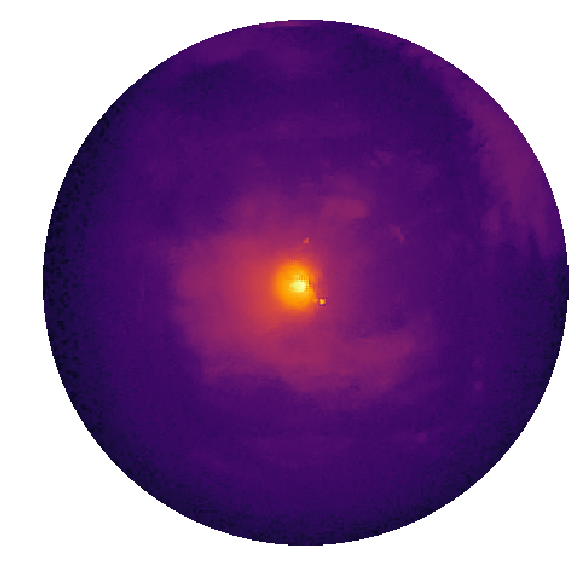}
    \end{subfigure}
    \begin{subfigure}{0.245\textwidth}
        \centering
        \includegraphics[scale = 0.20]{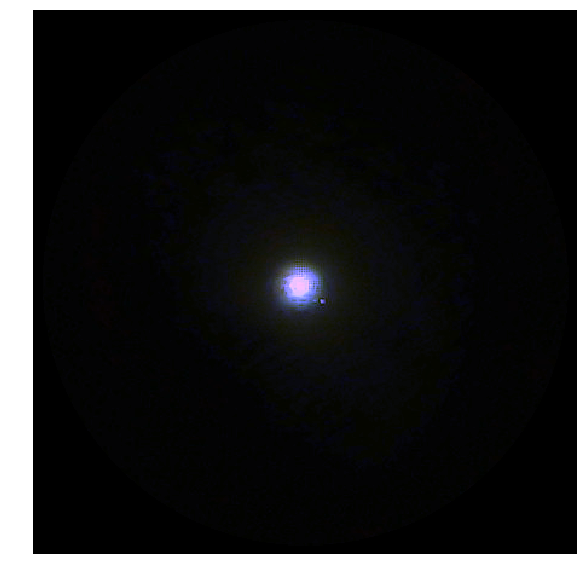}
        \includegraphics[scale = 0.20]{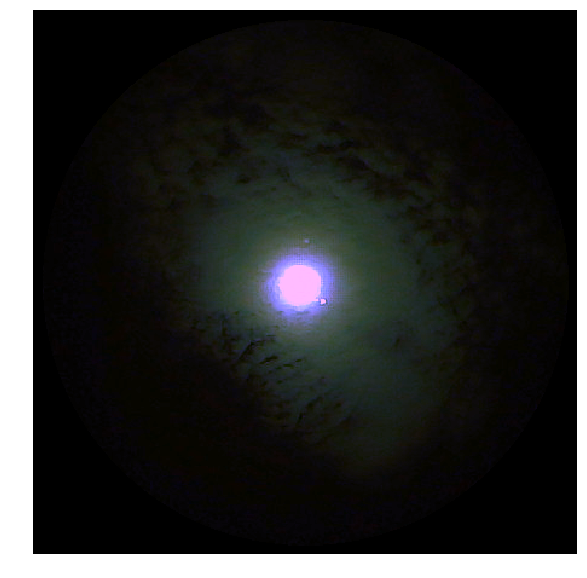}
        \includegraphics[scale = 0.20]{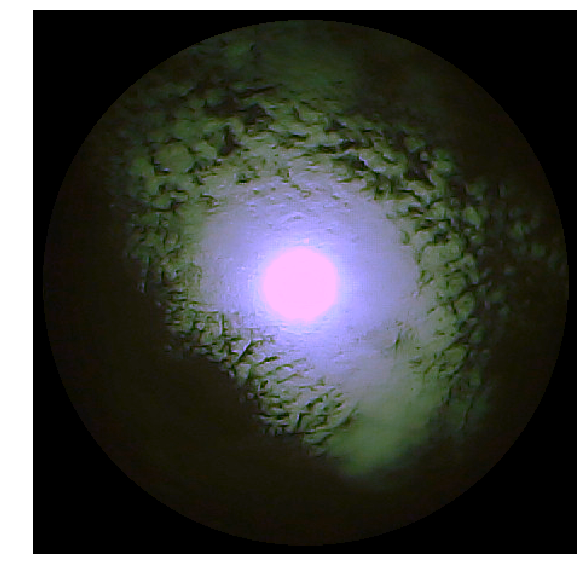}
        \includegraphics[scale = 0.20]{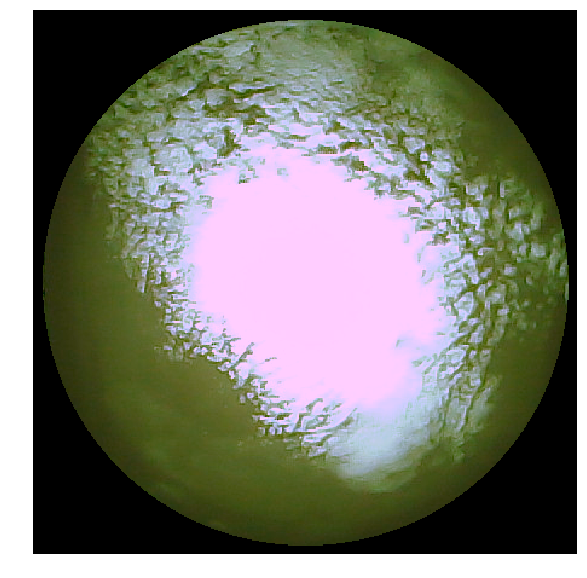}
        \includegraphics[scale = 0.20]{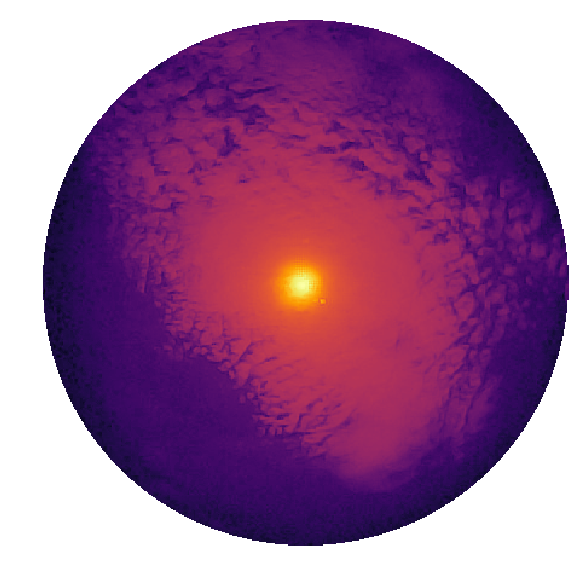}
    \end{subfigure}
    \begin{subfigure}{0.245\textwidth} 
        \centering
        \includegraphics[scale = 0.20]{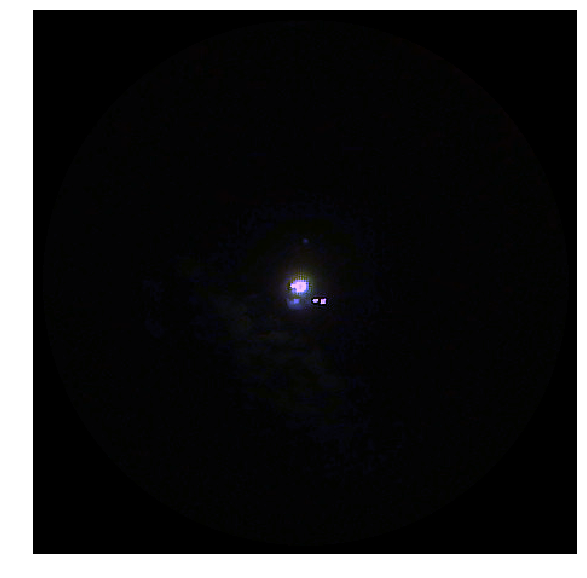}
        \includegraphics[scale = 0.20]{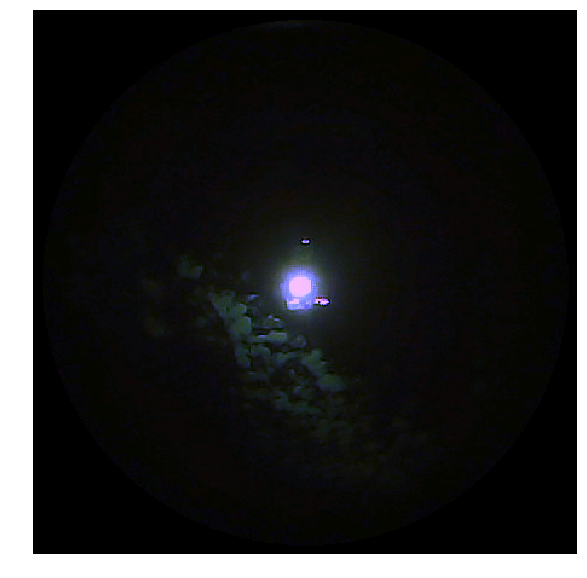}
        \includegraphics[scale = 0.20]{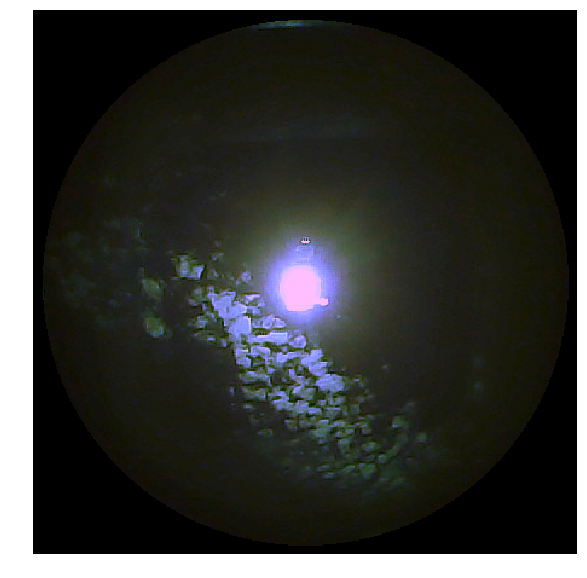}
        \includegraphics[scale = 0.20]{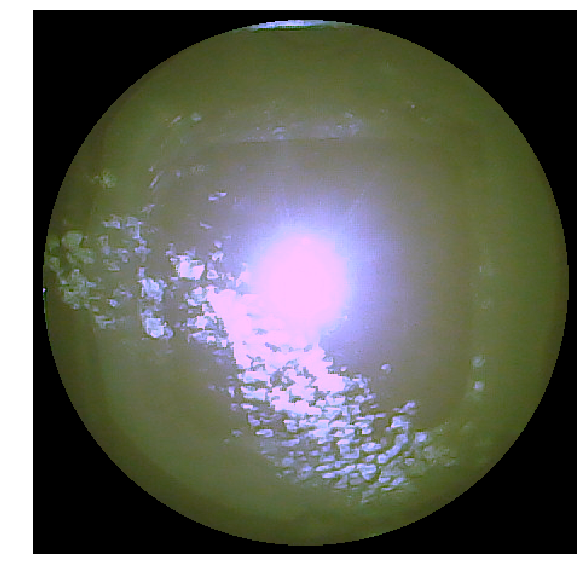}
        \includegraphics[scale = 0.20]{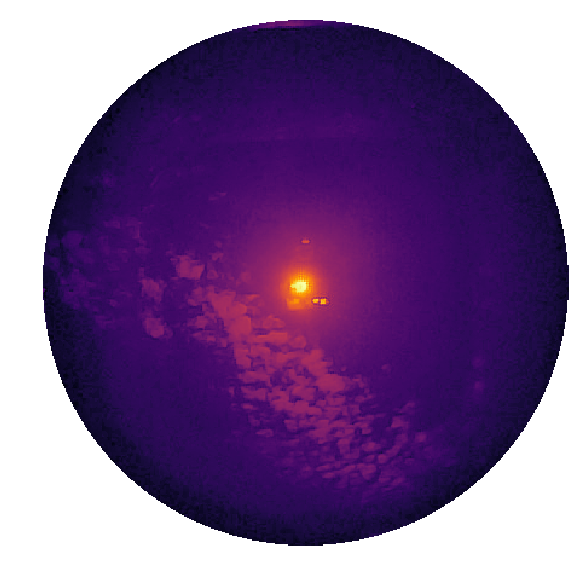}
    \end{subfigure}
    \caption{Each column shows an different image that were captured using different exposure times, that are displayed by column. The last row shows the images after computing the fusion algorithm. The resulting images are displayed in logarithmic-scale because of the large intensity of the pixels in the circumsolar region. Their size is $450 \times 450$, pixels without information are not displayed.}
\label{fig:exposure}
\end{figure}

\subsection{Visible Image Fusion}

We introduce a method to fusion images with different exposition to generate a High Dynamic Range (HDR) of 16 bits, that combines information of the clouds from captures with different exposure time. The images noise was attenuated before hand as explained. To merge the images we first regularized the RGB components in each exposition to avoid division by zero when light is too dim,
\begin{align}
    \mathbf{I}^k_{e, c} + \lambda, \quad \mathbf{I}^k_{e, c} \in \mathbb{R}^{D \times D}
\end{align}
and we convert the images to gray scale by a weighted sum of the image RGB channels with corresponding Luma coding system coefficients that are $\beta_c = \{0.299, 0.587, 0.114 \}$,
\begin{align}
    \mathbf{I}^k_e = \sum_{c = 1}^C \mathbf{I}^k_{e, c} \cdot \beta_c, \quad \forall c = 1, 2, 3.
\end{align}

The method propose is based on the distance from a pixel to the Sun. We know that the Sun is always centered in the frames thereby we define our fusion mask as a function of the radial distance of the Sun to a pixel such as,
\begin{align}
    \mathcal{M} \left( r\right) = \left[ \left( \mathbf{X} - x_{sun} \right)^2 + \left( \mathbf{Y} - y_{sun}\right)^2 \right]^{1/2} \leq r, \quad  \mathcal{M}  \in \mathbb{B},
\end{align}
notice that we are using boolean algebra. We define a mask $\mathbf{M}_e$ for each exposition time $r$ that as a constant distance to the Sun $\rho_e$. 
\begin{align}
     \mathbf{M}_e  = \mathcal{M}  \left( \rho_e 
\right), \quad \forall \rho_e = \{\rho_2, \cdots, \rho_E\}
\end{align}
the set of fusion mask is $\mathbf{M}^m = \{\mathbf{0}, \mathbf{M}^2, \cdots, \mathbf{M}^E, \mathbf{1}\}$. We apply a Gaussian function in sliding window,
\begin{align}
    \mathcal{G} \left( x, y\right) = \frac{1}{2\pi\sigma^2}\exp^{-\frac{x^2 - y^2}{2\sigma^2}},
\end{align}
to blur the masks' pixel in edges with the objective of smoothing out the transition between images regions in the resulting merged image, 
\begin{align}
    \mathbf{\tilde{M}}^m_{i,j}= \frac{1}{C} \sum_x \sum_y \left[ \mathbf{M}^m_{x, y} \odot \mathbf{G}_{x, y} \right], \quad \forall x,y \in \mathbf{W}_{N \times N}, \quad \forall m \in \{2, \dots, E\},
\end{align}
where normalization constant $C = \sum_{i = x}^{x + N} \sum_{j = y}^{y + N} \mathbf{G}_{x, y}$. Notice that the masks $\mathbf{\tilde{M}}$ are not longer Boolean $\mathbf{\tilde{M}}^m \in \mathbb{R}^{D \times D}$.

We define ring-shape masks $\mathbf{R^1_e}$ and $\mathbf{R^2_e}$. We use $\mathcal{M} \left( r \right)$ function as we did before, but applying logical-sum operation with a circular mask with a bit smaller and larger radii $\epsilon = 2$, to produce the ring such as,
\begin{align}
     \mathbf{R}^1_e &= \mathcal{M}\left( \rho_e 
\right) \oplus \mathcal{M}  \left( \rho_e + \epsilon 
\right) , \quad \forall \rho_e = \{\rho_1, \cdots, \rho_E\}\\
     \mathbf{R}^2_e &= \mathcal{M}\left( \rho_e 
\right) \oplus \mathcal{M}  \left( \rho_e - \epsilon 
\right) , \quad \forall \rho_e = \{\rho_1, \cdots, \rho_E\}.
\end{align}

The objective of the ring-shape masks is to select only adjacent pixels between regions different regions to weigh each image and produce smooth transitions. The weights are computed for each image applying the ring shape masks,
\begin{align}
    \alpha^k_{e + 1} = \alpha^k_e \cdot \frac{\sum_i \sum_j \mathbf{I}^k_e \odot \mathbf{R}^2_e }{\sum_i \sum_j \mathbf{I}^k_{e + 1} \odot \mathbf{R}^1_e }, \quad \forall e \in \{ 1, \cdots, E - 1 \}, \quad \alpha^k_{e}  \in \mathbb{R}.
\end{align}
We obtain a set of weights $\alpha^k_e = \{1, \alpha^k_2, \cdots, \alpha^k_E\}$ for each frame $k$. And, we proceed to merge the images with this formula,
\begin{align}
   \mathbf{X}^k =  \sum_{e = 1}^E  \frac{1}{e} \cdot \left[ \tilde{\mathbf{M}}_{e + 1} \odot \left( \mathbf{1} - \tilde{\mathbf{M}}_{e}\right) \odot \left( \sum_{i = 1}^e \alpha^k_i \cdot \mathbf{I}^k_i \right) \right], \quad \mathbf{X}^k \in \mathbb{R}^{D \times D}
\end{align}

The image obtained in not normalized, so we normalize it by diving the image by a large number and multiply by the number of bits we want in the the image.
\begin{align}
    \mathbf{I}^k = \left( \frac{\mathbf{X}^k}{225} \right) \cdot 2^{16}, \quad \mathbf{I}^k \in \mathbb{R}^{D \times D}
\end{align}

as we want a 16 bits image, we have to make sure there are not pixel above $2^16$, and we convert the real number of the pixels to the natural numbers encounter on a 16 bits image, such as
\begin{align}
    \mathbf{I}^k_{16bits} = 
    \begin{cases} 
    I^k_{i,j} \quad I^k_{i,j} \leq 2^{16} \\
    2^{16} \quad I^k_{i,j} > 2^{16}
    \end{cases} \quad \forall i,j = 1, \cdots D, \quad \mathbf{I}^k_{16bits} \in \mathbb{N}^{\leq 2^{16}}.
\end{align}

The outer circular region in the fusioned image is set to an intensity value of 0, It displays artifacts produce by own lens' support and case. Therefore has not relevant information for the prediction.


\section{Girasol Datasets Description}

The repository has available recordings from 240 days of 3 years days, that amount for 120Gb of data in total. The sampling interval of the cameras is 15 seconds when the Sun's elevation angles is $>15^\circ$, so there are approximately 1,200 to 2,400 captures per day from each camera depending of the year day.

\begin{itemize}
    \item VI images are 16 bits with resolution 450x450 with only intensity channel. Approx. 240Kb per frame. Between 200Mb to 400Mb per day depending on amount of clouds. Images are save in .png compress-less format in the directory */infrared.
    
    \item IR images are 16 bits with resolution 60x80 only intensity channel. Approx. 8Kb per frame. 20Mb per day roughly constant. Images are save in .png compress-less format in the directory */infrared.
    
    \item The pyranometer samples as many measures as it can in a second. This ranges from 4 to 6 samples per seconds. The measures are save in the directory */pyranometer in a .csv file with their date as name (yyyy-mm-dd), approx. 4,500Kb to 7,500Kb per file. The files contain unix time in the first column and GSI in $W/m^2$ in the second column.
    
    \item The Sun position files are generate out of the time recorded in the pyranometer files. The positions are save in the directory */position in a .csv file with their date as name (yyyy-mm-dd), approx. 6,500Kb to 11,500Kb per file. The files contain the unix time in the first column, elevation angle in the second column, and azimuth angle in the third column.
    
\end{itemize}

\section{Infrared Radiometry}

The Lepton\footnote{https://www.flir.com/} camera with Radiometry has a wavelength from 8 to 14 micros, and provides an uniform thermal image as the output. When applying the radiometry functionality, the camera recordings are turned into temperature measurements. To obtain the corresponding values of temperature from the intensity of a pixels, they have to be operated such as,
\begin{align}
    T = \frac{I}{100} \ [^{\circ}K]
\end{align}
for international system units, or as
\begin{align}
    T = \frac{I}{100} - 273.15 \ [^{\circ}C],
\end{align}
to obtain the temperature in Celsius degrees.

\subsection{Moist-Adiabatic Lapse Rate}

The temperature of a particle in the Troposphere is a function of the height. In contrast with the temperature in the Tropopause that can be considered approximately constant \cite{MURALIKRISHNA2017}. There is also a height of slightly increasing in temperature in the Stratosphere that will be depreciated in this investigation, as cloud phenomena is restricted to the Troposphere. Above the Stratosphere range of height to the Exosphere the content of water vapor particles will be considered zero.

The presence of water in the Troposphere implies that, in the a process of convection, a parcel of air which rises at the same time cools, and eventually saturates \cite{WALLACE2006}. At this point, as the temperature continues decreasing with the height the water vapor condenses and forms clouds realising heat \cite{MECHOSO2015}. This process is known as the water vapor adiabatic lapse, and while the dry adiabatic lapse rate is constant, the  moist adiabatic lapse is not \cite{ZHANG2015, HOLTON2013}. This fact is important because in most climates the relative humidity is nonzero. The equation for the moist adiabatic lapse is,
\begin{align}
    \Gamma_{MALP} = g \frac{ 1 + \frac{L_r \cdot r_r}{R \cdot T}}{c_{pd} + \frac{L_p^{2} r_p \epsilon}{R T^2}} = g \frac{ R_{sd} T^2 + H_v r_v T}{c_{pd} R_{sd} T^2 + H_{v}^2 r_v \epsilon} \ [^{\circ}K/m]
\end{align}
where:
\begin{itemize}
    \item Earth's gravitational acceleration, $g = 9.8076 \ m/s^2 $.
    \item Heat of vaporization of water $ H_v= 2501000 \ J/kg $.
    \item Specific gas constant of dry air $ R_{sd}= 287 \ J/kg ^{\circ}K $.
    \item Specific gas constant of water vapour $ R_{sw} = 461.5 \ J/kg ^{\circ}K$.
    \item Dimensionless ratio of the specific gas constant of dry air to the specific gas constant for water vapour $ \epsilon = \frac{R_{sd}}{R_{sw}} = 0.622$.
    \item Water vapour pressure of the saturated air, $ e = \epsilon \cdot \exp( \frac{7.5 \cdot T_{dew}}{273.3 + T_{dew}} ) \cdot 100 $. The original formula is in hPa but this is in Pa.
    \item Mixing ratio of the mass of water vapour to the mass of dry air $r_v = \frac{\epsilon \cdot e}{p - e}$.
    \item Specific heat of dry air at constant pressure, $c_{pd} = 1003.5 \ J/kg ^{\circ}K $.
\end{itemize}

Therefore, the moist adiabatic lapse rate is fully described knowing the air temperature $T$, the atmospheric pressure $p$, and the dew point $T_{dew}$.

\subsection{Cloud Base}

The base of a cloud can be estimated from ground measurements of air temperature, and dew point or relative humidity. This calculation of the lifted condensation level is based on the spread, that is the difference between the air temperature $T$ and the dew point $T_{dew}$,
\begin{align}
h_{base} = 304.8 \cdot \frac{(T - T_{dew})}{2.5} \ [m].
\end{align}
The spread has to be divided by 4.4 (if temperatures are in $^{\circ}F$) or 2.5 (if temperatures are in $^{\circ}C$), and multiplied by 304.8 (if the heights is sought in meters) or by 1000 (if it is sought in feet). Some networks of weather stations can be remotely assessed to read real-time measurements. In fact, there are available either time series\footnote{https://www.wunderground.com}, or spatio-temporal weather data\footnote{https://www.windy.com}.
        
\section{Atmospheric Radiation Model}

The techniques to estimate the motion vectors in an image are sensitive to gradient in the pixels' intensity. Therefore, it is necessary to remove the gradient which is produced by the solar direct radiation, and the atmospheric scatter radiation in order to extract dynamic features from the cloud. The diffuse or scatter radiation are emitted by water molecules or dust particles in the atmosphere. In particular, the scatter radiation from the atmosphere dome has systematical pattern that appear on the images in the course of a year. 

\subsection{Solar Radiation Models}

The Sun direct radiation is scattered by aerosol and water particles floating in the atmosphere \cite{LAMB2011}. The scatter radiation depends on the atmosphere air mass and clearness conditions. The air mass varies with the altitude, because the gravity attracts particles concentrating them near the surfaces. The clearness depends on the weather conditions such as wind that disperse the aerosol particles. The effects produced by the scatter radiation from atmosphere, and the direct radiation, appear on a IR image. 

The aim is to remove the scatter and direct radiation from the IR images, so that only the reflected radiation from clouds appears on the images. We define each pixel in a frame, as a point $\mathbf{x} = \{ x, y \}$ in a Cartesian coordinates system such $\mathbf{X} = \{ \left(1, \dots, N \right) \ | \ x_{,} \in \mathbb{N}, \ i = 1, \dots, M, \ j = 1, \dots, N \}$, and
$\mathbf{Y} = \{ \left(1, \dots, M \right) \ | \ y_{i,j} \in \mathbb{N}, \ i = 1, \dots, M, \ j = 1, \dots, N \}$. We propose to model effect from the background atmospheric radiation using an exponential function with scale $\sigma_1$ and length-scale $\lambda_1$, centered in the Sun's coordinates $\mathbf{x}_0 = \{x_0, y_0\}$ such as 
\begin{align}
    \mathcal{S} \left( y_{i,j}; y_0, \sigma_1, \lambda_1 \right) = \sigma_1 \exp \left( \frac{y_{i,j} - y_0}{\lambda_1} \right), \quad \sigma_1, \lambda_1 \in \mathbb{R}.
\end{align}

The effect produced on the image by Sun's direct radiation can be modeled by a bivariate quadratic exponential function centered in $\mathbf{x}_0$. However, an inverse-quadratic bivariate function, such as the Cauchy distribution with length-scale parameter $\lambda_2$, is more flexible in the function height and tails. The multivariate Cauchy distribution function is
\begin{align}
    \mathbf{Ca} \left( x_{i,j}, y_{i,j}; \mathbf{x}_0, \lambda_2 \right) = \frac{1}{2\pi} \cdot \left\{ \frac{\lambda_2}{\left[ \left( x_{i,j} - x_0 \right)^2 + \left( y_{i,j} - y_0 \right)^2 + \lambda_2^2\right]^{\frac{3}{2}}} \right\}, \quad \lambda_2 \in \mathbb{R}.
\end{align}

This distribution is transformed to a Lorentzian function multiplying by a scale parameter $\sigma_2$, that controls the density height. So, the function is now
\begin{align}
    \mathcal{D} \left( x_{i,j}, y_{i,j}; \mathbf{x}_0, \sigma_2, \lambda_2 \right) = \sigma_2 \left\{ \frac{\lambda_2^2}{ \left[ \left( x_{i,j} - x_0 \right)^2 + \left( y_{i,j} - y_0 \right)^2 + \lambda_2^2\right]^{\frac{3}{2}} } \right\}, \ \sigma_2, \lambda_2 \in \mathbb{R}.
\end{align}

This function is not a probability distribution, as its integral does not add to 1, but we do not intent to modeling probabilities, this is not important for us.

Combining both functions for the scatter radiation $\mathcal{S}$, and the direct radiation from the Sun $\mathcal{D}$, we obtain this other function,
\begin{align}
    \mathcal{A} \left( x_{i,j}, y_{i,j}; \mathbf{x}_0, \boldsymbol{\theta} \right) &= \theta_1 \exp \left\{\frac{y_{i,j} - y_0}{\theta_2}\right\} + \ldots \\ & \ldots + \theta_3 \left\{ \frac{ \theta_4^2}{\left[ \left( x_{i,j} - x_0 \right)^2 + \left( y_{i,j} - y_0 \right)^2 + \theta_4^2\right]^{\frac{3}{2}}} \right\},
\end{align}
that is a function of parameters' set $\boldsymbol{\theta} = \{\sigma_1, \lambda_1, \sigma_2, \lambda_2\}$. It models the deterministic component of the radiation in the IR. 

\subsection{Optimal Parameters}

The atmospheric model has to fit the radiation in frames recorded during all the seasons of a year, and at different times of the day. Because that, the parameters of the atmospheric radiation model are expected to be variables of a function that depends on the weather conditions, the date, and the Sun's position in the horizon. In order to approximate the function that models the parameters, it is necessary to find which were the optimal parameters in frames previously recorded during different days of year, but it is also necessary that all of the frames were recorded in clear-sky conditions, so that only the radiation appears on the images. 

If the intensity in a pixel is defined as $\mathbf{I}^{d,k} = \{ i^{d,k}_{i,j} \ | \ i^{d,k}_{i,j} \in \mathbb{N}^{[1, 2^{16}]}, \ i = 1, \dots, M, \ j = 1, \dots, N \}$, the Root Mean Squared Error (RMSE) function for sample frame $k$ in a day $d$ is,
\begin{align}
    \mathcal{E} \left( \boldsymbol{\theta}_{d,k} \right) = \frac{1}{MN} \cdot \sum_{i = 1}^M \sum_{j = 1}^N \sqrt{ \left( i^{d,k}_{i.j} - \mathcal{A} \left( x_{i,j}, y_{i,j}; \mathbf{x}_0^{d,k}, \boldsymbol{\theta}_{d,k} \right) \right)^2}.
\end{align}
The aim is to find the optimal set of parameters $\boldsymbol{\theta}_k$ that minimizes the loss function,
\begin{align}
    \hat{\boldsymbol{\theta}}_{d,k} =
    \underset{\boldsymbol{\theta}_{d,k}}{\operatorname{argmin}}
     \ \mathcal{E} \left(\boldsymbol{\theta}_{d,k} \right), \quad \forall k = 1, \dots, K, \ \forall d = 1, \dots, D.
\end{align}
The gradient of error function $\mathcal{E} \left(\boldsymbol{\theta}_{d,k} \right)$ w.r.t. the parameters $\theta^{d,k}_i$ for any sample $d,k$ is,
\begin{align}
    \frac{\partial \mathcal{E} \left( \boldsymbol{\theta}_{d,k} \right)}{\partial \theta^{d,k}_i} = - \frac{1}{MN} \sum_{\ell = 1}^{MN} \left[ \frac{\tilde{i}_\ell^{d,k} - \mathcal{A} \left( \tilde{x}_\ell, \tilde{y}_\ell; \mathbf{x}_0^{d,k}, \boldsymbol{\theta}_k \right) }{\left|\tilde{i}^{d,k}_\ell - \mathcal{A} \left( \tilde{x}_\ell, \tilde{y}_\ell; \mathbf{x}_0^{d,k}, \boldsymbol{\theta}_k \right)  \right| } \frac{\partial \mathcal{A} \left( \tilde{x}_\ell, \tilde{y}_\ell; \mathbf{x}_0^{d,k}, \boldsymbol{\theta}_{d,k} \right)}{\partial \theta^{d,k}_i} \right],
\end{align}
where $\tilde{\mathbf{i}}^{d,k} = \mathbf{Vec} \left( \mathbf{I}^{d,k} \right)$, $\tilde{\mathbf{y}} = \mathbf{Vec} \left( \mathbf{Y} \right)$, and $\tilde{\mathbf{x}} = \mathbf{Vec} \left( \mathbf{X} \right)$ for notation simplification. The optimal parameters were found for a dataset consisting of 51 clear-sky conditions days selected out of a year long of samples. The number of sample-frames available in each day is different, because the number of day-light hours in each day are also different.
\begin{figure}[!ht]
    \begin{subfigure}{0.495\textwidth}
        \centering
        \includegraphics[scale = 0.20]{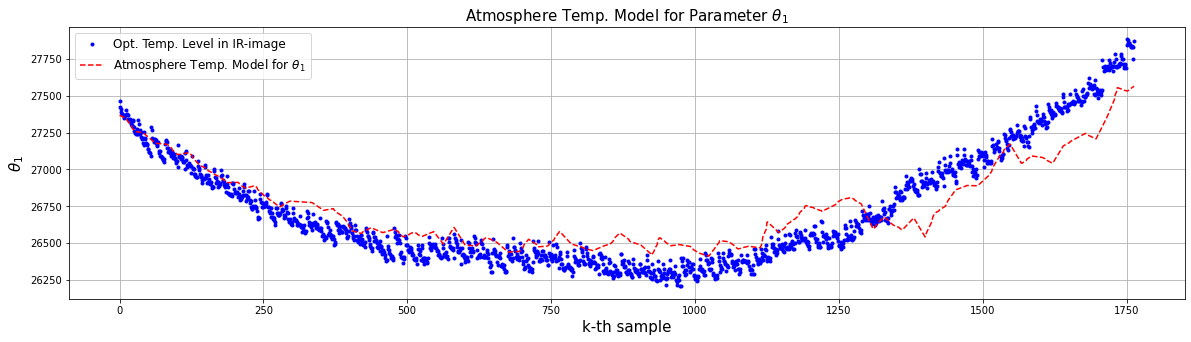}
        \includegraphics[scale = 0.20]{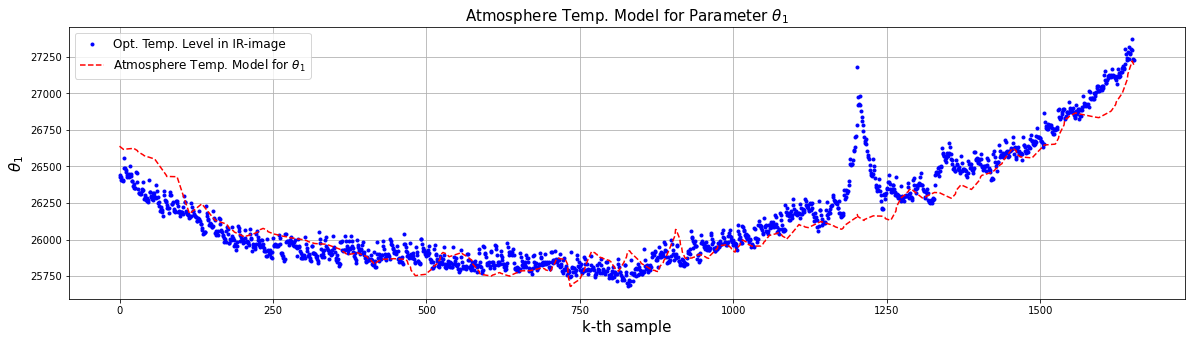}
    \end{subfigure}
    \begin{subfigure}{0.495\textwidth}
        \centering
        \includegraphics[scale = 0.20]{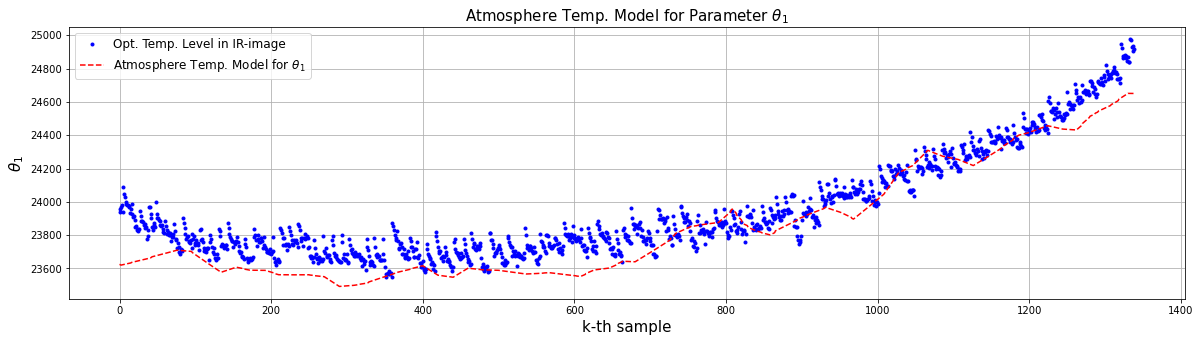}
        \includegraphics[scale = 0.20]{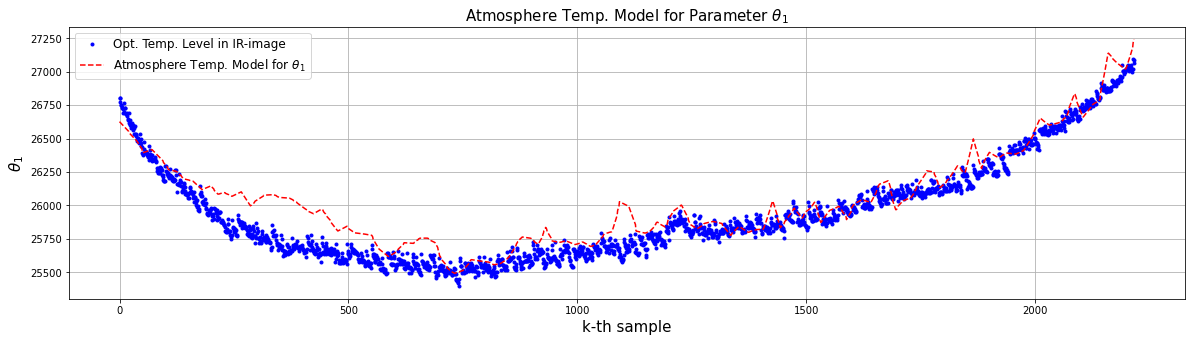}
    \end{subfigure}
\caption{The graphs optimization results for the parameter $\theta^{(1)}$ in four different days $d$. The optimal $\theta^{(1)}_{d,k}$ for a frame $k$ is displayed in color blue. The predicted $\hat{\theta}^{(1)}_{d,k}$ by the polynomial model is displayed in color red.}
\label{fig:theta_1_model}
\end{figure}
\begin{figure}[ht]
    \begin{subfigure}{0.495\textwidth}
        \centering
        \includegraphics[scale = 0.20]{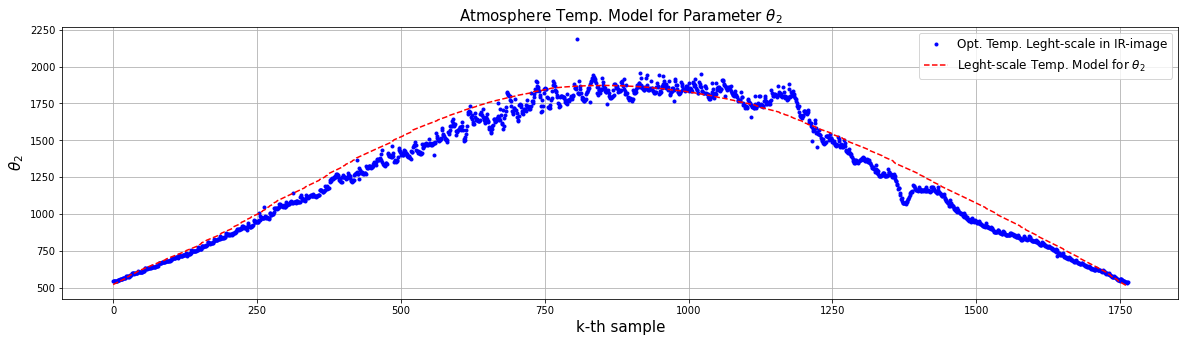}
        \includegraphics[scale = 0.20]{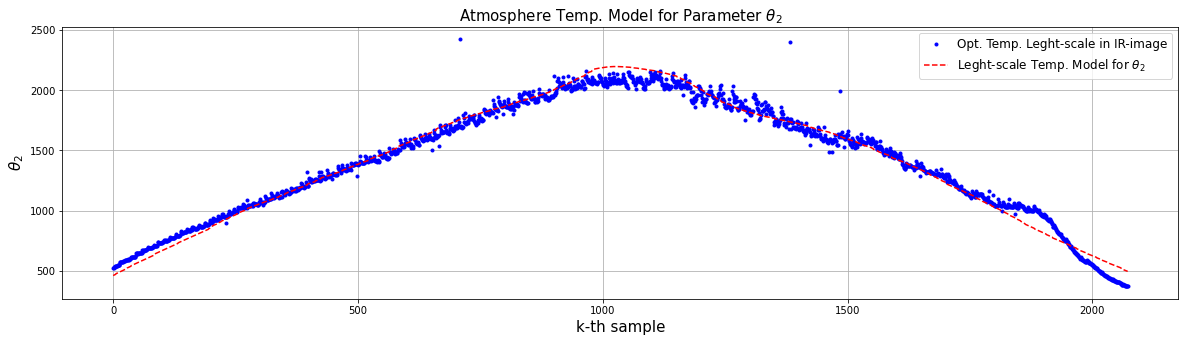}
    \end{subfigure}
    \begin{subfigure}{0.495\textwidth}
        \centering
        \includegraphics[scale = 0.20]{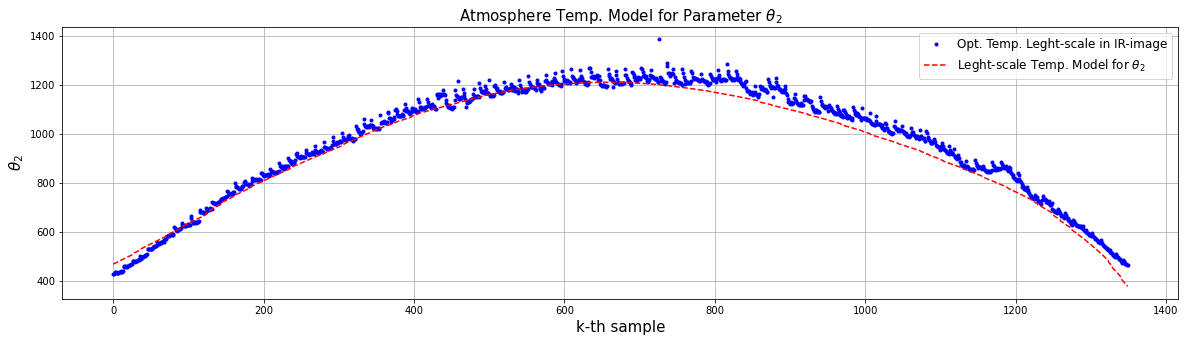}
        \includegraphics[scale = 0.20]{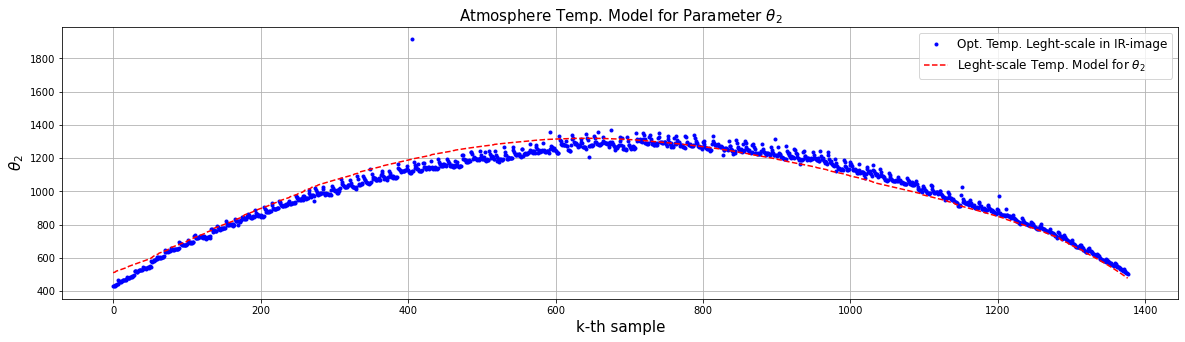}
    \end{subfigure}
\caption{The graphs show the optimal parameter $\theta^{(2)}$ in four different days $d$. The $\theta^{(2)}_{d,k}$ frame $k$ is displayed in color blue. The predicted $\hat{\theta}^{(2)}_{d,k}$ by the polynomial model is displayed in color red.}
\label{fig:theta_2_model}
\end{figure}

\subsection{Model of the Parameters}

Once the optimal set of parameters $\hat{\boldsymbol{\theta}}_{d,k}$ are found for each sample $k$ in day $d$, the aim is to model these optimal sets of parameters $\hat{\boldsymbol{\Theta}} = \{ \hat{\boldsymbol{\theta}}_k \mid \hat{\boldsymbol{\theta}}_k \in \mathbb{R}^{4}, \ k = 1, \ldots,  D K \}$ for any possible moment. In fact, these set of parameters has a physical interpretation. The parameter $\theta^{(1)}$ is the average height of the Tropopause in the IR image. This height varies along the year as a function of the latitude and the weather conditions. The parameter $\theta^{(2)}$ models the cycle-stationary pattern of the Earth curvature cross-section. The great circle of Tropopause is not perfectly spherical and influence by local and global climatic condition, so these patters are not easy to model as the IR images are spatial-temporal measurements. 

The model's parameters of the Sun direct radiation are considered constant as the Sun's irradiance does not varies along the day in the IR images. The optimal values of $\theta^{(3)}$ follows a normal distribution, and the ones of $\theta^{(4)}$ are a uniform distribution. Therefore, the optimal value is computed as the sample mean,
\begin{align}
    \hat{\theta} = \frac{1}{D \cdot K_d} \sum_{d = 1}^D \sum_{k = 1}^{K_d} \theta_{d,k}.
\end{align}

In order to model $\theta^{(1)}$ and $\theta^{(2)}$, we know the Sun's position is a function of the azimuth $\boldsymbol{\alpha} = \{ \alpha_{k} \mid  \alpha_{k} \in \mathbb{R}^{\left[1, 2\pi \right]}, \ k = 1, \ldots, D K \}$, and the elevation angle $\boldsymbol{\varepsilon} = \left\{ \varepsilon_{k} \mid \varepsilon_{k} \in \mathbb{R}^{[1,\frac{\pi}{2}]}, \ k = 1, \ldots, D K \right\}$. The cycle-stationary pattern correspond to the temporal sequence of days in a year $\mathbf{d} = \left\{ d_{k} \mid d_{k} \in \mathbb{N}^{[1, 365]}, \ k = 1, \ldots, D K \right\}$. The height of the Tropopause can be approximate by ground measurements, which can be obtained from a weather station that has available measurements of air temperature $\mathbf{T}^{air} = \left\{ T^{air}_k \mid T^{air}_{k} \in \mathbb{R}, \ k = 1, \ldots, D K \right\}$ and dew point $\mathbf{T}^{dew} = \left\{ T^{dew}_k \mid T^{dew}_{k} \in \mathbb{R}, \ k = 1, \ldots, D K \right\}$. Despite the fact that other variables, such as relative humidity or atmospheric pressure, can be available, no correlation was found with $\theta_1$ or $\theta_2$, consequently no further experimentation was carried out with them. The consistent features in the pattern vector for each parameters were found to be different, these feature vectors are $\mathbf{x}^{(1)}_k = \left[ T^{air}_k \ T^{dew}_k \ \varepsilon_k \ \alpha_k \right]^\top$ and $\mathbf{x}^{(2)}_k = \left[ d_k \ \varepsilon_k \ \alpha_k \right]^\top$ for any frame $k$.

We propose to fit a model that applies a polynomial expansion of order $n$ in the feature $\mathcal{P} \left( \mathbf{x}^{(i)}_k \right)$, and that has coefficients $\mathbf{w}^{(i)}$. The polynomial expansion formula implemented in the feature vector $\mathbf{x}^{(1)}_k$ is,
\begin{align}
    \mathcal{P}\left(\mathbf{x}^{(1)}_k\right) = \sum_{i,j,n,m}^P  \mathbf{w}^{i,j,n,m}_k \cdot T^{air \ i}_k \cdot T^{air \ j}_k \cdot \varepsilon_k^n \cdot \alpha_k^m, \quad \forall k = 1, \ldots, K \cdot D.
\end{align}
The expansion implemented in the feature vector $\mathbf{x}^{(2)}_k$ has this formula,
\begin{align}
    \mathcal{P}\left(\mathbf{x}^{(2)}_k\right) = \sum_{i,j,n}^P \mathbf{w}^{i,j,n}_k \cdot d_k^i \cdot \varepsilon^j_k \cdot \alpha_k^n, \quad \forall k = 1, \ldots, K \cdot D,
\end{align}

The formulation of our regression problem with a loss function applied to the polynomial expansion coefficients, which is a simplification of the Tikhonov's regularization \cite{TIKHONOV1977}, it is known as Ridge Regression (RR) when is implemented with the L-2 norm. This model's objective function is,
\begin{align}
    \min_{\mathbf{w}} \ \sum_{k = 1}^{DK} \left[ \mathbf{y}_k - \mathcal{P}\left(\mathbf{x}_k\right) \right]^2 + \lambda \cdot \| \mathbf{w} \|_2, \quad \forall k = 1, \ldots, K \cdot D.
\end{align}
The parameters $\mathbf{w}$ that minimize loss function can be found analytically applying least squares,
\begin{align}
    0 &= \frac{\partial}{\partial \mathbf{w}} \cdot \left[ \left( \mathbf{y}_k - \mathbf{w}^\top\mathbf{\Phi} \right)^\top \left( \mathbf{y}_k - \mathbf{w}^\top\mathbf{\Phi} \right) + \lambda \cdot \mathbf{tr} \left( \mathbf{w}^\top \mathbf{w} \right) \right] \\
    0 &= 2 \cdot \left[ \mathbf{\Phi}^\top \left( \mathbf{\Phi} \mathbf{w} - \mathbf{y}_k \right) + \lambda \mathbf{w} \right]\\
    \bar{\mathbf{w}} &= \left( \mathbf{\Phi}^\top \mathbf{\Phi} + \lambda \cdot \mathbf{I}\right)^{-1} \mathbf{\Phi}^\top \mathbf{y}.
\end{align}

The polynomial transformation of the features vectors is $\varphi : \mathcal{X} \mapsto \mathcal{P}^n$, where $n$ is the order of the polynomial expansion and the number of terms is $\mathcal{P}^n =  \frac{ \left( n + \left( d - 1 \right) \right) !}{  n! \cdot \left( d - 1 \right)}$. When applying the transformation to a feature vector such as $\mathbf{x}_k \mapsto \varphi \left(\mathbf{x}_k\right), \ \forall k \in \left(0, K\right]$, the resultant vector is transformed to a space of higher dimensions $\Phi_k = \left[ \varphi \left( \mathbf{x}_k \right)_{1} \ \ldots \ \varphi \left( \mathbf{x}_k \right)_{\mathcal{P}^n} \right] \in \mathbb{R}^{\mathcal{P}^n}$. The feature vectors $\mathbf{x}^{(1)}_k $ or $\mathbf{x}^{(2)}_k $ have a scalar dependent variable that is either $\theta^{(1)}_k $ or $\theta^{(2)}_k $. 

The definition of the problem in matrix form contains the all training samples. The set of matrices is different for each model $\mathcal{M}^{(i)}$, and those are,
\begin{align}
    \begingroup 
    \setlength\arraycolsep{1.25pt}
    \mathbf{y}^{(i)} = 
      \begin{bmatrix}
        \theta_{1}^{(i)} \\
        \vdots  \\
        \theta_{K}^{(i)}  \\
      \end{bmatrix}, \
    \mathbf{\Phi}^{(i)} = 
      \begin{bmatrix}
        \varphi \left( \mathbf{x}_1^{(i)} \right)_{1} & \hdots & \varphi \left( \mathbf{x}_1^{(i)} \right)_{\mathcal{P}^n}\\
        \vdots & \ddots & \vdots \\
        \varphi \left( \mathbf{x}_K^{(i)} \right)_{1} & \hdots & \varphi \left( \mathbf{x}_K^{(i)} \right)_{\mathcal{P}^n}\\
      \end{bmatrix}, \
    \mathbf{w}^{(i)} = 
      \begin{bmatrix}
        w_1^{(i)}   \\
        \vdots   \\
        w_{\mathcal{P}^n}^{(i)} \\
      \end{bmatrix}.
      \endgroup
\end{align}

The prediction of parameter $\hat{\theta}^{(i)}_*$ for a new observation $\mathbf{x}^{(i)}_*$ is computed independently using each models parameters such as,
\begin{align}
    \hat{\theta}_*^{(i)} = \bar{\mathbf{w}}^{(i),\top} \varphi \left( \mathbf{x}^{(i)}_*  \right).
\end{align}

Leave-One-Out (LOO) Cross-Validation (CV) method was implemented to find the optimal polynomial expansion order $n$, and regularization parameter $\lambda$. This routine was independently computed for $\theta_1$ and $\theta_2$, since their are modelled using different feature vectors. The dataset contains 51 days with no drops of CSI. These days were selected from a year long set of weather and IR image recordings. 20\% of those days were reserved for testing proposes, whilst the reaming set were the individual samples left aside for LOO validation during the training of the models. Notice that each sample set is a whole day $d$ that itself has $K_d$ recorded weather and inferred samples.
\begin{figure}[!ht]
    \includegraphics[scale = 0.40]{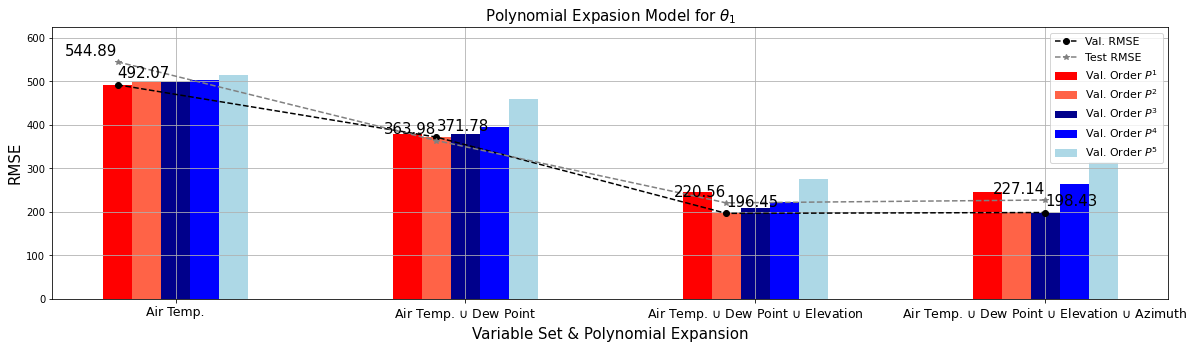}
    \centering
    \caption{The validation of the model for $\theta_1$ shows that, the minimum error $\mathrm{RMSE} \left( \theta_1 \right) = 220.56$, is accomplish by a polynomial expansion of order $n = 2$, which uses as features the air temperature, dew point, and elevation angle.}
    \label{fig:theta_1_cv}
\end{figure}
\begin{figure}[!ht]
    \includegraphics[scale = 0.40]{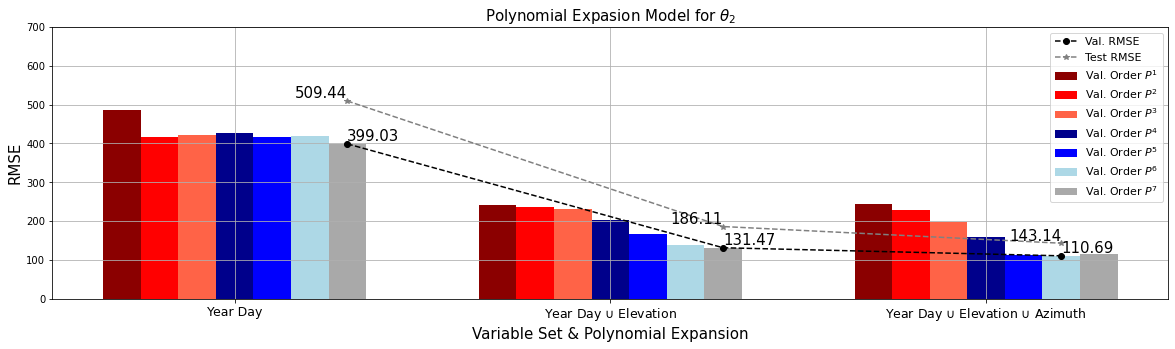}
    \centering
    \caption{The results of the validation perform in the model for $\theta_2$. The minimum error in test is $\mathrm{RMSE} \left( \theta_2 \right) = 143.14$ for a polynomial expansion of order $n = 6$. The optimal features were day of the year, elevation, and azimuth angles.}
    \label{fig:theta_2_cv}
\end{figure}

The atmospheric radiation can be approximated for a new observation $\mathbf{x}_*$, knowing the set of parameters $\hat{\boldsymbol{\theta}}_*$ and the Sun's position is $\mathbf{x}_0$, with the model $\hat{\mathbf{I}}_* = \mathcal{A} \left( \mathbf{X}, \mathbf{Y}; \mathbf{x}_0^k, \hat{\boldsymbol{\theta}}_* \right)$, where $\hat{\mathbf{I}}_* \in \mathbb{R}^{[1, 2^{16}]}$. 

\begin{figure}[!ht]
    \begin{subfigure}{0.245\textwidth}
        \centering
        \includegraphics[scale = 0.15]{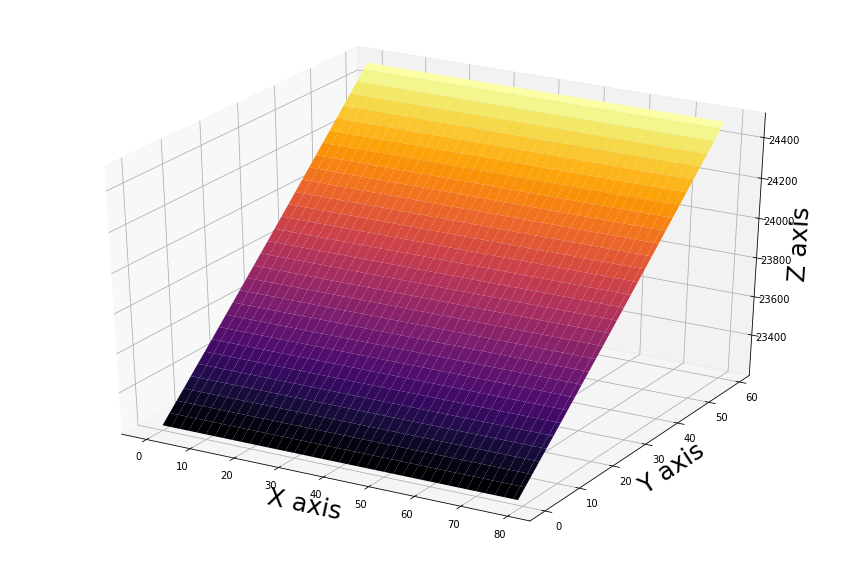}
        \includegraphics[scale = 0.15]{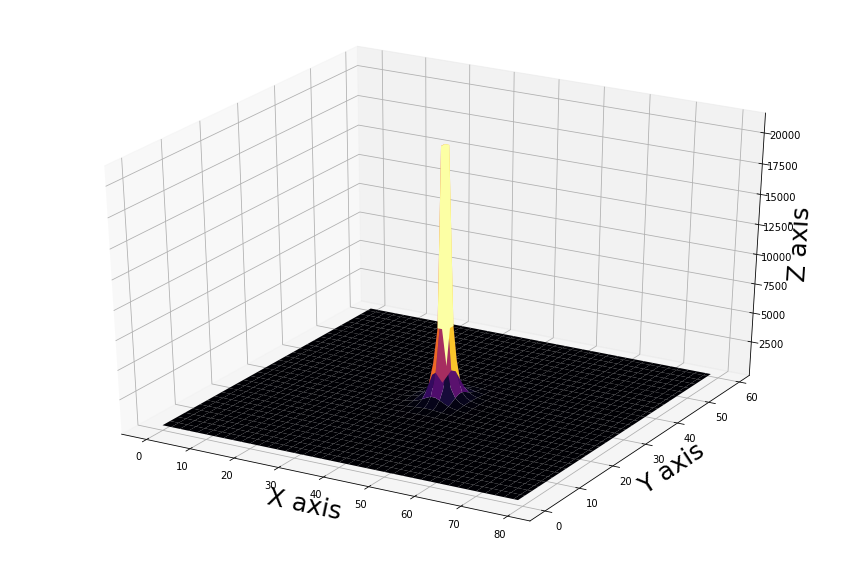}
    \end{subfigure}
    \begin{subfigure}{0.245\textwidth}
        \centering
        \includegraphics[scale = 0.15]{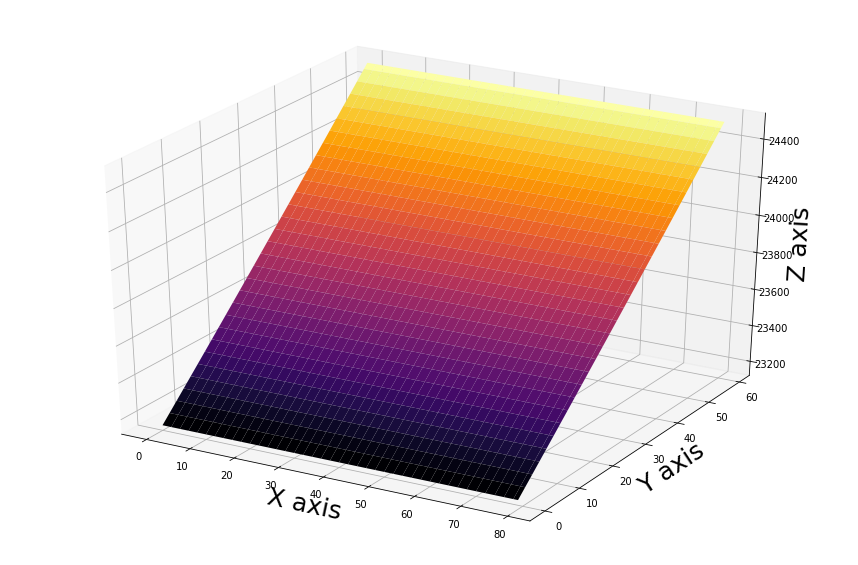}
        \includegraphics[scale = 0.15]{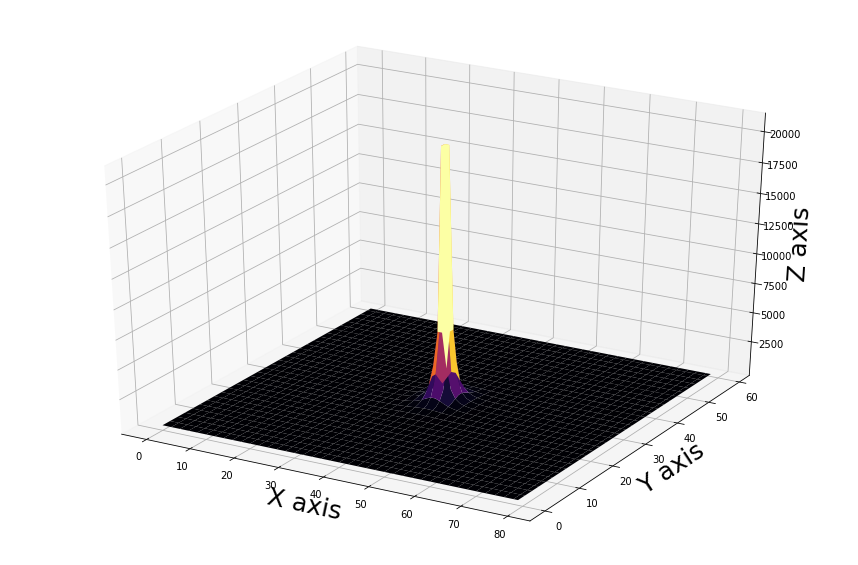}
    \end{subfigure}
    \begin{subfigure}{0.245\textwidth}
        \centering
        \includegraphics[scale = 0.15]{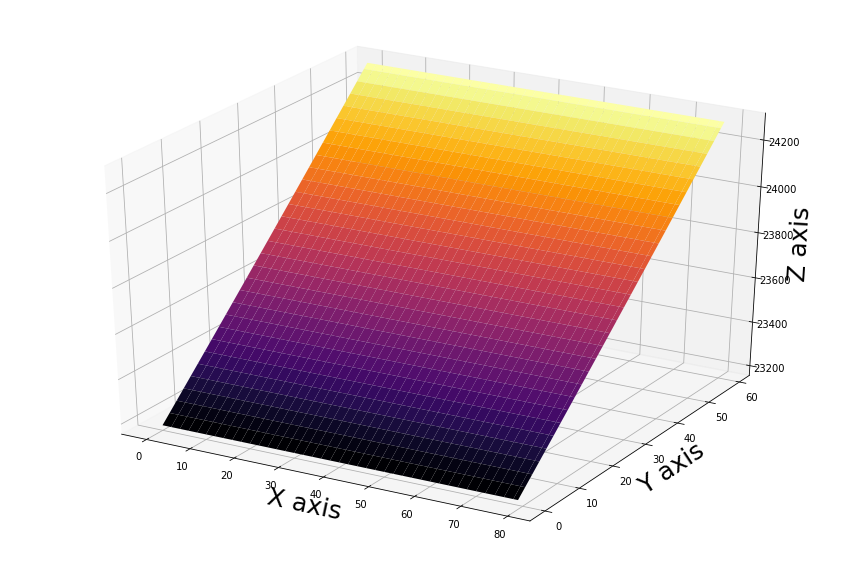}
        \includegraphics[scale = 0.15]{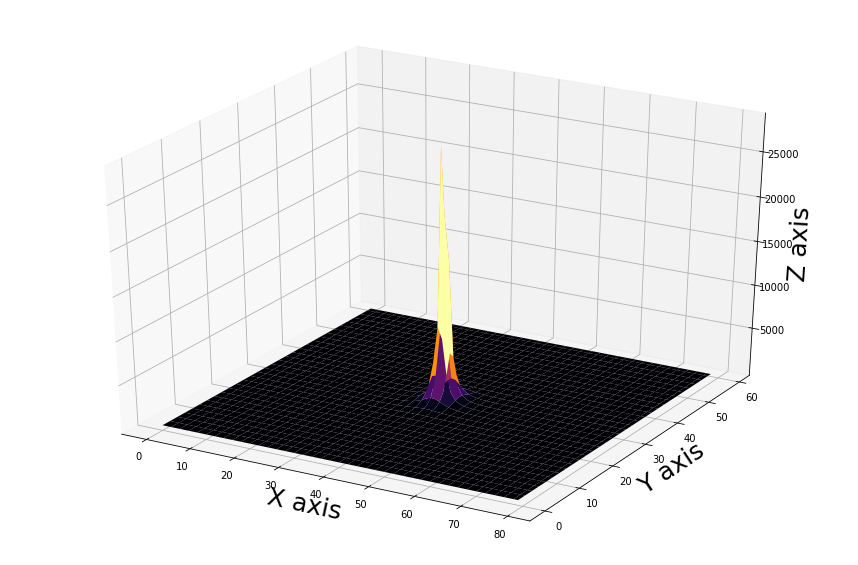}
    \end{subfigure}
    \begin{subfigure}{0.245\textwidth}
        \centering
        \includegraphics[scale = 0.15]{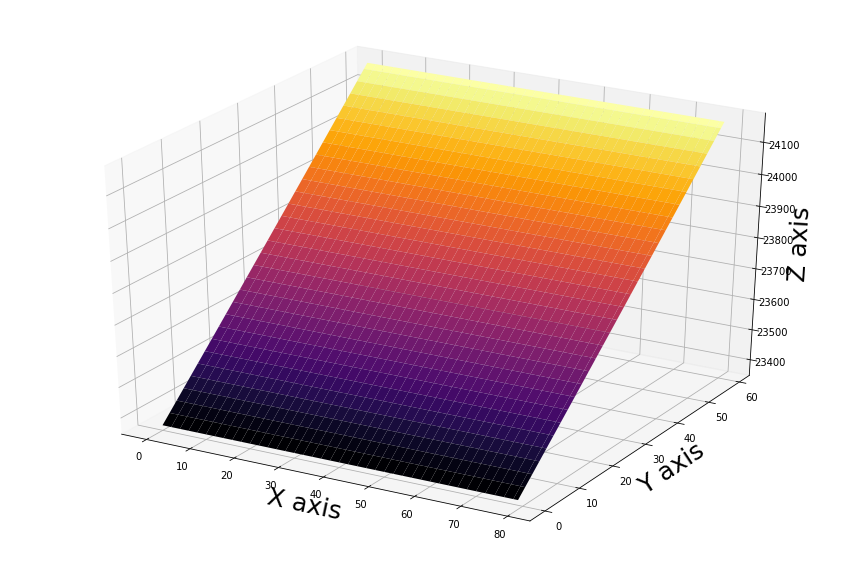}
        \includegraphics[scale = 0.15]{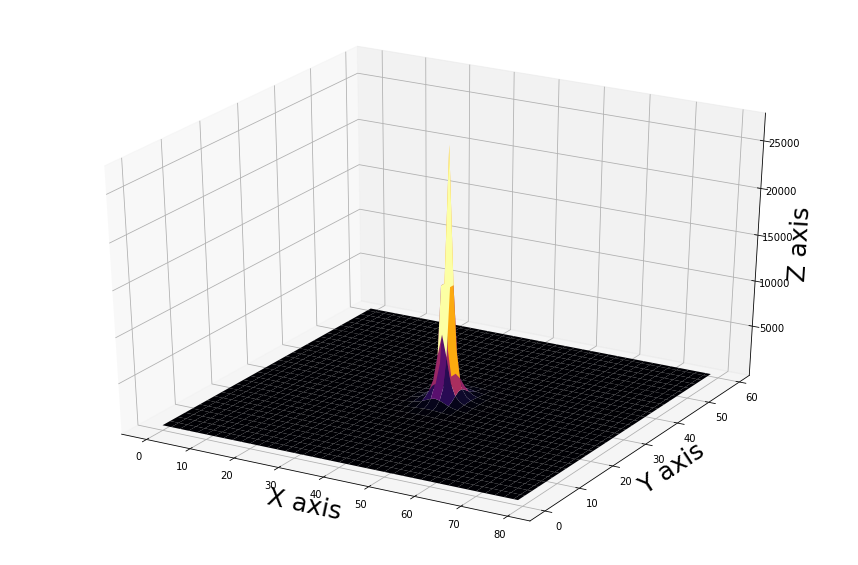}
    \end{subfigure}
\caption{In the upper row shows the model $\mathcal{S}\left( x_{,}, y_{i,j}\right)$, and in the bottom shows the model $\mathcal{D}\left( x_{i,j}, y_{i,j} \right)$. Each column displays the models for a different frame. Notice the scale of the z-axis is different in each graph.}
\label{fig:background_fit}
\end{figure}

When a sequence of images is defined as space-time series such as $\mathbf{I}_k = \left\{ \mathbf{I}_k : k \in \left[1, \infty \right) \right\} \in \mathbb{R}^{\left[1, 2^{16}\right]}$, the atmospheric model for scatter and direct radiation is the deterministic component of the solar irradiance,
\begin{align}
    \hat{\mathbf{I}}_k = \left\{ \mathcal{A} \left( \mathbf{X}, \mathbf{Y}; \mathbf{x}_0^k, \hat{\boldsymbol{\theta}}_k \right) : k \in \left[1, \infty \right) \right\} \in \mathbb{R}^{\left[1, 2^{16}\right]}.
\end{align}
Therefore, we propose to implement this model to detrend the 2D-dimensional time series, so it can be obtained an image that contains only information from the reflected radiation emitted by the clouds, see figure \ref{fig:background_fit}.

\subsection{Atmospheric Model Application}

\begin{figure}[!htbp]
    \begin{subfigure}{0.3275\textwidth}
        \centering
        \includegraphics[scale = 0.195]{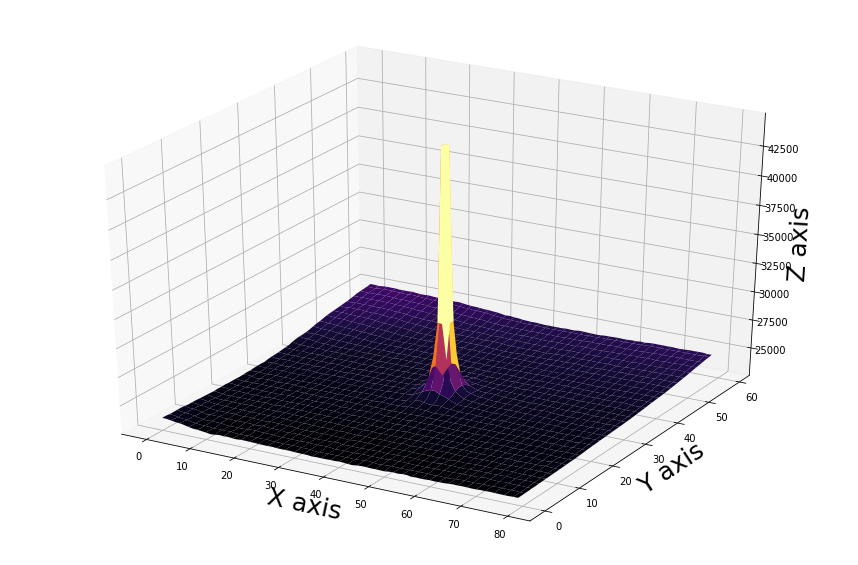}
        \includegraphics[scale = 0.195]{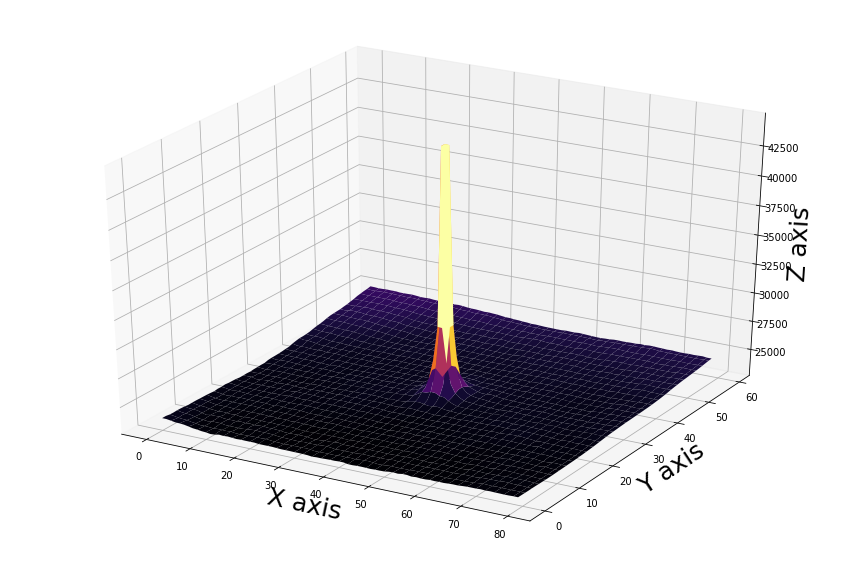}
        \includegraphics[scale = 0.195]{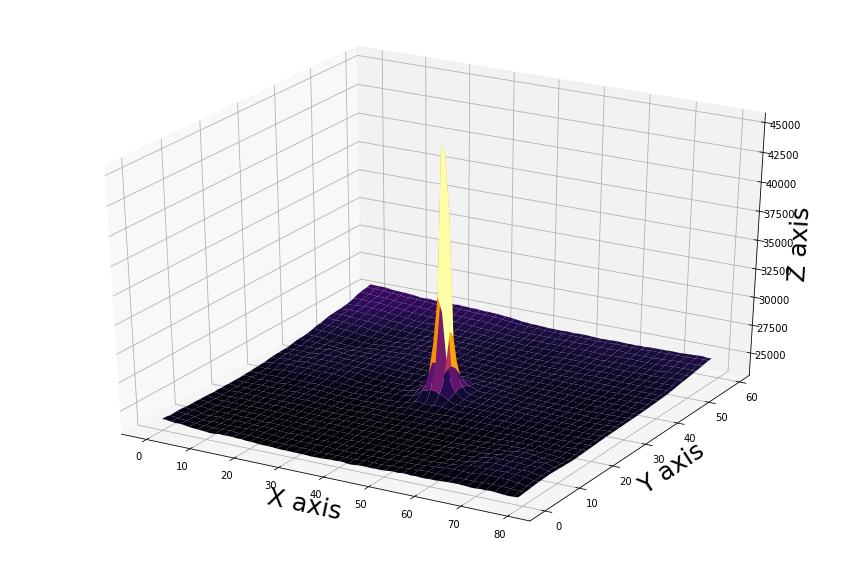}
        \includegraphics[scale = 0.195]{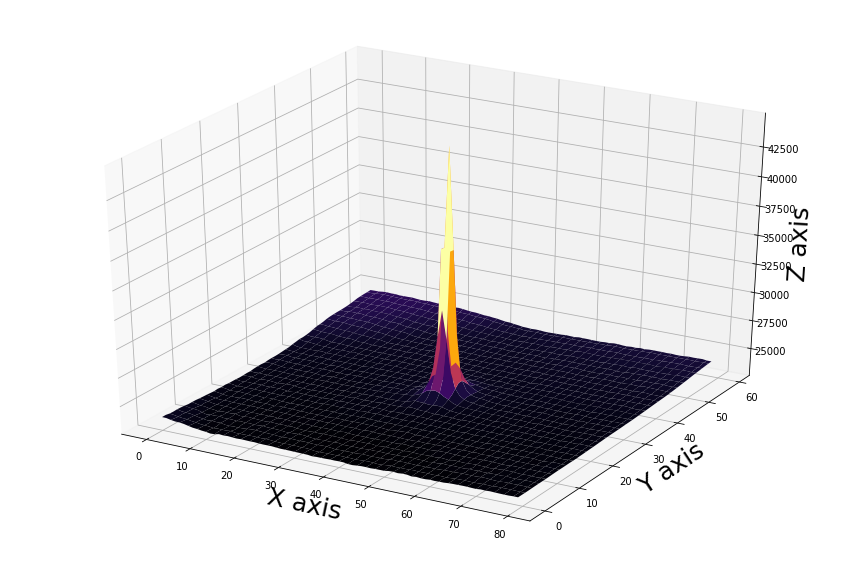}
    \end{subfigure}
    \begin{subfigure}{0.3275\textwidth}
        \centering
        \includegraphics[scale = 0.195]{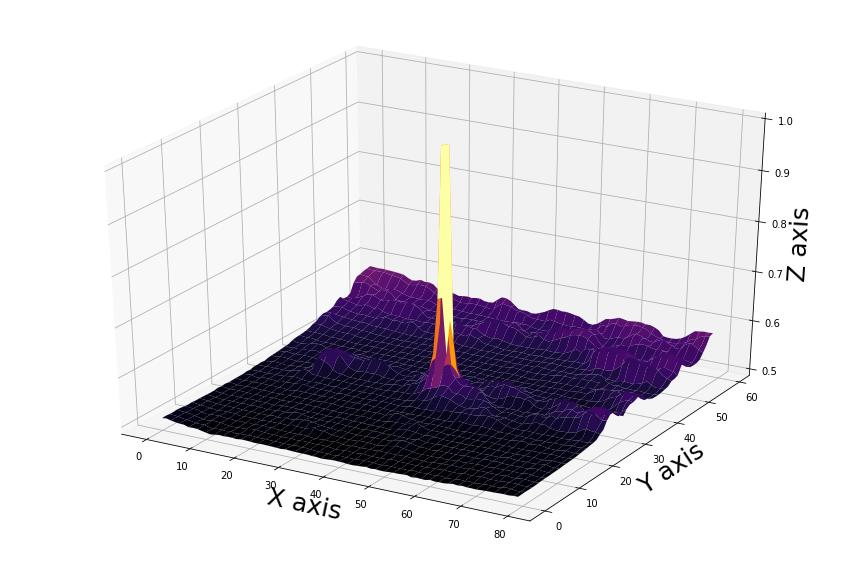}
        \includegraphics[scale = 0.195]{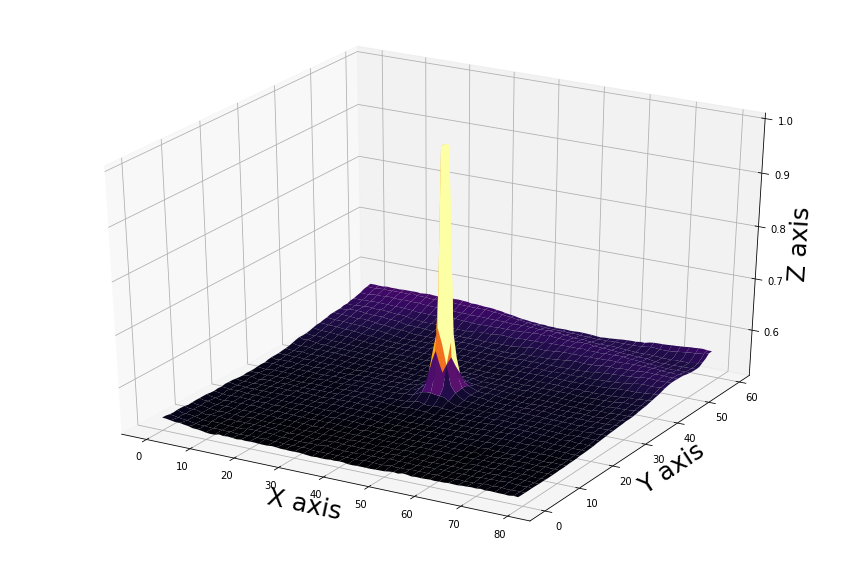}
        \includegraphics[scale = 0.195]{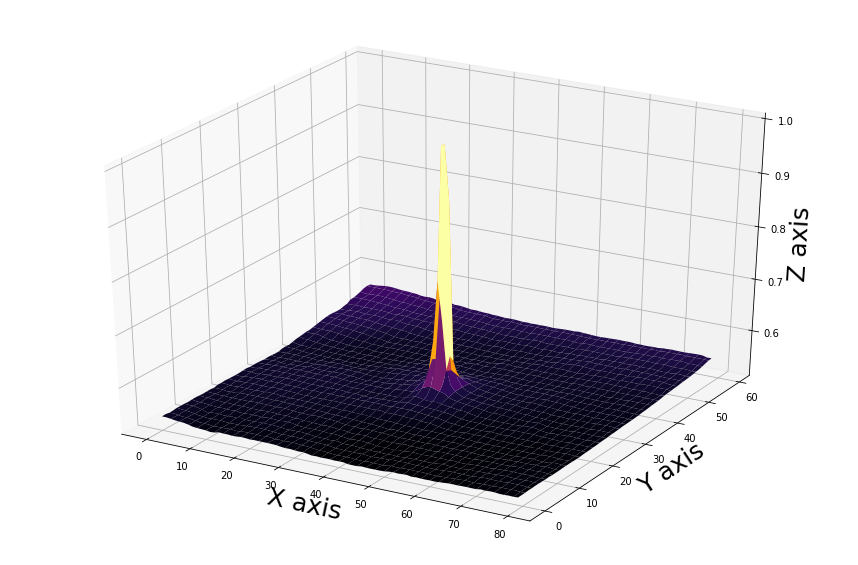}
        \includegraphics[scale = 0.195]{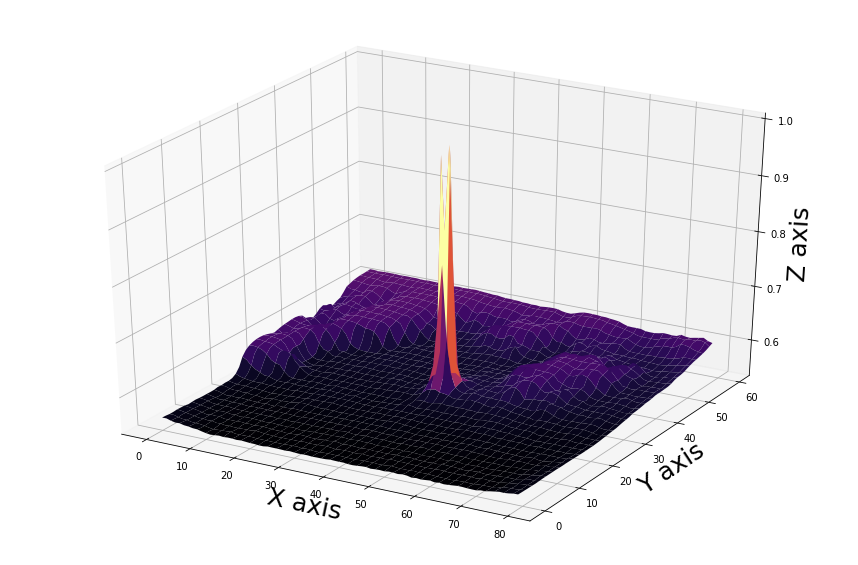}
    \end{subfigure}
    \begin{subfigure}{0.3275\textwidth}
        \centering
        \includegraphics[scale = 0.195]{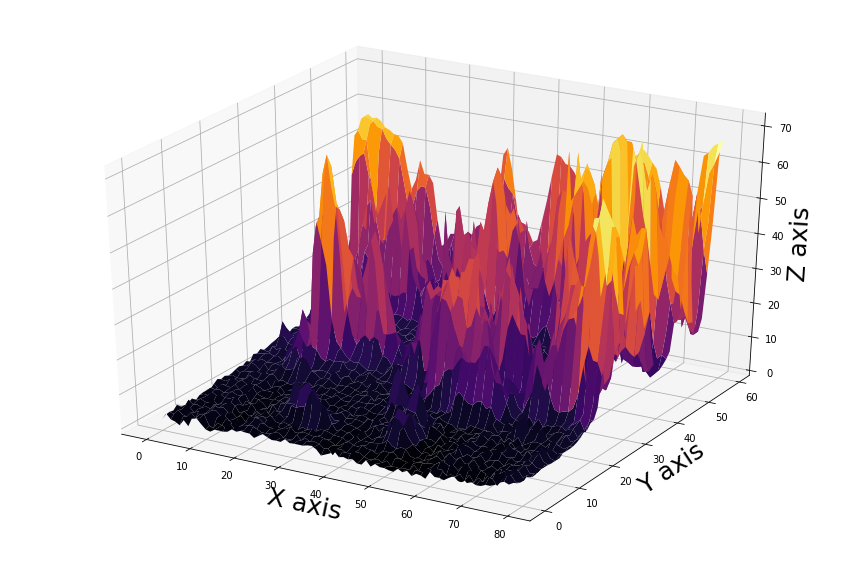}
        \includegraphics[scale = 0.195]{radiometry/norm_1.png}
        \includegraphics[scale = 0.195]{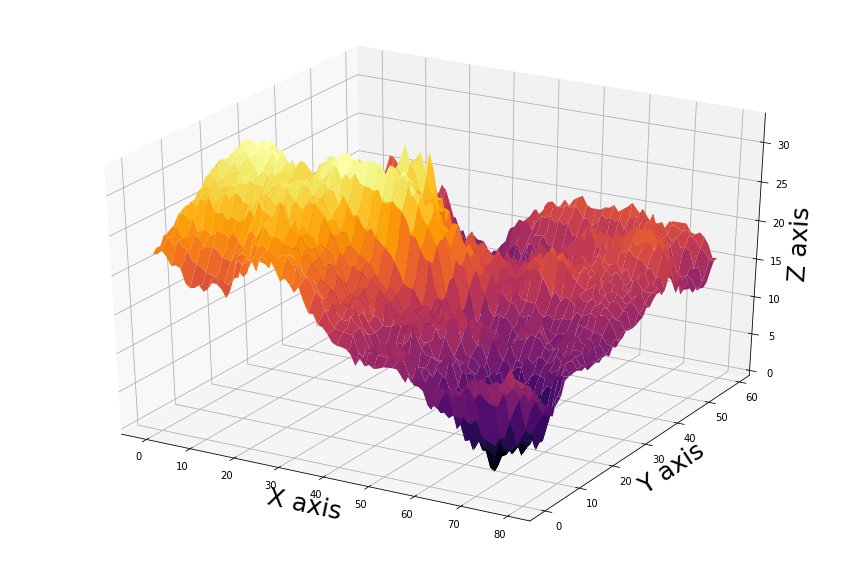}
        \includegraphics[scale = 0.195]{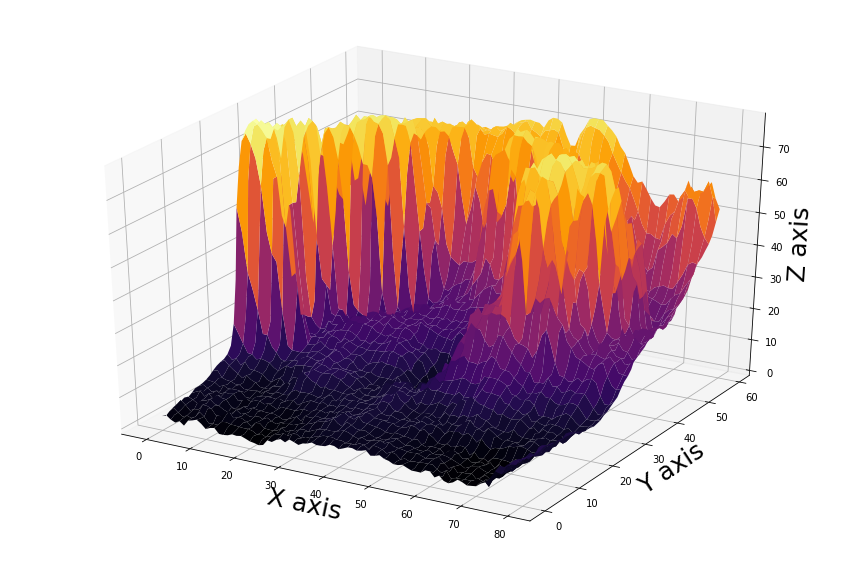}
    \end{subfigure}
\caption{The graphs in each row corresponds to a different day. In the left column, the graphs of the background model. The graphs in the middle display the raw pixels image. The right column graphs show the results after removing the constant background from the raw image.}
\label{fig:decorrelation}
\end{figure}

The Sun has always constant intensity $\tau_0$ in the images unless a cloud is occluding it, so that the direct radiation has to be removed only when it is on the images. Therefore, when the Sun is detected on the image, the atmospheric radiation model $\mathcal{A}\left(\mathbf{x}_0, \boldsymbol{\theta}^*\right)$ is applied, otherwise the scatter radiation model $\mathcal{S}\left(y_0, \boldsymbol{\theta}^*\right)$ is applied,
\begin{align}
\mathbf{\bar{I}}_k = 
\begin{cases} 
    \mathbf{I}^k - \mathcal{S} \left(y_0, \boldsymbol{\theta}^*\right) \ & \max \left[ \mathbf{I}_k \right] < \tau_0 \\
    \mathbf{I}^k - \mathcal{A} \left(\mathbf{x}_0, \boldsymbol{\theta}^*\right) \ & \mathrm{Otherwise}
\end{cases} \quad \mathbf{\bar{I}}_k\in \mathbb{N}^{[1, 2^{13}]}.
\end{align}
When the model $\mathcal{A}\left(\mathbf{x}_0, \boldsymbol{\theta}^*\right)$ is subtracted to the frames can be errors, see figure \ref{fig:decorrelation}. The inaccuracies are due to the low resolution of the IR sensor, plus any other systematic errors added by the tracking system. 

The interpolation of the pixels' intensity which Euclidean distance is $\leq r_0$ from the Sun $\mathbf{x}_0 = \{ x_0, y_0  \}$, can attenuate the effects produced by errors. The interpolation is performed using the remaining pixels in the image $\mathbf{x} = \{ \left( x_{i,j}, y_{i,j} \right) \in \mathbb{N} \ | \ i = 1, \ldots, M, \ j = 1, \ldots,\ r_{i,j} \geq r_0 \}$, to find which one of these are the nearest neighbors,
\begin{align}
    \hat{x}_{i,j} &= \operatorname{argmin} \ \left| \mathbf{x}^1 - x_{i,j} \right|, \\  \hat{y}_{i,j} &= \operatorname{argmin} \ \left|\mathbf{x}^2 - y_{i,j} \right|, \quad \forall i,j \in \mathbb{N}^{x_{i,j} 
    \land y_{i,j} \leq r_0}.
\end{align}
To the intensity of those pixels, that are in the circumsolar region defined as $x_{i,j} 
\land y_{i,j} \leq r_0$, is assigned their nearest neighbour's intensity \cite{ANDREWS1976},
\begin{align}
    \mathbf{\bar{I}}_k \left(x_{i,j}, y_{i,j} \right) = \mathbf{\bar{I}}_k \left( \hat{x}_{i,j}, \hat{y}_{i,j} \right), \quad \forall i,j \in \mathbb{N}^{x_{i,j} 
\land y_{i,j} \leq r_0}.
\end{align}
where $r_0 = 3$ is the radius that we apply in the interpolation, see the results in figure \ref{fig:decorrelation}.

The frame, after removing the direct and scatter atmospheric radiation, is a measure of the reflected radiation. This reflected component is the radiation emitted by objects such as clouds, streetlights, or building structures, but it emitted by particles attached to the window protecting the IR camera. The window is made out of germanium because is a material that allows the transmission of the IR radiation, In next section, we will explore how to numerically model the transmittance of a germanium window.

\section{Window Model}

The window of the camera gets dirty with water spots produced by dried droplets containing dust particles after rainy days, and dust particles suspended in the air that accumulate on the window over time. The window cannot be routinely cleaned because of difficult access to the DAQ localization. Therefore, we propose a persistent model to remove the effects produced by the dust particles and water spots on the window.

\subsection{Atmospheric Condition Model}

The aim is to classify the sky conditions in four possible categories: clear sky, cumulus, stratus, or nimbus cloud. This classification determine when the image has to be segmented, when the sky is clear, we do not need to segment, and when there are stratus or nimbus, the sky is entirely cover. Therefore, the segmentation has to be applied only if there are cumulus clouds. The classification is necessary when thin stratus cover the sky. In this situation, optimal segmentation is difficult to achieve, however we know that the entire image is cover with a thin layer a cloud. 

We propose a persistent linear Support Vector for Classification (SVC) for this tasks due to the large number of clouds samples \cite{FAN2008, HSU2010}. The formulation of the unconstrained optimization problem for a dataset $\mathcal{D} = \left\{ \mathbf{X}, \ \mathbf{y} \right\}$, where $\mathbf{X} = \left\{ \mathbf{x}_i \in \mathbb{R}^D \mid \forall i = 1, \ldots, N  \right\}$, and $\mathbf{y} = \left\{ y_i \in \left[ -1,+1 \right] \mid \ \forall i = 1, \ldots, N \right\}$ is,
\begin{align}
    \min _{\mathbf{w}} \frac{1}{2} \|\mathbf{w} \| +C \sum_{i=1}^N \xi \left( \mathbf{w} ; \mathbf{x}_i, y_i \right)
\end{align}
where $C > 0$ is the penalty term, and $\xi\left( \mathbf{w} ; \mathbf{x}_i, y_i\right)$ is the loss function that in our model is the L-2 for a SVC, so we have this,
\begin{align}
    \min _{\mathbf{w}} \frac{1}{2} \| \mathbf{w}\|_2+C \sum_{i=1}^{N} \left( \max \left(0,1-y_i \mathbf{w}^\top \mathbf{x}_i \right) \right)^2.
\end{align}

The solution in the dual formulation is the following constrained quadratic programming problem,
\begin{align}
    & \min_\mathbf{\alpha} \ \frac{1}{2} \mathbf{\alpha}^\top \mathbf{Q} \mathbf{\alpha} - \mathbf{1}^\top \mathbf{\alpha} \\ 
    \mathrm{s.t.} \quad & 0 \leq \alpha_i \leq C, \ i = 1, \ldots, N
\end{align}
where $Q_{i,j} = y_i y_j \mathbf{x}_i^\top \mathbf{x}_j$, $\mathbf{1} = \left[ 1_1 \ldots, 1_N \right]^\top$, and $N$ is the number of samples.

The standard solution of a liner SVC is defined for two possible categories. We have four categories, so this is a multi-class classification problem. We propose to solve it implementing the one-versus-all scheme. Hence, there will have a different set of labels for each one of the linear SVC that is,
\begin{align}
    \tilde{\mathbf{y}}_{i,\ell} = 
    \begin{cases}
    +1 \ & y_i = \ell \\
    -1 \ & \mathrm{Otherwise}
    \end{cases} \quad \forall i = 1, \ldots, N, \ \forall \ell = 1, \ldots, L.
\end{align}

Notice that the bias is included in the weights, so for a new sample $k$ we obtain a classification such as,
\begin{align}
    \hat{y}_k = \underset{\ell \in \mathcal{L}}{\operatorname{argmax}} \ \mathrm{sign} \left( \mathbf{w}_\ell^\top \mathbf{x}_k \right).
\end{align}

The features vector includes the atmospheric pressure $p_k$ and the CSI $i_k$ collected by a weather station. In addition to a set of statistical metrics that are the mean, variance, kurtosis, and skewness. These were computed using the pixels' of the radiometric measures of temperature in the IR image $\mathbf{T}_k$, and the magnitude of the velocity vectors $\mathbf{M}_k = \sqrt{\mathbf{U}_k^2 + \mathbf{V}_k^2}$. The feature vector for each sample $k$ is,
\begin{align}
    \mathbf{x}_k = \left[ p_k \ i_k \ \mathbb{E} \left( \mathbf{T}_k \right) \ \mathbb{V} \left( \mathbf{T}_k \right) \ \mathbb{S} \left( \mathbf{T}_k \right) \ \mathbb{K} \left( \mathbf{T}_k  \right) \ \mathbb{E} \left( \mathbf{M}_k \right) \ \mathbb{V} \left( \mathbf{M}_k \right) \ \mathbb{K} \left( \mathbf{M}_k \right) \ 1 \right]^\top,
\end{align}
the skewness of the magnitude is not included as the model under-performs when is included.

A polynomial expansion is applied to each vector such as $\mathbf{x}_k \to \varphi \left( \mathbf{x}_k \right) \in \mathbb{R}^{\mathcal{P}^n}$, where $n$ is the polynomial expansion of order, and the atmospheric condition classes are such as $y_k \in \{ 0, 1, 2, 3 \}$. The dataset in matrix form is,
\begin{align}
    \mathbf{X} = 
    \begin{bmatrix}
    \varphi \left( \mathbf{x}_1 \right)  \\
    \vdots \\
    \varphi \left( \mathbf{x}_N \right)
    \end{bmatrix}, \ \mathbf{X} \in \mathbb{R}^{N \times \mathcal{P}^n},
    \quad \mathbf{y} = 
    \begin{bmatrix}
    y_1 \\
    \vdots \\
    y_N
    \end{bmatrix}, \ \mathbf{y} \in \mathbb{N}^N.
\end{align}

In the experiments, we aim to find the polynomial expiation order $n$ and together with the combination of weather features that produces the highest accuracy in the atmospheric condition classification. We implement a leave-one-out (LOO) Cross-Validation (CV) to find $C$ for each $n$, and we compare the accuracy in validation of the linear SVC different models.

The different pre-classified atmospheric condition sequences that composed the dataset were divided into batches of the same size to be each of the samples in LOO-CV implementation. The samples were not scramble to preserve the time structure in the sequences. The sequences of samples belong to nonconsecutive 21 days out of a year of recordings, that were previously selected as the most suitable for the training due to containing artifacts (trees, structures, buildings, light posts, water marks in the window, etc...).

\begin{figure}[htbp]
    \includegraphics[scale = 0.4]{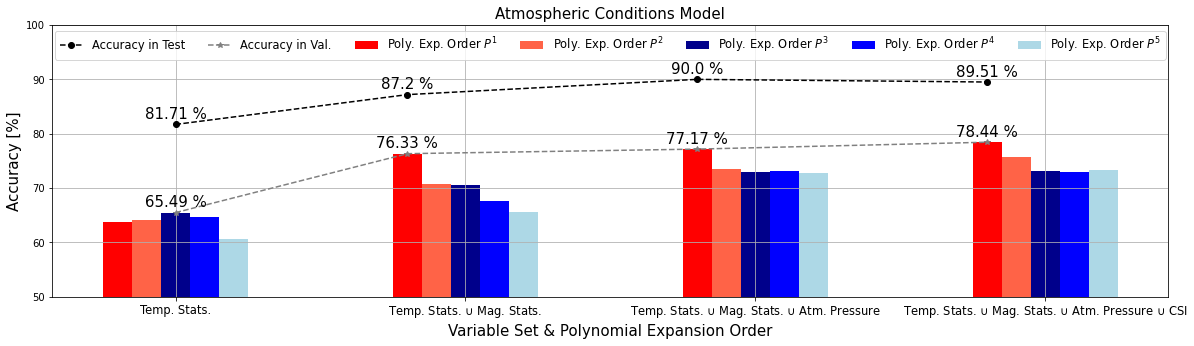}
    \centering
\centering
\caption{The graph shows the accuracy obtained in the classification during the validation of the polynomial expansion order and the weather features. The SVC without polynomial expansion and that uses all the weather features has 89.51\% accuracy.}
\label{fig:condition_cv}
\end{figure}

Histograms instead of the statistics from the pixels' temperature and magnitude was found to under-perform and consequently discarded from further experiments. The weather station measures that were found to be uncorrelated with the classification are: air temperature, dew point, relative humidity, the Sun's elevation and azimuth.
\begin{figure}[htbp]
    \includegraphics[scale = 0.45]{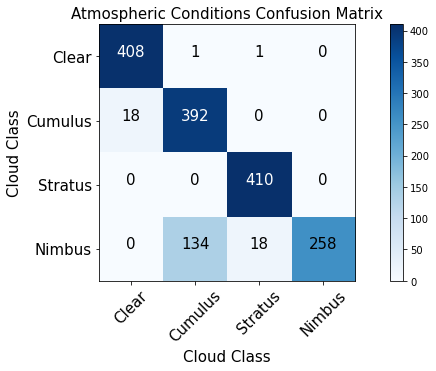}
    \centering
\centering
\caption{The graph shows the confusion matrix obtained in the test of the SVC without polynomial expansion, and using all the features.}
\label{fig:condition_matrix}
\end{figure}

The accuracy of the classification model implies that there will be miss-classifications, but as transition between type of sky conditions is slow, we propose to implement a persistent classification so sporadic errors due to artifacts can be avoided,
\begin{align}
    \mathbf{\hat{y}}_k = \left[\hat{y}_k, \ldots, \hat{y}_{k - T} \right],
\end{align}
where $T$ is the lag in the persistent time series of consecutive classifications. We classify frame $k$ such as,
\begin{align}
    \hat{y}^{\prime}_k = \mathrm{Mode} \left( \mathbf{\hat{y}}_k \right).
\end{align}
Notice that the persistent classification vector of classes does not contain the modes, is updated with the output of the model classification.

\subsection{Persistent Window Model}

We have the set $\mathcal{I}$ that contains up to $L$ selected frames. These frames had previously applied the atmospheric model so that is only remaining the reflected radiation from the clouds, and other particles, a part from other objects that can be on the images and that were not accounted for,
\begin{align}
    \mathcal{I}_k = \left\{ \mathbf{\bar{I}}_1, \ldots, \mathbf{\bar{I}}_L \right\}, \quad \mathbf{\bar{I}}_k \in \mathbb{R}^{D \times N}.
\end{align}
The objective is that as the tracking system rotates the particles accumulated on the surface of the window stay constant over short periods of time, and consequently is possible to model them. 

In addition, we have a vector $\mathbf{i}_k$ that contains the sequence of last $\ell$ measures of CSI,
\begin{align}
    \mathbf{i}_k = \left[ i_k, \ldots, i_{k-\ell} \right], \quad i_k \in \mathbb{R}^+.
\end{align}
When the Sun is totally or partially occluded, the absolute value of the drop in CSI $i_k$ is expected to be high, 
\begin{align}
    r_k  = \left| 1 - \frac{1}{\ell}\sum_{i = 0}^{\ell} i_{k - i} \right|, \quad r_k \in \mathbb{R}^+.
\end{align}
The total amount of CSI $r_k$ is another indicator of when there was not stratus, nimbus, or any other type of large cloud. The aim is to add robustness to atmospheric conditions  classification, so that a new frame is added to the set iff the classification is clear sky and the drop on the CSI was under a threshold,
\begin{align}
    \mathcal{I}_k  = 
    \begin{cases} 
    \mathcal{I}_k \cup \mathbf{\bar{I}}_k \quad & \mathcal{C}_k = 0 \land r_k \leq 0.05 \\
    \mathcal{I}_k \cap \mathbf{\bar{I}}_k \quad & \mathrm{Otherwise,}
    \end{cases}
\end{align}
so that $\mathcal{C}_k = 0 $ and the robustness condition $ r_k \leq 0.05$ are always fulfill. The window's artifacts model is updated when there are enough frames $L$ with clear sky conditions to attenuate the variance of the noise, 
\begin{align}
    \omega_{i,j} = \mathrm{Median} \left( \mathcal{I}_k \right), \quad \mathrm{iff} \ \left| \mathcal{I}_k \right| \geq L,
\end{align}
where $|\cdot|$ the cardinally of a set. When the DAQ session is finished for the day, an algorithm randomly selects $L$ elements in the set form sampling an uniform distribution to initialize the set that will be used in the next day.
\begin{figure}[ht]
    \begin{subfigure}{0.245\textwidth}
        \centering
        \includegraphics[scale = 0.15]{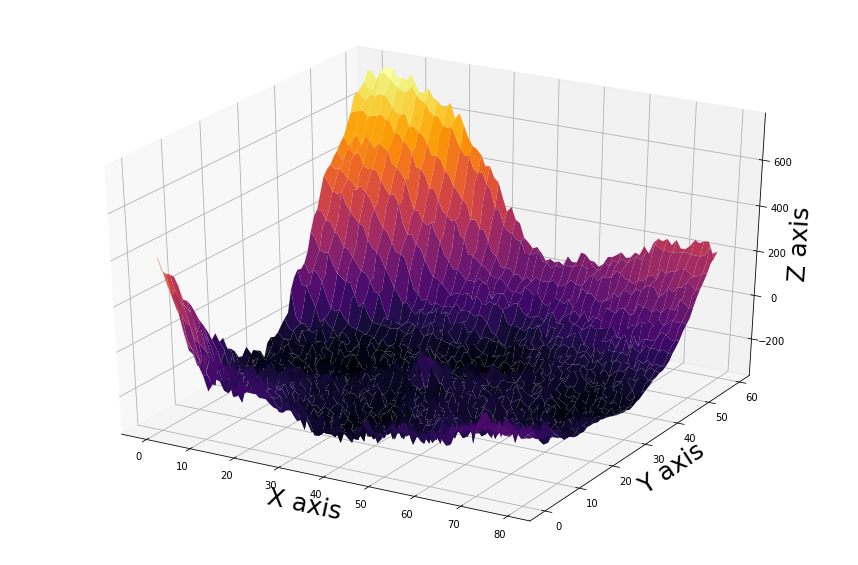}
    \end{subfigure}
    \begin{subfigure}{0.245\textwidth}
        \centering
        \includegraphics[scale = 0.15]{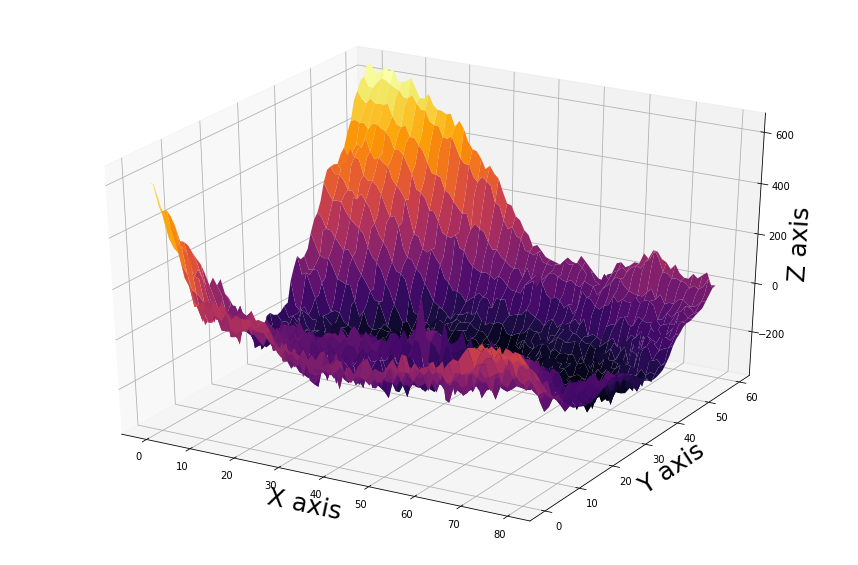}
    \end{subfigure}
    \begin{subfigure}{0.245\textwidth}
        \centering
        \includegraphics[scale = 0.15]{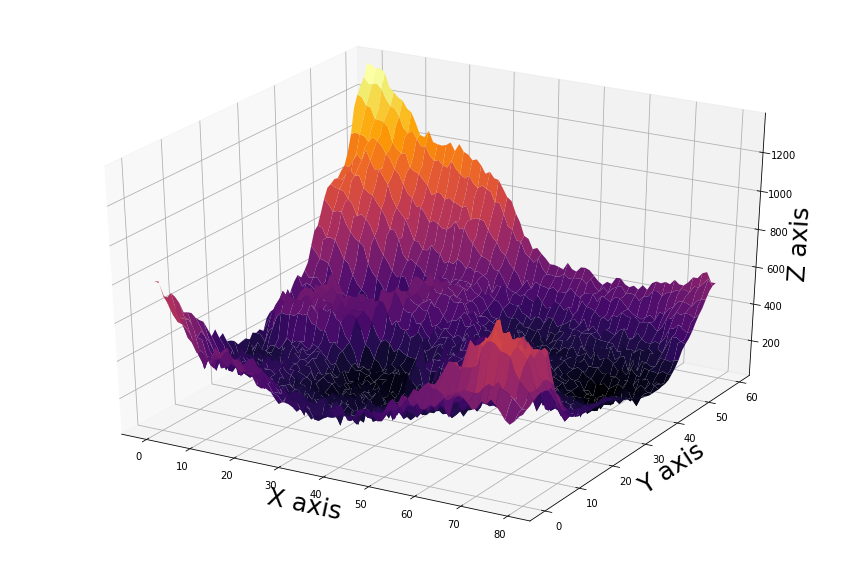}
    \end{subfigure}
    \begin{subfigure}{0.245\textwidth}
        \centering
        \includegraphics[scale = 0.15]{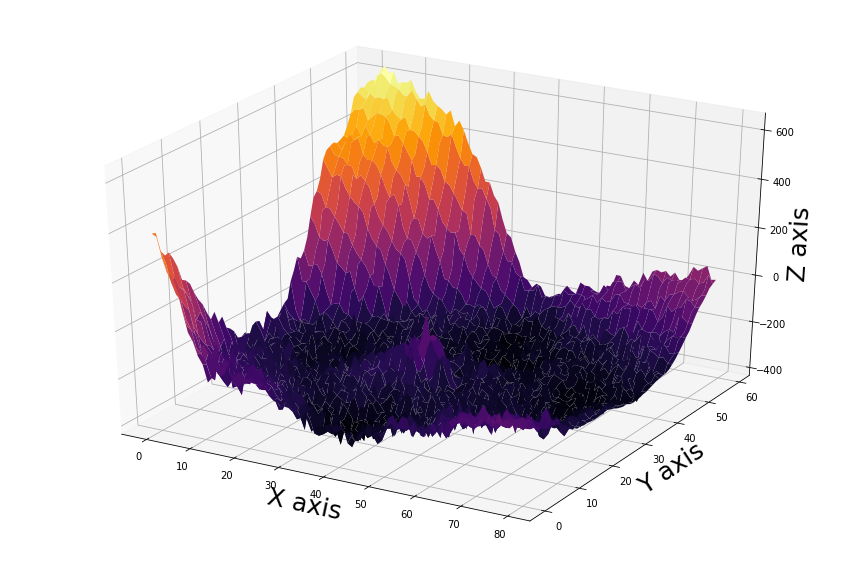}
    \end{subfigure}
\caption{The graphs show the effects of window produced in the radiation measure by the IR sensor. Although theses effect varies along the day, the changes are most visually appreciable after heavy rain. This algorithm requires $L \approx 250$ selected frames to learn the artifacts on the window. This task can be performed either during the night or day time.}
\label{fig:window_model}
\end{figure}

\subsection{Window Model Application}

In order to avoid any window artifacts that can affect computation of the clouds motion vector, and the quality of the segmentation, the window model is applied to image after the atmospheric model subtraction, 
\begin{align}
\mathbf{\bar{I}}_k = \mathbf{\bar{I}}_k - 
\omega_{i,j}, \quad \forall i = 1, \ldots, M, \ \forall j = 1, \ldots, N, \quad \mathbf{\bar{I}}_k\in \mathbb{N}^{[1, 9700]}.
\end{align}
The resulting image is the reflect radiation from the clouds.

The fact of having the cloud's intensity in the same interval facilitates the segmentation, and thereby the clouds labelling. To normalize an infrared image of 16 bits to 8 bits is necessary to consider that the average distance from the sea level to the Tropopause is $\sim 12 \ km$ at $36^\circ$ latitude north\footnote{https://www2.acom.ucar.edu/news/cloud-tops-and-tropopause}, but Albuquerque is at $1,641 \ km$ above the sea level. Then, if the air temperature decreases by $9.8 \frac{\circ}{km}$, the maximum feasible intensity that a cloud can have, should be at around: $9,8 \cdot (11,5 - 1,6) \cdot 100 = 9,7 \times 10^3$. The interval where the clouds are recognizable in the images $\mathbf{\bar{I}}_k$ is,

\begin{equation}
    \mathbf{\tilde{I}}_k = \frac{\mathbf{\bar{I}}_k - \min \left[ \mathbf{\bar{I}}_k \right]}{ 9,7 \times 10^3} \cdot 2^8, \quad \mathbf{\tilde{I}}^k \in \mathbb{N}^{[1, 2^{8}]}.
\end{equation}
The minimum intensity $\min \left[\mathbf{\bar{I}}_k\right]$
can vary between consecutive images, and it can have a value either above zero (this is most in cloudy condition) or below zero. It is a offset or a floating point in the signal cause by the noise introduced in errors of the atmospheric and window models. Nevertheless, as the normalization is only for segmentation purposes, not for feature extraction, this is not an issue.

\begin{figure}[htbp]
    \begin{subfigure}{0.245\textwidth}
        \centering
        \includegraphics[scale = 0.135]{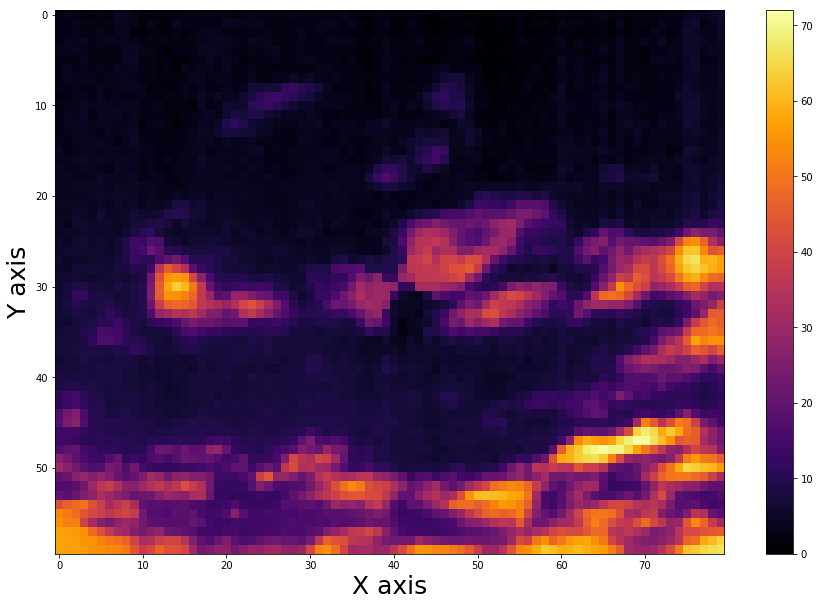}
    \end{subfigure}
    \begin{subfigure}{0.245\textwidth}
        \centering
        \includegraphics[scale = 0.135]{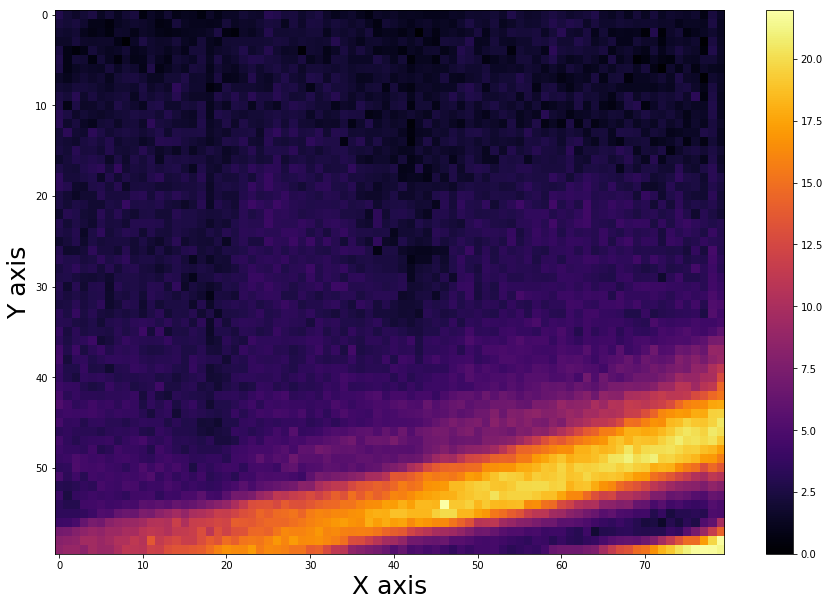}
    \end{subfigure}
    \begin{subfigure}{0.245\textwidth}
        \centering
        \includegraphics[scale = 0.135]{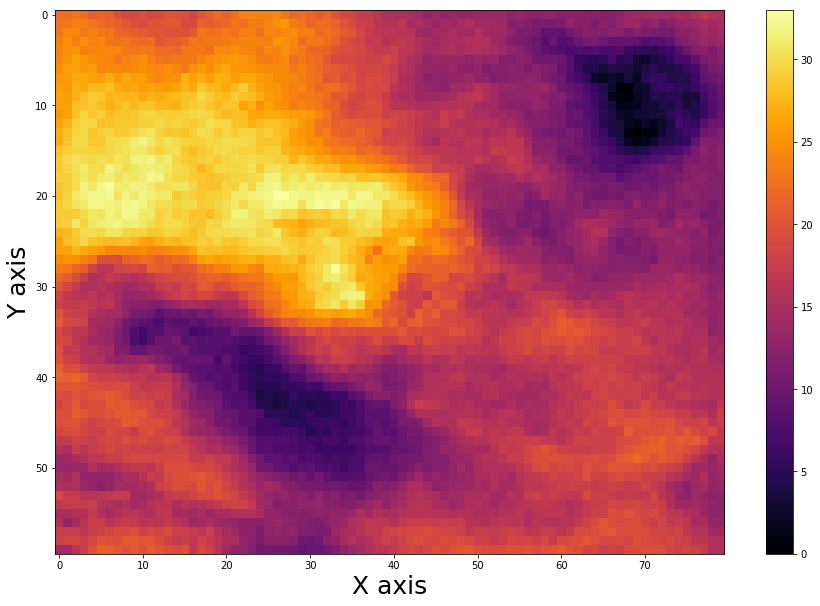}
    \end{subfigure}
    \begin{subfigure}{0.245\textwidth}
        \centering
        \includegraphics[scale = 0.135]{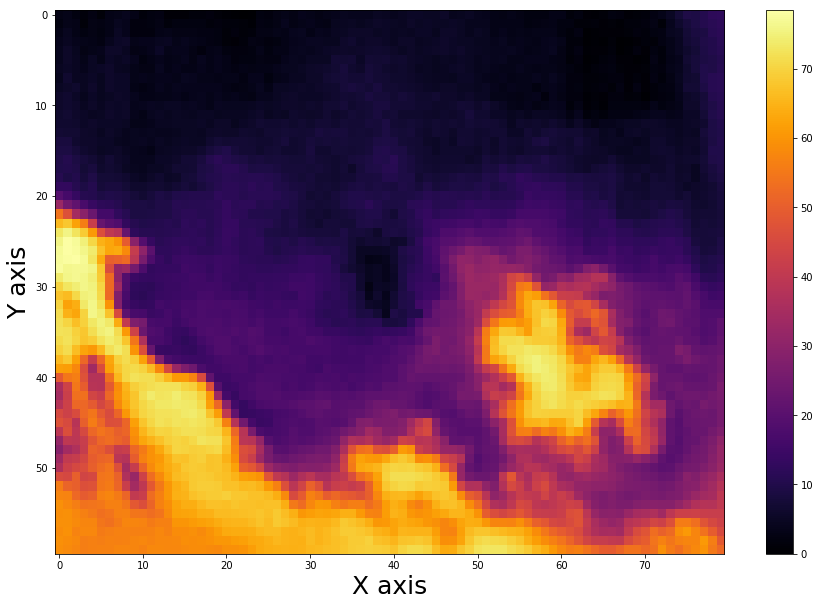}
    \end{subfigure}
\caption{View of the images shown in previous figures after applying the models. The images in this figure from left to right corresponds to the graphs left top to right as well.}
\label{fig:normalization}
\end{figure}

\section{Perspective Transformation}

The images are taken at an angle from the normal position of the camera in relation to the ground. This angle is the Sun's zenith angle. This causes, that the relative distance of a object in horizon, to increase from the top to the bottom of an image. In order to account for this effect, we propose a geometrical transformation from the original euclidean frame of reference, to a non-linear frame of reference.

\subsection{Rectilinear Lens}

The image captured by an IR camera is a refraction of the black body radiation emitted by objects in a converging point. We consider that the radiant objects, which are the Sun and the clouds, are at an infinity distance from the camera's germanium lens, so the radiation rays converge at the focal length. This can be shown by the thin lens equation, which is
\begin{align}
    \frac{1}{f} = \frac{1}{z} + \frac{1}{d_{i}},
\end{align}
where $f$ is the focal length, $z$ is the distance from the lens to the clouds, and $d_{i}$ is the distance from the lens to converging point. If the equation factor is $0 \approx \frac{1}{z}$, then the rays converge at the focal length $d_{i} = f$.

The relation between the diagonal Field of View ($\mathrm{FOV}$), and the focal length $f$ for a standard rectilinear lens is,
\begin{align}
    \tan \frac{\mathrm{FOV}}{2} = \frac{N_{diag}}{2f}
\end{align}
where $N_{diag}$ is the number of pixels in the diagonal of the sensor. As we know from the camera's manufacture: the diagonal $\mathrm{FOV} $ is $ 63.75^\circ$, the horizontal $\mathrm{FOV}_x$ is $51^\circ$, and the size of a pixel is $ \delta = 17\mu m$. Therefore, the focal length of our camera is,
\begin{align}
    f = \frac{\sqrt{ \left(  N \delta \right)^2 + \left( M \delta \right)^2 }}{2\tan \frac{\mathrm{FOV}}{2}}
\end{align}
that is a camera with resolution of $N \times M$ pixels, and thus has a vertical $\mathrm{FOV}_y$ of $38.25^\circ$.

\subsection{Flat Earth Approximation}

\begin{figure}[!htbp]
    \centering
    \resizebox{12.5cm}{!}{\scalebox{1.}{
\begin{tikzpicture}[thick,help lines/.style={ultra thin, draw=black!50}]
        \def\A{Ground} 
        \def\B{$ $}
        \def\C{Cloud} 
        \def\D{$ $}
        \def\E{$ $} 
        \def\F{\textcolor{red}{Lens}}
        \def\G{$y_0^\prime$} 
        \def\I{$ $}
        \def\H{$ $}
        \def\J{$ $} 
        \def\K{$ $}
        \def\L{$ $} 
        \def\M{$ $}
        \def\N{$y_0$} 
        \def\O{$ $}
        \def\Q{\textcolor{orange}{Sensor Plane}}
        \def\T{$ $}
        \def\R{$ $} 
        \def\Z{$ $}
        \def\W{$ $} 
        \def\X{$ $}
        \def\P{Cross-Section Plane} 
        \def\S{$ $} 

        \def\AA{$ $} 
        \def\BB{$ $}
        \def\CC{$ $} 
        \def\DD{$ $}
        \def\EE{$ $} 
        \def\FF{$ $}
        \def\GG{$x_0^\prime$} 
        \def\HH{\textcolor{orange}{Sensor Plane}}
        \def\II{$ $}
        \def\JJ{$x_0$} 
        \def\KK{\textcolor{red}{Lens}}
        \def\LL{$ $}
        \def\MM{$ $}
        \def\NN{$ $}
        \def\OO{$ $}
        \def\PP{$ $}
        \def\QQ{$ $}
        \def\RR{$ $}
        \def\SS{$ $}
        \def\TT{$ $}
        \def\UU{$ $}
        \def\WW{$ $}
        \def\XX{$ $}
        \def\YY{$ $}
        \def\ZZ{$ $}
        
        \def\AAA{$ $}
        \def\BBB{$ $}
        \def\CCC{\textcolor{orange}{Camera Plane}}
        \def\DDD{Cross-Section Plane}
        \def\EEE{$ $}
        \def\FFF{$ $}
        \def\GGG{$ $}
        \def\HHH{$ $}
        \def\III{$ $}
        \def\JJJ{$ $}
        \def\KKK{$ $}
        \def\LLL{$ $}
        \def\MMM{$ $}
        
        \def\NNN{$ $}
        \def\OOO{$ $}

        \colorlet{ground}{brown!70!black} \colorlet{cloud}{blue!85!black}
        \colorlet{sensor}{orange!} 
        \colorlet{fov}{black!60!black!} 
        \colorlet{sun}{yellow!} 
        
        \coordinate [label=left:\A] (A) at ($ (0, 2) $);
        \coordinate [label=left:\B] (B) at ($ (10, 2) $);
        \coordinate [label=left:\C] (C) at ($ (0, 5) $);
        \coordinate [label=left:\D] (D) at ($ (10, 5) $);
        
        \draw [ultra thick, ground] (A) -- (B);
        \draw [thick, cloud] (C) -- (D);
        
        \coordinate [label=above:\W] (W) at ($ (5.5, 5) $);
        \coordinate [label=below:\O] (O) at ($ (2., 1) $);
        \coordinate [label=above:\H] (H) at ($ (7.675, 7.) $);
        \coordinate [label=above:\K] (K) at ($ (10.05, 5.5) $);
        
        \coordinate [label={[xshift=-.6cm, yshift=0.05cm]\F}] (F) at ($ (2.87,2) $);
        \coordinate [label={[xshift=-0.05cm, yshift=-0.55cm]\N}] (N) at ($ (1.8, 1.23) $);
        \coordinate [label={[xshift=.1cm, yshift=0.05cm]\G}] (G) at ($ (7, 5) $);
        \coordinate [label={[xshift=-1.cm, yshift=-0.1cm]\Q}] (Q) at ($ (1.65, 1.45) $);
        
        \draw [ultra thick, sensor] (O) -- (Q);
        \draw [thin, fov] (H) -- (O);
        \draw [thin, fov] (K) -- (Q);
        
        \draw [->] (-2, 1) -- (0, 1) node [midway, align=right, xshift=0.cm, yshift=-0.2cm]{$y-axis$};
        \draw [->] (-2, 1) -- (-2, 3) node [midway, align=right, xshift=-0.2cm, yshift= 0.cm, rotate = 90]{$z-axis$};
        \draw [->] (-2, -6) -- (0, -6) node [midway, align=right, xshift=0.cm, yshift=-0.2cm]{$y-axis$};
        \draw [->] (-2, -6) -- (-2, -4) node [midway, align=right, xshift=-0.2cm, yshift= 0.cm, rotate = 90]{$x-axis$};
        
        \coordinate [label=left:\R] (R) at ($ (0, 3.7) $);
        \coordinate [label=right:\Z] (Z) at ($ (0, 5.85) $);
        \coordinate [label=left:\L] (L) at ($ (2.75, 0) $);
        \coordinate [label=right:\M] (M) at ($ (4.35, 0) $);
        \coordinate [label=right:\AA] (AA) at ($ (8.2, 6) $);
        \coordinate [label=right:\BB] (BB) at ($ (10, 3.75) $);
        \coordinate [label=below:\CC] (CC) at ($ (0.1, 0) $);
        \coordinate [label=above:\DD] (DD) at ($ (8.35, 6) $);
        
        \draw [thick, sun] (CC) -- (DD);
        
        \coordinate [label=above:\S] (S) at ($ (5.75, 5) $);
        \coordinate [label={[xshift=-4.75cm, yshift=0.1cm]\P}] (P) at ($ (9, 5) $);
        
        \draw [ultra thick] (S) -- (P);
        
        \coordinate [label=above:\NNN] (NNN) at ($ (7.55, 6.85) $);
        \coordinate [label=above:\OOO] (OOO) at ($ (6.65, 3.85) $);
        
        \coordinate [label=above:\BBB] (BBB) at ($ (7.6, 4.3) $);
        \coordinate [label={[xshift=-1.cm, yshift=0.cm]\CCC}] (CCC) at ($ (6.425, 5.71) $);
        
        \draw [ultra thick, sensor] (BBB) -- (CCC);
        \draw [sensor] (NNN) -- (P);
        \draw [sensor] (S) -- (OOO);

        \coordinate [label=below:\EE] (EE) at ($ (0, -4) $);
        \coordinate [label=above:\FF] (FF) at ($ (10, -4) $);
        
        \coordinate [label={[xshift=.0cm, yshift=.0cm]\GG}] (GG) at ($ (7, -4) $);
        \coordinate [label={[xshift=-.9cm, yshift=0.25cm]\HH}] (HH) at ($ (1.65, -4) $);
        \coordinate [label={[xshift=-.3cm, yshift=-0.4cm]\JJ}] (JJ) at ($ (1.8, -4) $);
        \coordinate [label={[xshift=0.cm, yshift=0.1cm]\KK}] (KK) at ($ (2.87, -4) $);

        \coordinate [label=below:\LL] (LL) at ($ (7., -7) $);
        \coordinate [label=above:\MM] (MM) at ($ (1.65, -7) $);
        \coordinate [label=below:\NN] (NN) at ($ (1.8, -7) $);
        \coordinate [label=above:\OO] (OO) at ($ (2.87, -7) $);
        \coordinate [label=below:\QQ] (QQ) at ($ (7, 7) $);
        \coordinate [label=below:\SS] (SS) at ($ (1.8, 7) $);
        \coordinate [label=above:\TT] (TT) at ($ (2.87, 7.) $);
        
        \draw [dashed, help lines] (LL) -- (QQ);
        \draw [dashed, help lines] (NN) -- (SS);
        \draw [dashed, help lines] (OO) -- (TT);
        
        \coordinate [label=above:\RR] (RR) at ($ (9., 7) $);
        \coordinate [label=above:\UU] (UU) at ($ (9, -7) $);
        \coordinate [label=above:\WW] (WW) at ($ (5.75, 7) $);
        \coordinate [label=above:\PP] (PP) at ($ (5.75, -7) $);

        \draw [dashed, help lines] (UU) -- (RR);
        \draw [dashed, help lines] (PP) -- (WW);

        \coordinate [label=above:\XX] (XX) at ($ (10, -1.5) $);
        \coordinate [label=above:\YY] (YY) at ($ (10, -6.5) $);
        \coordinate [label=above:\ZZ] (ZZ) at ($ (1.8, -3.6) $);
        \coordinate [label=above:\AAA] (AAA) at ($ (1.8, -4.4) $);

        \draw [thin, fov] (XX) -- (AAA);
        \draw [thin, fov] (YY) -- (ZZ);
        \draw [thick, sun] (EE) -- (FF);

        \draw [ultra thick, sensor] (ZZ) -- (AAA);
        
        \coordinate [label={[xshift=-1.25cm, yshift=0.1cm]\DDD}] (DDD) at ($ (9, -1.85) $);
        \coordinate [label=above:\EEE] (EEE) at ($ (5.75, -3.) $);
        \coordinate [label=above:\FFF] (FFF) at ($ (9, -6.15) $);
        \coordinate [label=above:\GGG] (GGG) at ($ (5.75, -5) $);   
        
        \draw [ultra thick] (DDD) -- (EEE);
        \draw [ultra thick] (DDD) -- (FFF);
        \draw [ultra thick] (EEE) -- (GGG);
        \draw [ultra thick] (FFF) -- (GGG);
        
        \coordinate [label=above:\I] (I) at ($ (-1, -1.85) $);   
        \coordinate [label=above:\II] (II) at ($ (10, -1.85) $);   
        \coordinate [label=above:\III] (III) at ($ (-1, -3.) $);   
        \coordinate [label=above:\HHH] (HHH) at ($ (10, -3.) $);   
        \coordinate [label=above:\JJJ] (JJJ) at ($ (-1, -2.55) $);   
        \coordinate [label=above:\KKK] (KKK) at ($ (10, -2.55) $);   
        
        \draw [dashed, help lines] (I) -- (II);
        \draw [dashed, help lines] (III) -- (HHH);
        \draw [dashed, help lines] (JJJ) -- (KKK);

        \draw [<->] (9.5, -1.85) -- (9.5, -3.) node [midway, align=right, xshift=0.55cm, yshift=0.cm]{$\Delta \mathbf{x}_j/2$};
        \draw [<->] (9, -4) -- (5.75, -4) node [midway, align=right, xshift=0.2cm, yshift=-0.25cm]{$\Delta \mathbf{y}_i$};

        \foreach \point in {N, G, GG, JJ}
        \fill [black, opacity = .75] (\point) circle (1.5pt);
        \foreach \point in {F, KK}
        \fill [red, opacity = .75] (\point) circle (2.5pt);
        
        \draw [<->] (1.85, 1.25) -- (2.80, 1.95) node [midway, align=right, xshift=0.2cm, yshift=-0.2cm]{$f$};
        \draw [<->] (1,2.05) -- (1, 4.95) node [midway, align=left, xshift=-0.2cm]{$h$};
        \draw [<->] (2.95, 2.05) -- (6.95, 4.95) node [midway, yshift=0.3cm]{$d_i$};
        \draw [<->] (4.5,2) arc (0.5:39.5:1.5) node [midway, align=right, xshift = 0.15cm]{$\varepsilon_i$};
        \draw [<->] (5.75, 3.4) arc (0:67.5:1.) node [midway, align=right, xshift = 0.4cm, yshift = -0.75cm]{$\alpha_y$};
        \draw [<->] (6.25, 3.65) arc (5:42.5:1.) node [midway, align=right, xshift = 0.5cm, yshift = -0.4cm]{$\alpha_y/2$};
        
        \draw [<->] (4.65, -4.575) arc (-35:35:1.) node [midway, align=right, xshift = 0.35cm, yshift = 0.cm]{$\alpha_x$};
        \draw [<->] (7.5, -3.) arc (-18:18:1.) node [midway, align=right, xshift = 0.5cm, yshift = 0.cm]{$\alpha_x/2$};
        
        \coordinate [label=above:\X] (X) at ($ (8.25, -5.25) $);
        \coordinate [label=above:\LLL] (LLL) at ($ (8.25, 5.) $);
        
        \coordinate [label=above:\MMM] (MMM) at ($ (7.45, 4.5) $);
       
        \foreach \point in {X, LLL, MMM}
        \fill [green!60!black, opacity = .75] (\point) circle (2pt);
        
        \draw[green!60!black, line width=1.75pt,-stealth, ](X)--(8.25, -4.65) node[anchor=south west, midway, align=right, xshift = 0.cm, yshift = -0.4cm]{$\mathbf{v}^{\prime \prime}_x$};
        \draw[green!60!black, line width=1.75pt,-stealth, ](X)--(7.65, -5.25) node[anchor=south west, midway, align=right, xshift = -0.25cm, yshift = -0.65cm]{$\mathbf{v}^{\prime \prime}_y$};
        \draw[green!60!black, line width=1.75pt,-stealth, ](LLL)--(7.65, 5.) node[anchor=south west, midway, align=right, xshift = -0.25cm, yshift = 0.cm]{$\mathbf{v}^{\prime \prime}_y$};
        \draw[green!60!black, line width=1.75pt,-stealth, ](MMM)--(7.15, 4.85) node[anchor=south west, midway, align=right, xshift = 0.cm, yshift = -0.35cm]{$\mathbf{v}^\prime_y$};
\end{tikzpicture}}}
    \caption{Schematic of the geometric transformation assuming that the recorded portion of the atmosphere is small enough to be viable to approximate it as flat. The upper plot shows the transformation in the $\mathrm{FOV}$ y-axis relative to the distance from the camera to a cloud at a height $h$. The bottom plot shows the x-axis and y-axis differential increments calculate by the proposed transformation w.r.t. the camera plane.}
\label{fig:flat_earth}
\end{figure}

An approximate of perspective transformation can be easily found with the assumption that the Earth's surface is flat. This consideration is practicable without large error due to small proportion recorded of the Earth's atmosphere segment. In this approximation, the perspective transformation from the sensor plane to the camera plane, as described in figure \ref{fig:flat_earth}, is a function of a cloud layer height $h$ and the Sun's elevation angle $\varepsilon$,
\begin{align}
    d_i = \frac{h}{\sin \varepsilon_i}.
\end{align}
Applying the rectilinear lens approximation, the distance of the atmosphere projected into a pixel is $x^{\prime} = x \cdot \frac{\delta}{f} \cdot d$. Therefore, knowing the pixels' size and the focal length of the camera, the set of coordinates in the camera plane are given by the following formulas,
\begin{align}
    x^\prime_{i,j} &= x_{i,j} \cdot \frac{\delta}{f} \cdot d_i = x_{i,j} \cdot \frac{\delta}{f \cdot \sin \varepsilon_i} \cdot h
\end{align}
and,
\begin{align}
    y^\prime_{i, j} &= y_{i, j} \cdot \frac{\delta}{f \cdot \sin \varepsilon_i} \cdot h 
\end{align}

In order to transform the coordinates of the frame of reference from the camera plane to the cross-section plane, imagine a virtual pyramid delineated by the $\mathrm{FOV}$ of the camera when intercepts a cloud's layer. The pyramid defines a triangle when is projected into a 2D plane, see upper schematic in figure \ref{fig:flat_earth}, so we propose to compute the distance $d_i$ for each row of pixels in a frame. Knowing that the $\mathrm{FOV}$ of a pixel is $\Delta \alpha_y = \frac{\mathrm{FOV}_{y}}{M}$ in our camera , and that these are the same for the x-axis and the y-axis, $\Delta \alpha_y = \Delta \alpha_x = \Delta \alpha$. The vector of angular differential increments for each row of pixels is,
\begin{align}
    \boldsymbol{\alpha} = \left\{ \left( i \cdot \frac{\Delta \alpha}{2} \right) \ \middle| \ \alpha_i \in \mathbb{R}^{(0, 2 \pi]}, \ \forall i = 1, \ \ldots, \ \frac{M}{2} \right\},
\end{align}
and when it is calculated w.r.t the Sun's elevation angle $\varepsilon_0$, the elevation of each row of pixels in a frame is $\varepsilon_i = \varepsilon_0 + \left( \alpha_i - \alpha_0 \right)$. 

The origin of coordinates is defined as the current position of the Sun for each vector in the x-axis,
\begin{align}
    \mathbf{x}^{\prime\prime}_i = \mathbf{x}_i^{\prime\prime} - x^{\prime\prime}_{i, x_0}, \ \forall  i = 1, \ldots, M,
\end{align}
and the y-axis is,
\begin{align}
     \mathbf{Y}^{\prime\prime} = \mathbf{Y}^{\prime\prime} - y^{\prime\prime}_{x_0, y_0},
\end{align}
where $x_0$ and $y_0$, are the Sun's position index in the new coordinate system. Notice that the transformation of the camera to the cross-section plane are a linear function of the height.

\begin{figure}[!ht]
    \centering
    \includegraphics[scale = 0.195]{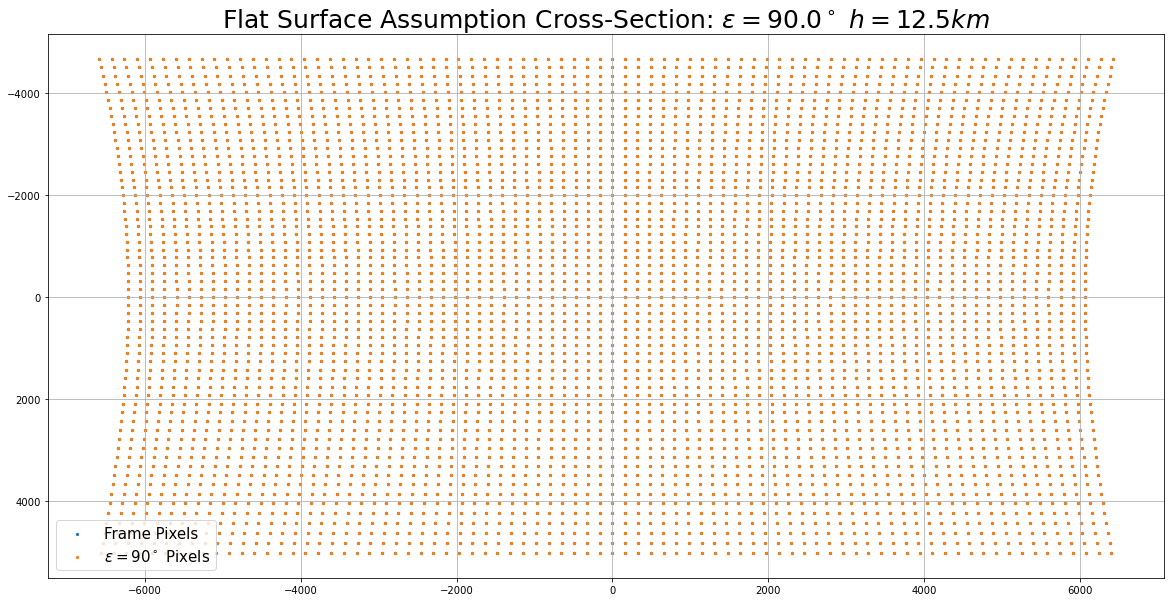}
    \includegraphics[scale = 0.195]{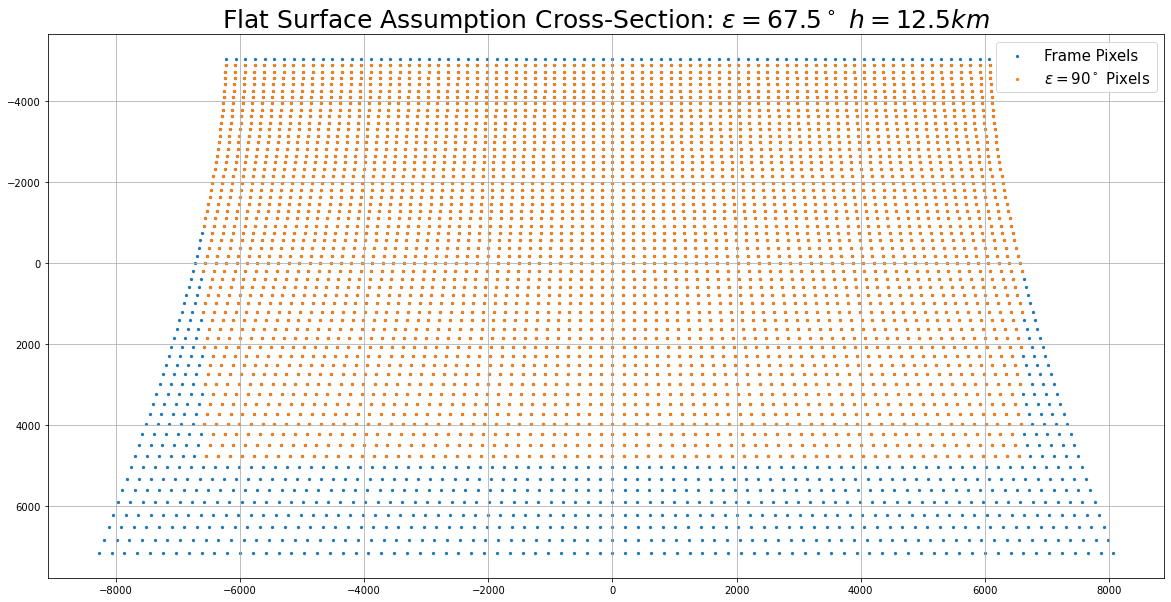}
    \includegraphics[scale = 0.195]{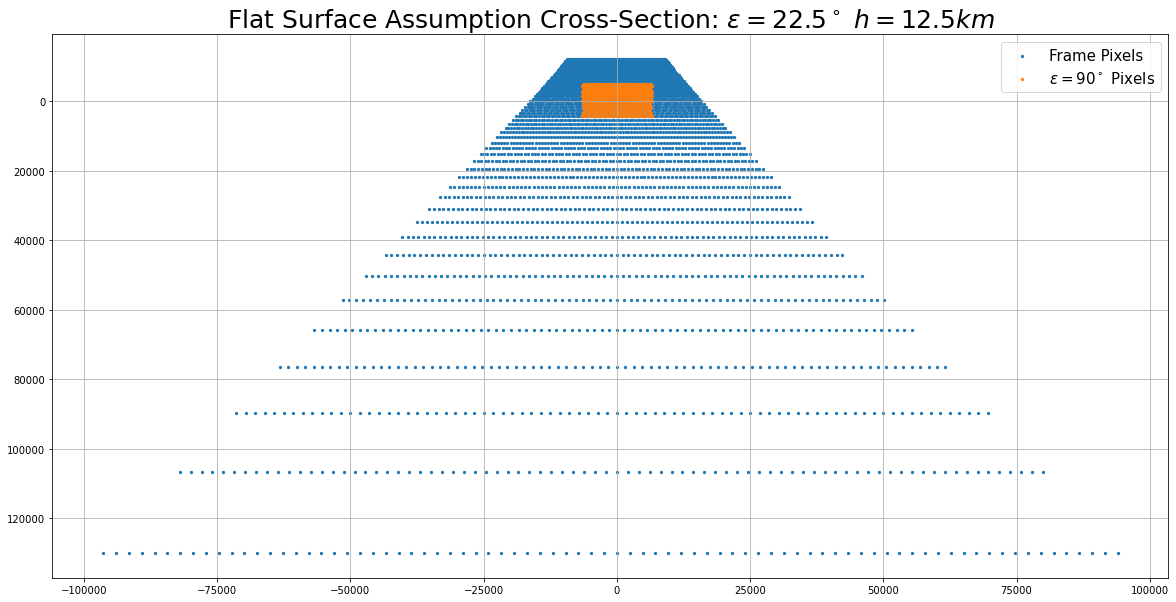}
    \includegraphics[scale = 0.195]{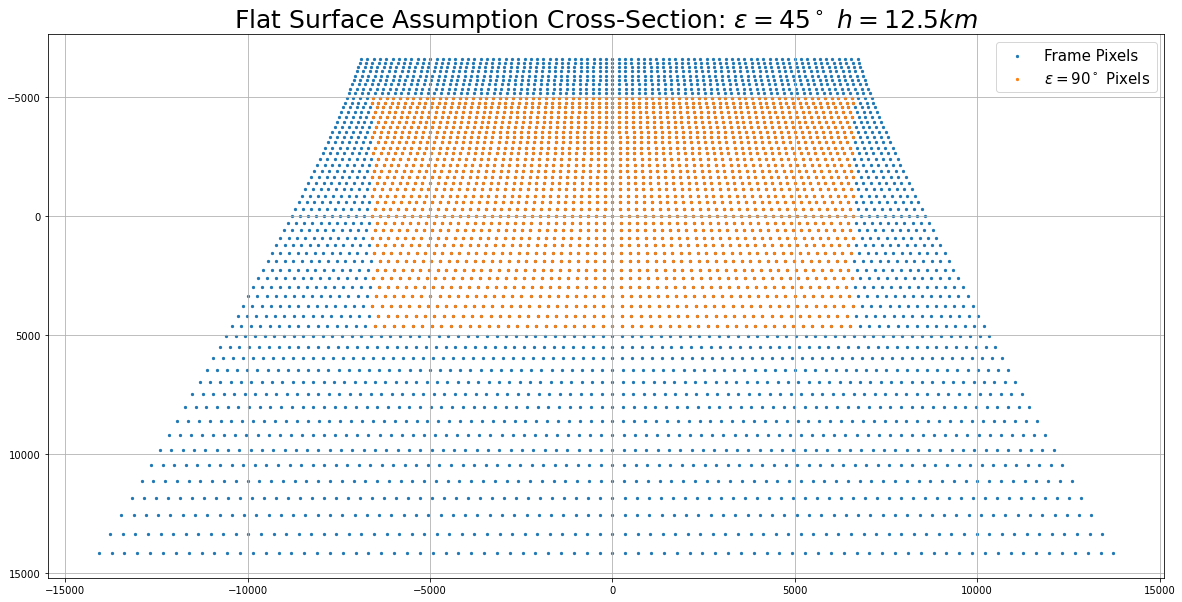}
\caption{These graphs show the transformation, when assuming that the surface is flat, of the original pixels' Euclidian coordinates system decreasing with Elevation angle clockwise. The points in orange color are the pixels within the cross-section length when the camera is completely normal to the ground, which is at $\varepsilon = \frac{\pi}{2}$ rad. The tropopause is considered static at $12.5$km in all the graphs.}
\end{figure}

\subsection{Great Circle Approach}

The clouds phenomena is restricted to the troposphere, so the cloud bases can be found at any height from sea level, fog or mist, to the tropopause, alto-stratos or cirrus \cite{HOUZE1993}. The height from the ground to the tropopause varies along the year, and it depends in the latitude. However, due to the fact that the cloud bases are found in a very small range compared to total atmosphere vertical distance from the ground up to the exosphere, we assume that the layers of clouds, are at constant height $h$ from the Earth surface, and we also assume that their surface is flat, for simplicity in finding the transformation that gives us the relative distance from a cloud to the camera on a frame. Therefore, the transformation from the sensor plane to the camera plane, as described in figure \ref{fig:cross_section}, is a function of a cloud layer height and the Sun's elevation angle. 

\begin{figure}[htbp]
    \centering
    \resizebox{12.5cm}{!}{\input{transformation/cross_section.tikz}}
    \caption{Diagrams of the geometrical transformation in the y-axis and the x-axis respectively. $Sensor$ and $Lens$ are parts of the IR camera, and are physically separated by a focal length $f$. The velocity decomposition shows that the components of a cloud velocity has a perspective distortion either in the x-axis as in the y-axis due to the camera plane inclination of $\varepsilon$ degrees with respect to the normal. The distance $z$ from the camera to a layer of clouds at height $h$ is depicted in the upper graph. The $\mathrm{FOV}_x$ and $\mathrm{FOV}_y$ are $\alpha_x$ and $\alpha_y$.}
\label{fig:cross_section}
\end{figure}

In order to transform the coordinates system from the camera plane to the cross-section plane, we have to compute the effects in each pixel due to the prospective in the image in differential increments as function of the distance. For that, we know that each pixels' field of view is $\Delta \alpha_y = \frac{\mathrm{FOV_{y}}}{M}$, and as the lens in our camera camera is a rectilinear, a pixel's field of view is the same either in the x-axis, or in the y-axis, so that $\Delta \alpha_y = \Delta \alpha_x = \Delta \alpha$.

\begin{figure}[htbp]
    \centering
    \resizebox{6.5cm}{!}{\begin{tikzpicture}

    \shade[ball color = blue, opacity = 0.25] (0,0) circle (5.cm);
    \shade[ball color = brown, opacity = 0.75] (0,0) circle (3.cm);
    
    \draw (0,0) circle (3 cm);
    \draw (0,0) circle (5 cm);
    \draw [dashed, opacity = 0.4] (-4, 3) -- (4, 3);
    
    \node at (-4, 3) [xshift=-.25cm, yshift=0.cm]{A};
    \node at (4, 3) [xshift=0.25cm, yshift=0.cm]{B};

    \node at (2.25, 4.5) [xshift=.25cm, yshift=0.cm]{D};
    \node at (-2.25, 4.5) [xshift=-0.25cm, yshift=0.cm]{C};
    
    \draw [dashed, opacity = 0.4] (-2.25, 4.5) -- (2.25, 4.5);
    
    \draw [<->] (0.25, 4.5) -- (0.25, 5) node [midway, align = right, xshift=0.25cm, yshift=0.cm]{$\ell_i$};
    
    \draw [<->] (-0.25, 3) -- (-0.25, 5) node [midway, align = right, xshift=0.25cm, yshift=0.cm]{$h$};
    
    \draw [<->] (0, 0) -- (3, 0) node [midway, align = right, xshift=0cm, yshift=0.25cm]{$r$};
    
    \draw [<->] (0, 3) -- (2.25, 3) node [midway, align = right, xshift=0.25cm, yshift=-0.18cm]{$x_i$};

    \draw (2.25, 4.5) -- (2.25, 3) node [midway, align = right, xshift=0.25cm, yshift=0.cm]{$y_i$};
    
    \draw (0, 3) -- (2.25, 4.5) node [midway, align = right, xshift=-0.15cm, yshift=0.15cm]{$z_i$};

    \draw [<->] (1., 3) arc (0:30:1) node [midway, align = right, xshift=0.25cm, yshift=0.05cm]{$\varepsilon_i$};
\end{tikzpicture}}
    \caption{$Great$ circle approach to find the distance $y$ between the chord of the small circle and small circle. The $small$ circle radius is $r = r_{earth} + \rho$, where $\rho$ is the altitude of the localization. The $great$ circle radius is $R = r + h$, where $h$ is the height of Troposphere, and it varies with respect to the latitude of the localization.}
\label{fig:great_circle}
\end{figure}
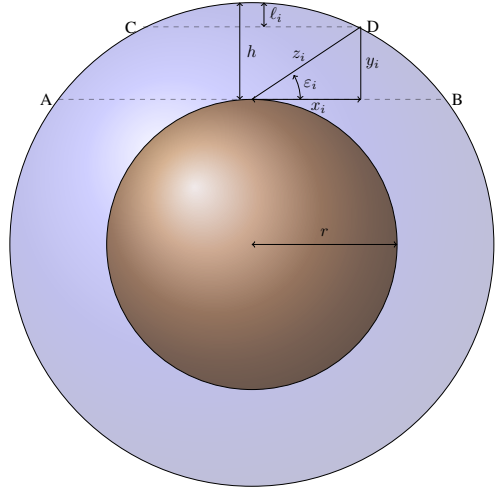

The pyramid formed by the camera's $\mathrm{FOV}$, defines symmetric triangles from the center of the image $\mathbf{x}_0 = \left(x_0, \ y_0 \right)$ to the sides of the plane when intercepts with a layer of clouds that is at a certain height $h$, see figure \ref{fig:cross_section}. Taking advantage of this fact, the actual cross-section plane can be estimated w.r.t. the camera plane, which is the projection of that plane in a frame. The differential increments in the y-axis of the camera plane are derived first, the differential increments in the x-axis are directly developed out from them. Notice that the x-axis in the Earth's cross-section plane is the y-axis in the camera plane, and the x-axis in the camera plane is normal to  Earth's the cross-section plane, and it is obtained from the $z_i$ distance, see figure \ref{fig:cross_section} and \ref{fig:great_circle}.

The great circle is the tropopause, and the small circle is the Earth. The tangent of Earth's surface that intercepts the troposphere, is the chord defined as $AB$, see figure \ref{fig:great_circle}. The radius from center of the Earth to the tropopause, and the sagitta, which is the length from chord $CD$ to the troposphere top, are,
\begin{align}
    R &= r + h \\
    \ell_i &= h - y_i.
\end{align}
The relation between the sides of the triangle formed by the line that connects the Earth's surface with the top of the tropopause with an inclination angle w.r.t. the normal $\frac{\pi}{2} - \varepsilon_i$ is, 
\begin{align}
    \tan \varepsilon_i &= \frac{y_i}{x_i} \\
    x_i \cdot \tan \varepsilon_i &= y_i
\end{align}
The sagitta of the chord $CD$ is related to the chord $AB$. The formula that describes the sagitta's length of the chord $CD$, is a function of the connecting triangle formed by the intercepting line from $AB$ to $CD$, when the vertex angle $\varepsilon_i$ varies,
\begin{align}
    \ell_i &= R - \sqrt{R^2 - x_i^2} \\
    h - y_i &= R - \sqrt{R^2 - x^2}\\
    R^2 - x_i^2 &= \left( x_i \cdot \tan \varepsilon + r \right)^2 \\
    R^2 - x_i^2 &= x_i^2 \cdot \tan^2 \varepsilon_{i} + r^2 + x_i \cdot r \cdot \tan \varepsilon_i  \\
    0 &=
    x_i^2 \cdot \left( 1 + \tan^2 \varepsilon_{i} \right) + x_i \cdot \left( r \cdot \tan \varepsilon_{i} \right) - h \cdot (1 + 2\cdot  r) 
\end{align}
The relation is explained by the quadratic equation that has the following coefficients,
\begin{align}
\boldsymbol{\beta}^{(y)}_i = 
\left\{\begin{array}{l}
    a_{i}= 1 + \tan^2 \varepsilon_{i} \\
    b_{i}= r \cdot \tan \varepsilon_{i} \\
    c_{i}= - h \cdot \left( 1 + 2 \cdot r \right)
\end{array}\right.
\end{align}

The side $x_i$ of the triangle sis the positive result obtained from solving the quadratic equation with this formula,
\begin{align}    
x_i = \frac{- b_i \pm \sqrt{b_i^2 - 4 \cdot a_i \cdot c_i}}{2 \cdot a_i}, \ x_i \in \mathbb{R}^+
\end{align}

The side $x_i$ of the triangle is the $y$-axis of the cross-section plane. To account for the curvature of the Earth, the side $x_i$ of the triangle, which is flat, is projected to the segment of the great circle with this formula,
\begin{align}
    y_i^\prime &= \theta_i \cdot R = \frac{\kappa_i}{2} \cdot \sin^{-1} \left( \frac{2 x_i}{\kappa_i} \right) \\
    \kappa_i &= \ell_i + \frac{x_i^2}{\ell_i},
\end{align}
This gives the length of the segment formed by the normal of the camera, with each one of the pixel's elevation angle. The distance of segment, from the normal of the camera to the cross-section plane, is subtracted from each pixel,
\begin{align}
    y_i^\prime = y_i^\prime - y_1,
\end{align}
to calculate the relative distance between the extremes of the cross-section plane.

The vector of differential increments that transforms the camera plane to the cross-section plane in the y-axis, is obtained via numerical differentiation,
\begin{align}    
    \Delta y_i^\prime &= \Delta y_i^\prime - \Delta y_{i - 1}^\prime
\end{align}
so that $\mathbf{Y}^\prime = \mathbf{y}^\prime \cdot \mathbf{1}^\top$, and analogously $\Delta \mathbf{Y}^\prime = \Delta \mathbf{y}^\prime \cdot \mathbf{1}^\top$, where $\mathbf{1}$ is a vector of ones with dimensions $M \times 1$.

The projection of the camera plane, in the x-axis, at each pixel's distance from the camera in the y-axis direction is, 
\begin{align}    
    z_i &= \sqrt{x_i^2 + y_i^2} \\
    \alpha_j &= \left(N + 2 - 2j \right) \cdot \nu \\
    \hat{x}_{i,j} &= 2 \cdot z_i \cdot \tan \frac{\alpha_j}{2}
\end{align}
where $\nu = \frac{\mathbf{FOV}}{\sqrt{N^2 + M^2}} \cdot \frac{\pi}{180^\circ}$ is radians per pixel ratio of the camera.

The relation between the length of an arc segment, and the chord, is found through the sagitta. The formula which describes the relation between these variables is,
\begin{align}   
    R^2 &= \left( \frac{\hat{x}_{i,j}}{2} \right)^2 + \left( R - \lambda_{i,j} \right)^2 \\
    - \left( \frac{\hat{x}_{i,j}}{2} \right)^2 &= \lambda_{i,j}^2 - 2 R \lambda_{i,j} \\
    - \lambda^2_{i,j} + 2 R \lambda_{i,j} - \left( \frac{\hat{x}_{i,j}}{2} \right)^2 &= 0.
\end{align}
The derived quadratic equation for computing the sagitta, has the following set of coefficients,
\begin{align}
\boldsymbol{\beta}^{(x)}_{i,j} = 
\left\{\begin{array}{l}
    a_{i,j}= - 1 \\
    b_{i,j}= 2 \cdot R  \\
    c_{i,j}= - \left( \frac{\hat{x}_{i,j}}{2} \right)^2.
\end{array}\right.
\end{align}
In the x-axis, the great circle's sagitta of the arc is the negative result obtained solving the quadratic equation,
\begin{align}    
    \lambda_{i,j} = \frac{- b_{i,j} \pm \sqrt{b_{i,j}^2 - 4 \cdot a_{i,j} \cdot c_{i,j}}}{2 \cdot a_{i,j}}, \quad \lambda_{i,j} \in \mathbb{R}^-.
\end{align}
The segment length, which is calculated with the found sagitta, is the actual length of the x-axis cross-selection,
\begin{align}    
    x_{i, j}^\prime &= \frac{\kappa_{i,j}}{2} \cdot \sin^{-1} \left( \frac{\hat{x}_{i,j}}{\kappa_{i,j}} \right) \\
    \kappa_{i,j} &= \lambda_{i,j} + \frac{\hat{x}_{i,j}^2}{4\lambda_{i,j}}
\end{align}

To compute the vector of differential increments that transform the camera plane to the cross-section plane in the x-axis, the numerical finite differential formula is applied only to the x-axis,
\begin{align}    
    \Delta x_{i, j}^\prime &= \Delta x_{i, j}^\prime - \Delta x_{i, j - 1}^\prime,
\end{align}
and it is symmetric to the center of the frame, so that $\mathbf{X}^\prime = \left[ \mathbf{X}^\prime \ \left( \boldsymbol{\Lambda} \cdot \mathbf{X}^\prime \right)^\top \right]$, and analogously $\Delta \mathbf{X}^\prime = \left[ \Delta \mathbf{X}^\prime \ \left( \boldsymbol{\Lambda} \cdot \Delta \mathbf{X}^\prime \right)^\top \right]$, where $\boldsymbol{\Lambda}$ is an anti-identity matrix of dimensions $M \times M$.

\begin{figure}[!ht]
    \centering
    \includegraphics[scale = 0.195]{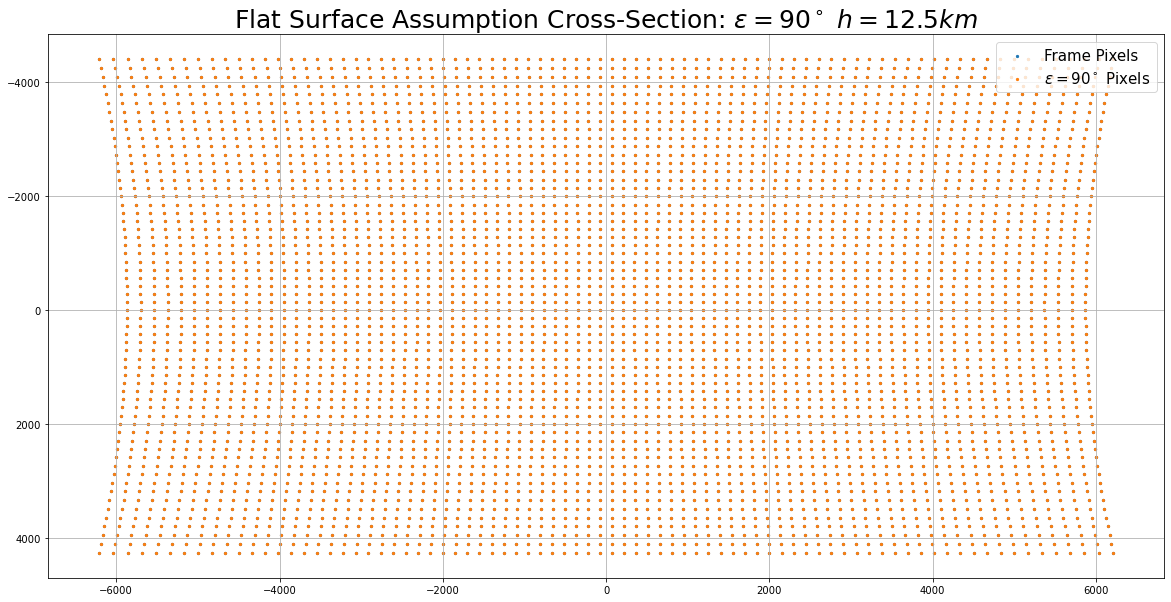}
    \includegraphics[scale = 0.195]{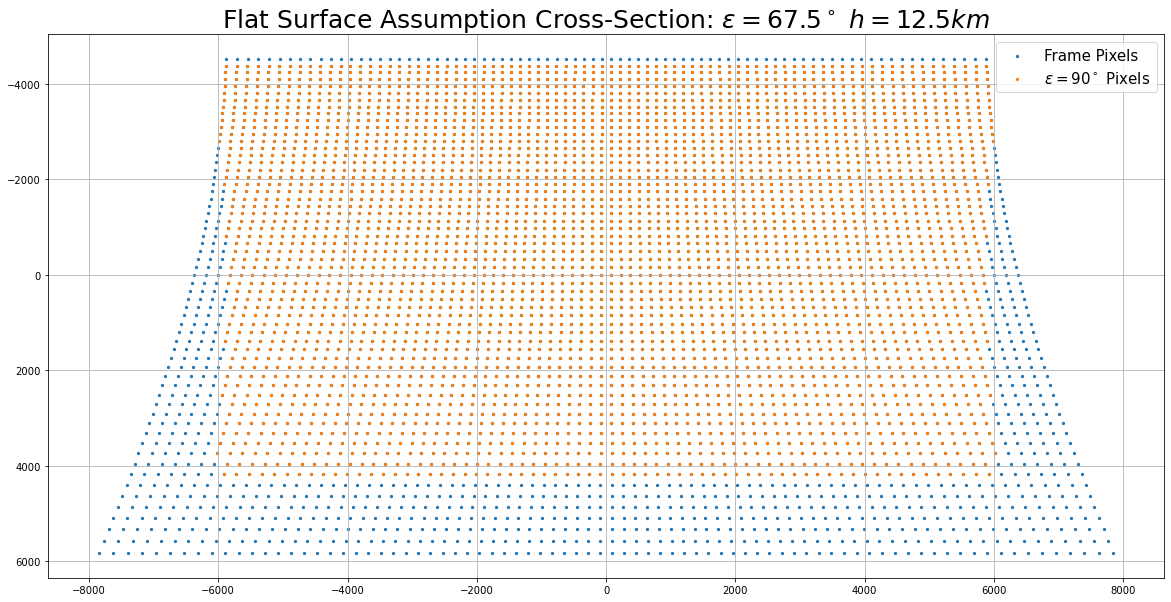}
    \includegraphics[scale = 0.195]{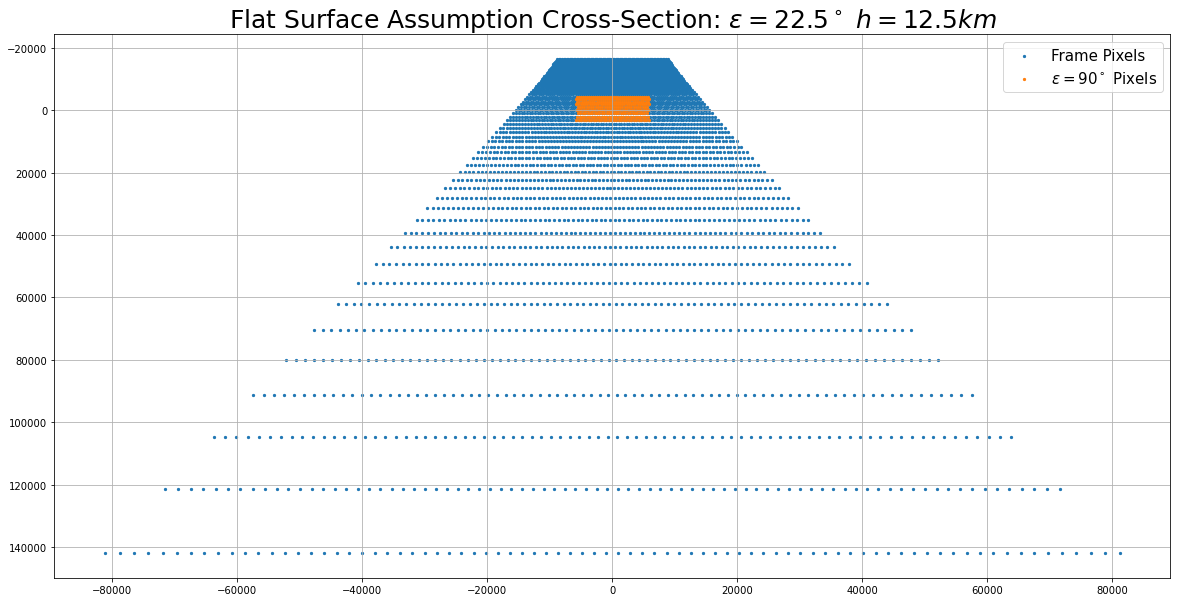}
    \includegraphics[scale = 0.195]{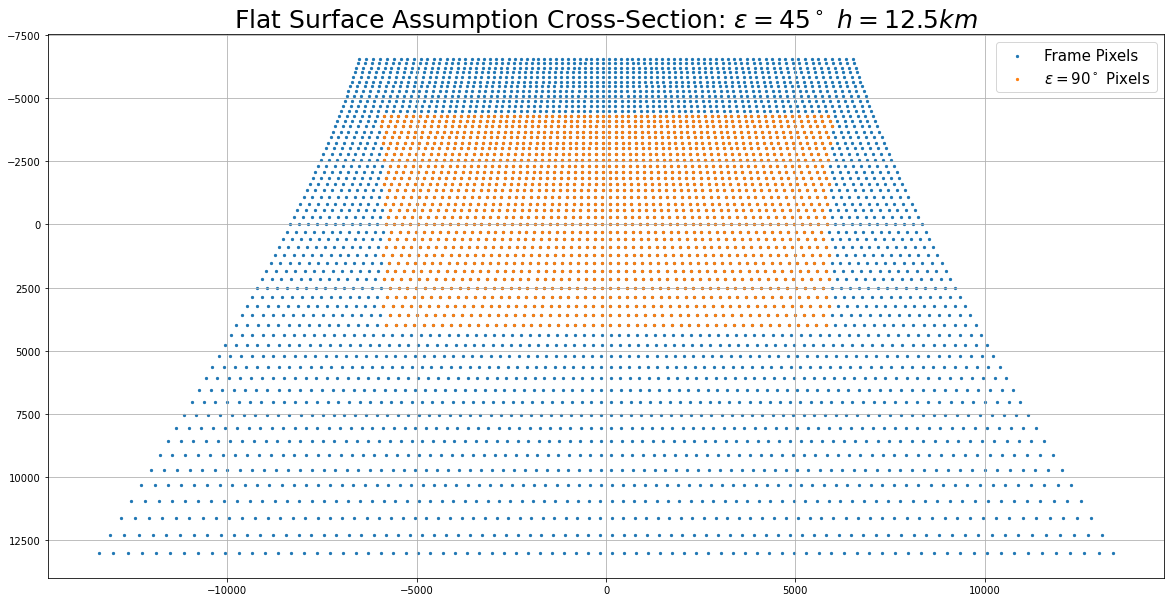}
\caption{These graphs show the non-linear transformation of the original pixels' Euclidian coordinates system decreasing with Elevation angle clockwise. The points in orange color are the pixels within the cross-section length when the camera is completely normal to the ground, which is at $\varepsilon = \frac{\pi}{2}$ rad. The tropopause is considered static at $12.5$km in all the graphs.}
\label{fig:cross-section}
\end{figure}

\begin{figure}[!ht]
    \centering
    \includegraphics[scale = 0.22]{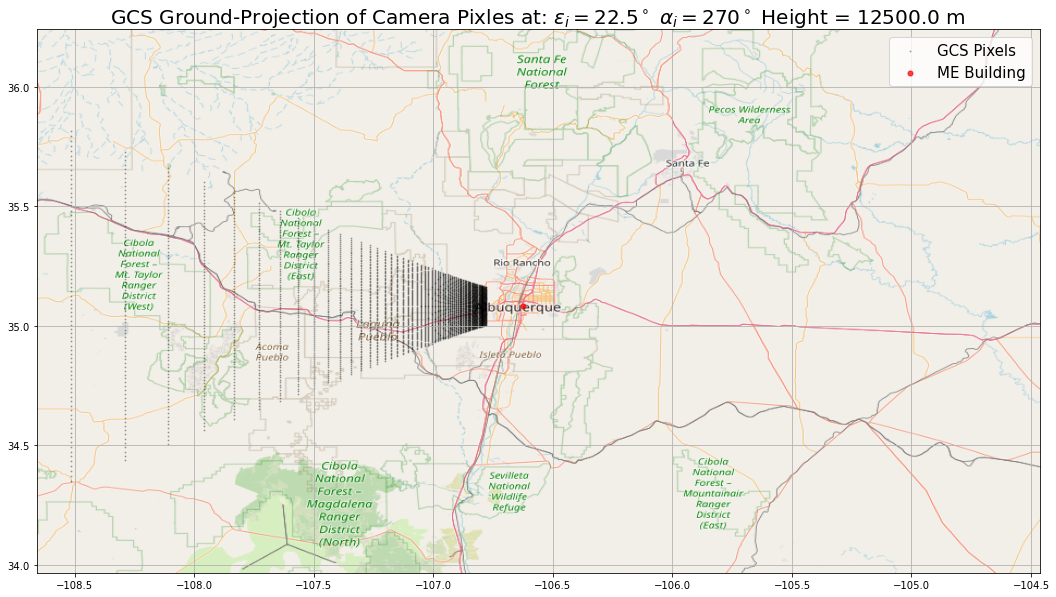}
    \includegraphics[scale = 0.22]{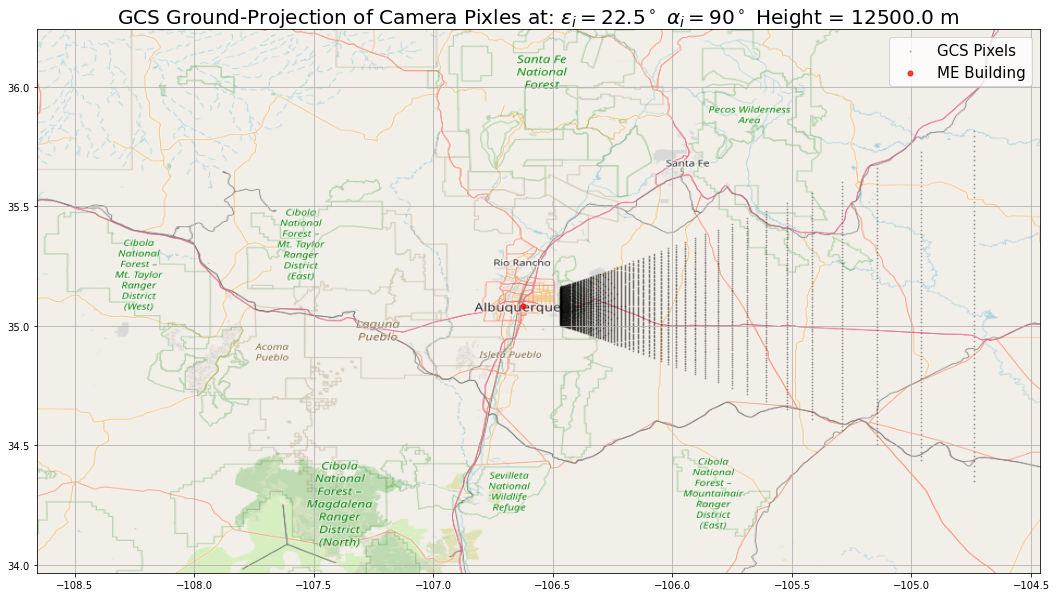}
    \includegraphics[scale = 0.22]{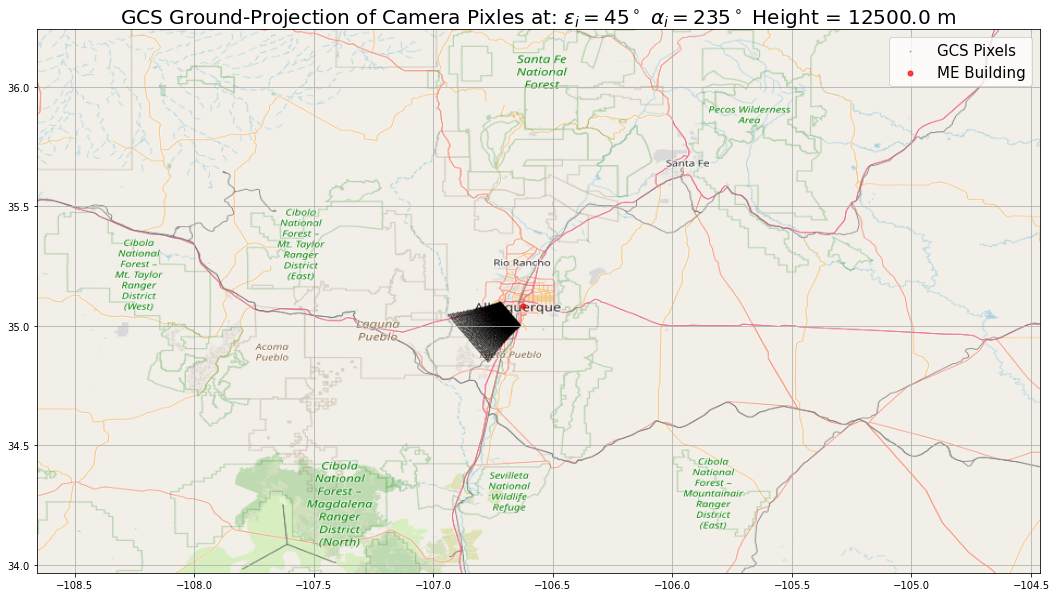}
    \includegraphics[scale = 0.22]{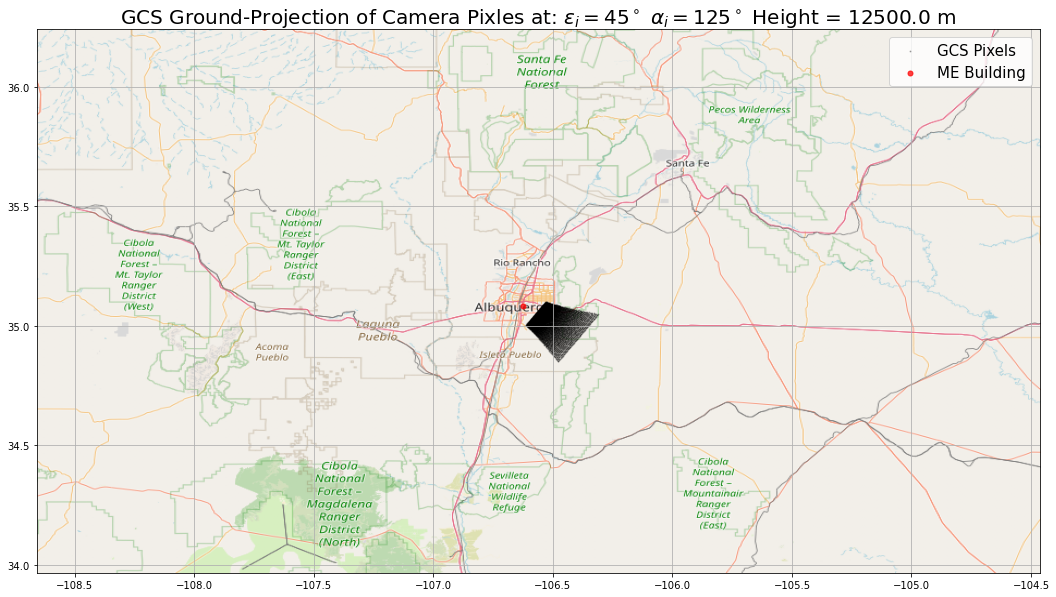}
    \includegraphics[scale = 0.22]{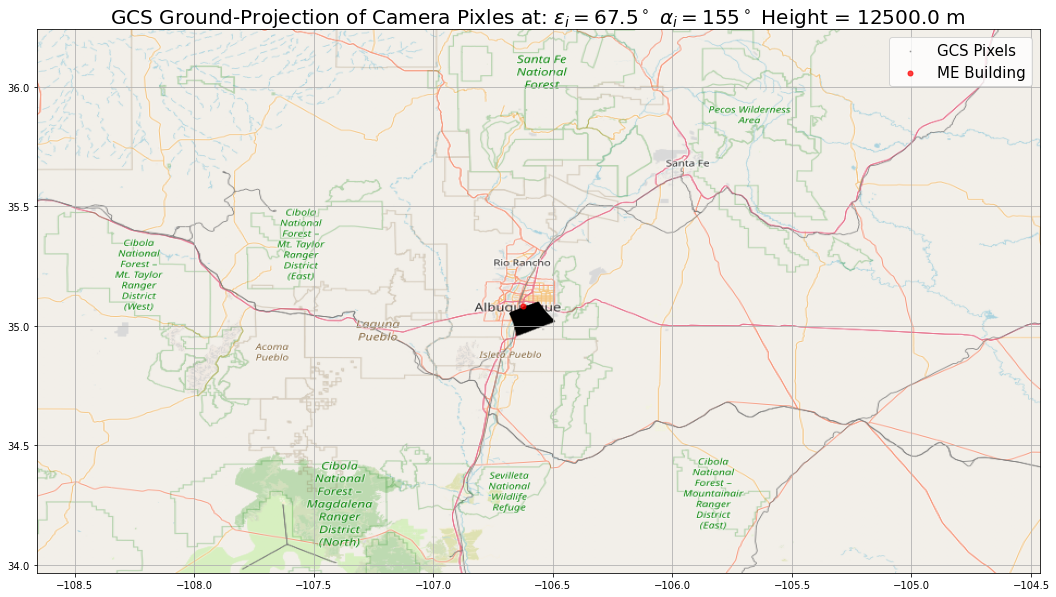}
    \includegraphics[scale = 0.22]{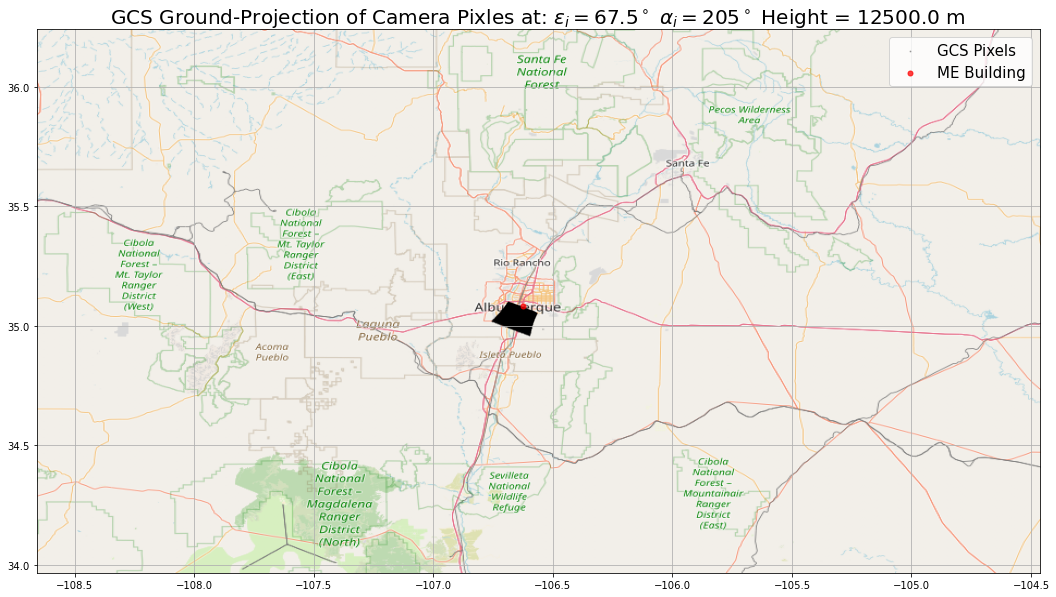}
\caption{Projection of the tropopause's segment into the Earth surface. The Geographic Coordinates System (GCS) is in degrees. The camera is localized at the ME Building in the UNM central campus. The New Mexico maps, that organized clockwise, are simulations of the frames projections tracking the Sun along a day.}
\label{fig:ground_projection}
\end{figure}

\begin{figure}[!ht]
    \centering
    \includegraphics[scale = 0.425]{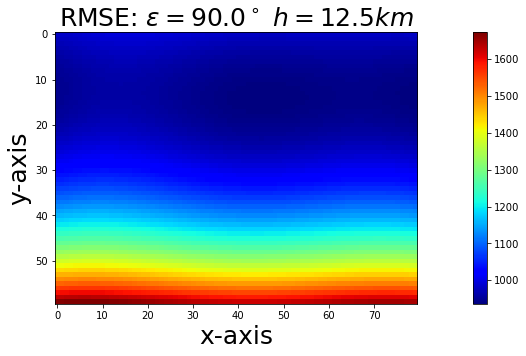}
    \includegraphics[scale = 0.425]{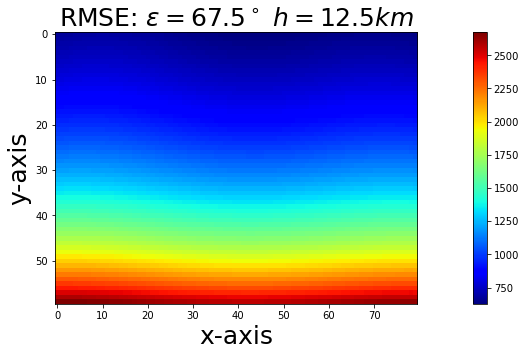}
    \includegraphics[scale = 0.425]{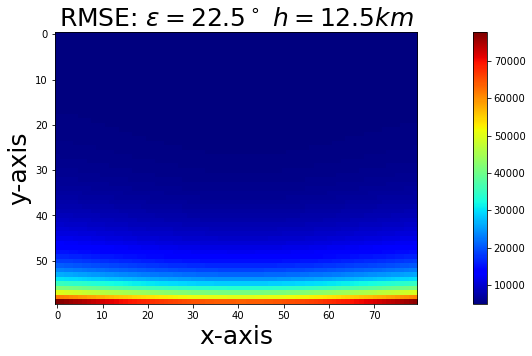}
    \includegraphics[scale = 0.425]{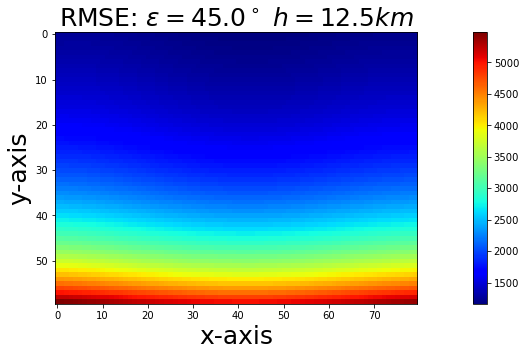}
\caption{This figure shows the error maps between the full-prospective transformations and the transformation with flat surface assumption for the previously displayed elevation angles. The total error maps were computed from the squared residuals of each coordinate $\mathbf{E} = \sqrt{ \frac{1}{2} \cdot \left[ \mathcal{E} \left(\mathbf{X}_1, \mathbf{X}_2 \right) + \mathcal{E} \left(\mathbf{Y}_1, \mathbf{Y}_2 \right) \right] }$, where the squared residuals are $\mathcal{E} \left(\mathbf{X}_1, \mathbf{X}_2 \right) = \left( \mathbf{X}_1^\prime - \mathbf{X}_2^\prime \right)^2$}.
\label{fig:error_maps}
\end{figure}

\subsubsection{Length of a Segment Approximation}

The segment of a circle formula dictates, that given the same angle of the $great$ and $small$ circle's arc, the length are directly proportional,
\begin{align}   
    s &= \theta \cdot R \\
    \frac{s_1}{R_1} &= \frac{s_2}{R_2}
\end{align}
so that, when $R >> h$, the grid can be approximated at different heights applying the segment formula,
\begin{align}   
    \mathbf{X}^\prime_1 &\approx \mathbf{X}^\prime_2 \cdot \frac{h_1}{h_2} \\
    \mathbf{Y}^\prime_1 &\approx \mathbf{Y}^\prime_2 \cdot \frac{h_1}{h_2}.
\end{align}
The origin of the non-linear coordinates system is the current position of the Sun, which is $\left( x_0, \ y_0 \right)$. The Sun's coordinates are given in the pixel's index position in this new coordinates system.

\section{Cloud Velocity Vectors}

The cloud velocity vectors are used to approximate the streamlines and potential lines of the wind velocity field in a frame, as the clouds are the only elements out at our disposal for the visualization of the wind velocity field. This section aims to find the most suitable method available on the literature to compute the velocity vectors between two consecutive frames.

For that, different wind velocity fields are simulated on a sequence of images. These fields are either linear or non-linear. A sequence of IR images from a cloud evolving over time, was selected beforehand. These images are added to the simulated wind velocity fields, so that the cloud flows in the simulated velocity field. Bayesian Optimization (BO) is implemented to optimize the parameters for each one of the methods in our analysis. The most effective method is selected looking for a trade-off between the error approximating the velocity vector, and the computing time.

\subsection{Techniques}

In current computer vision literature, there are three primary techniques to estimate the motion of objects in a sequence of images: the Lucas-Kanade \cite{LUCAS1981}, Horn-Schunck \cite{HORN1981} and Farneb{\"a}ck \cite{FARNEBACK2003} methods. These three methods are based upon the space-time partial derivatives of two consecutive frames. Taking a different disciplinary approach, the field of experimental fluid dynamics uses research methods that are based on the statistical principle of signal cross-correlation in the frequency domain \cite{ADRIAN2011}. The signal Cross-Correlation (CC) can either be normalized or unnormalized. Techniques such as these are called Particle Image Velocimetry (PIV). The techniques to estimate the motion vectors in an image are sensitive to the pixels' intensity gradient. We implemented a model that removes the gradient produced by the solar direct radiation, and atmospheric scattered radiation. Both of which routinely appear on the images in the course of the year. A persistent model of the window of the camera removes sporadic artifacts that appear in the image such as water spots, or dust particles. A series of sequences of images with clouds flowing different directions were simulated to cross-validate the set of parameters for each one of the mentioned methods. The investigation was searching for a dense implementation of a motion vector method to approximate the dynamics of a cloud.

\subsubsection{Optical Flow}

The optical flow equation considers that exists a small displacement in the $\Delta x$ and $\Delta y$ direction of an object in an image. The object is assumed to have constant intensity $\mathcal{I}$ between two consecutive frames, and the frames are separated in time by small time increment $\Delta t$,
\begin{align}
    \mathcal{I} \left( x, y, t \right) = \mathcal{I} \left( x + \Delta x, y + \Delta y, t + \Delta t \right).
\end{align}
If the difference in intensity between neighboring pixels is assumed smooth, and the brightness of a pixel between consecutive frames is assumed constant, the Taylor series expansion can be applied to obtained the following equation,
\begin{align}
    \mathcal{I} \left( x + \Delta x, y + \Delta y, t + \Delta t \right) = \mathcal{I}  \left( x, y, t \right) + \frac{\partial \mathcal{I} }{\partial x} \Delta x+\frac{\partial \mathcal{I} }{\partial y} \Delta y+\frac{\partial \mathcal{I} }{\partial t} \Delta t.
\end{align}
The factors are simplified combining the last two equation,
\begin{align}
\frac{\partial \mathcal{I} }{\partial x} \Delta x + \frac{\partial \mathcal{I} }{\partial y} \Delta y+\frac{\partial \mathcal{I} }{\partial t} \Delta t = 0.
\end{align}
The velocity of an object can be derived, 
\begin{align}
\frac{\partial \mathcal{I} }{\partial x} \frac{\Delta x}{\Delta t} + \frac{\partial \mathcal{I} }{\partial y} \frac{\Delta y}{\Delta t} + \frac{\partial \mathcal{I} }{\partial t} \frac{\Delta t}{\Delta t} = 0,
\end{align}
dividing the terms of the displacement by the increment of time $\Delta t$. The velocity components are defined as $v_x$ and $v_y$.
\begin{align}
\frac{\partial \mathcal{I} }{\partial x} v_x + \frac{\partial \mathcal{I}}{\partial y} v_y + \frac{\partial \mathcal{I} }{\partial t} = 0.
\end{align}
This equation is known as the aperture problem,
\begin{align}
    \mathcal{I}_x v_x + \mathcal{I}_y v_y = - \mathcal{I}_t,
\end{align}
where the derivatives are $\mathcal{I}_x = \partial \mathcal{I} / \partial x$, $\mathcal{I}_y = \partial \mathcal{I} / \partial y $, and $\mathcal{I}_t = \partial \mathcal{I}/ \partial t $  for notation simplification. 

The 2-dimensional derivatives are approximated using convolutional filters in a image \cite{HAST2014}. The intensity of pixels a frame in time instant $t$ are $\mathbf{I}^{(t)} = \{ i^{(t)}_{i,j} \in \mathbb{R}^{[0, 2^8)} \mid i = 1, \dots, N, \ j = 1, \dots, M \}$. The Sobel derivatives \cite{FARID2005}, which were implemented in this research, are computed as,
\begin{align}
    \mathbf{I}_x^\prime &= \mathbf{I}^{\left(t - 1\right)} \star \mathbf{K}_x \\
    \mathbf{I}_y^\prime &= \mathbf{I}^{\left(t - 1\right)} \star \mathbf{K}_y \\
    \mathbf{I}_t^\prime &= \mathbf{I}^{\left(t - 1\right)} \star \left( \mathbf{K}_t \cdot \sigma \right) + \mathbf{I}^{\left(t\right)} \star \left( - \mathbf{K}_t \cdot \sigma \right)
\end{align}
where $\star$ represent a 2-dimensional convolution. $\mathbf{K}_x$, $\mathbf{K}_y$, and $\mathbf{K}_t$ are the derivative kernels in x, y and t direction respectively. $\mathbf{I}^{\left(t - 1\right)}$ and $\mathbf{I}^{\left(t\right)}$ are the first and second consecutive frames. $\sigma$ is amplitude of the temporal differential kernel. It may be cross-validated to match the magnitude of the velocity field when is known.

\begin{itemize}

\item Lucas-Kanade

The Lucas-Kanade method \cite{LUCAS1981}, proposes to find the solution for the optical flow equations via least-squares $\mathbf{A} \mathbf{x} = \mathbf{b}$ in a sliding windows that is defined as,
\begin{align}
    \mathbf{x} = \left\{ x_{i + k, j + m} \in \mathbb{R} \mid \ k = -w, \ldots, +w, \ m = -w, \ldots, +w, \right\} \in \mathbb{R}^W,
\end{align}
where $x_i,j$ refers to the pixel where the sliding windows is centered, and $W = 2w + 1$ is the size of the windows. The set of independent, and depended variables, in matrix form that is used to find the least-square solution is,
\begin{align}
    \mathbf{A} = \left[ 
    \begin{array}{cc}{\mathcal{I}_x \left( x_1 \right)} & {\mathcal{I}_y \left( x_1 \right)} \\
    {\mathcal{I}_x\left( x_2 \right)} & {\mathcal{I}_y \left( x_2 \right)} \\ {\vdots} & {\vdots} \\ 
    {\mathcal{I}_x \left( x_W \right)} & {\mathcal{I}_y \left( x_W \right)}
    \end{array} \right], \ \ 
    \mathbf{x} = \left[ 
    \begin{array}{c}{v_x} \\ {v_y}
    \end{array} \right], \ \ 
    \mathbf{b} =\left[ 
    \begin{array}{c}{-\mathcal{I}_t\left( x_1 \right)} \\ {-\mathcal{I}_t\left( x_2 \right)} \\ {\vdots} \\ {-\mathcal{I}_t\left( x_W \right)}
    \end{array} \right].
\end{align}
The solution to the least-square problem is,
\begin{align}
    \mathbf{A}^\top \mathbf{A} \mathbf{x} = \mathbf{A}^\top \mathbf{b} \\ 
    \mathbf{x} = \left( \mathbf{A}^\top \mathbf{A} \right)^{-1} \mathbf{A}^\top \mathbf{b},
\end{align}
notice that this can be an ill-conditioned problem. When we look at the eigenvalues of the covariance matrix that are,
\begin{align}
    \mathbf{A}^\top \mathbf{A} = \mathbf{1} \left[ 
    \begin{array}{cc}{\lambda_1} & {0} \\ {0} & {\lambda_2}
    \end{array} \right] \mathbf{1},
\end{align}
where $\mathbf{1}$ is an unitary matrix of size $2 \times 2$, $\lambda_1$ and $\lambda_2$ has to be non-zero. This is equivalent to reduce the noise in the estimation of velocities applying a threshold to the eigenvalues of $\mathbf{A}^\top \mathbf{A}$ such as $\lambda_1 \geq \lambda_2 > \tau $.
    
\item Horn-Schunk
    
The Horn-Schunck method \cite{HORN1981}, introduces a global Energy functional with a constrain applied to optical flow  equation that has an additional regularization term. It is know as the smoothness constrain,
\begin{align}
    \mathcal{E} \left( x, y \right) = \iint \left[ \left( \mathcal{I}_x v_x + \mathcal{I}_y v_y + \mathcal{I}_{t} \right)^2 + \alpha^2 \left( \| \nabla v_x \|^2 + \| \nabla v_y \|^2 \right) \right] d x \cdot d y
\end{align}
where $\alpha$ is the parameter of the regularization term. 

The aim is to minimize $ \mathcal{E} \left( x, y \right)$ via differentiating w.r.t. the variables $v_x$ and $v_y$. The solution for a pair of second order differential equations can be computed iteratively solving the Euler-Lagrange equations which are,
\begin{align}
    \frac{\partial \mathcal{E}}{\partial v_x} &= \frac{\partial}{\partial x} \frac{\partial \mathcal{E}}{\partial v_x} + \frac{\partial}{\partial y} \frac{\partial \mathcal{E}}{\partial v_y} \\ 
    \frac{\partial \mathcal{E}}{\partial v_y} &= \frac{\partial}{\partial x} \frac{\partial \mathcal{E}}{\partial v_x } + \frac{\partial}{\partial y} \frac{\partial \mathcal{E}}{\partial v_y }
\end{align}
The following set of equations are obtained after differentiating,
\begin{align}
    \mathcal{I}_x \left( \mathcal{I}_x v_x + \mathcal{I}_y v_y + \mathcal{I}_t \right) - \alpha^2 \Delta v_x = 0 \\
    \mathcal{I}_y \left( \mathcal{I}_x v_x + \mathcal{I}_y v_y + \mathcal{I}_t \right) - \alpha^2 \Delta v_y = 0,
\end{align}
where $\Delta = \frac{\partial^2}{\partial x^2} + \frac{\partial^2}{\partial y^2}$ is the Laplace operator that controls the smoothness. It can be differentiated numerically so that $\Delta v_i = 4 \left(\bar{v}_i - v_i\right)$, where $\bar{v}$ represents the weighted sample mean of the pixels in the neighborhood. This formula can rearranged to obtain,
\begin{align}
    \left( \mathcal{I}_x^2 + 4 \alpha^2 \right) v_x + \mathcal{I}_x \mathcal{I}_y v_y = 4 \alpha^2 \bar{v}_x - \mathcal{I}_x \mathcal{I}_t\\ 
    \mathcal{I}_x \mathcal{I}_y v_x + \left( \mathcal{I}_y^2 + 4 \alpha^2 \right) v_y = 4 \alpha^{2} \bar{v}_y - \mathcal{I}_y \mathcal{I}_t
\end{align}

The velocity components $v_x$ and $v_y$ are iteratively computed, and updated $\bar{v}_x$ and $\bar{v}_y$ for each pixels' neighborhood in the sliding window (similar to the approach implemented in Lucas-Kanade method). The velocity components are defined to be 0 at the beginning of the algorithm, and are updated following these set of equations,
\begin{align}
    v_x^{k + 1} &= \bar{v}_x^k - \frac{ \mathcal{I}_x \left( \mathcal{I}_x \bar{v}_x^k + \mathcal{I}_y \bar{v}_y^k + \mathcal{I}_t \right)}{4 \alpha^2 + \mathcal{I}_x^2 + \mathcal{I}_y^2} \\ v_y^{k + 1} &= \bar{v}_y^k - \frac{I_y \left( \mathcal{I}_x \bar{v}_x^k + \mathcal{I}_y \bar{v}_y^k + \mathcal{I}_t \right)}{4 \alpha^2 + \mathcal{I}_x^2 + \mathcal{I}_y^2}
\end{align}
where $\alpha$ is the parameters of the regularization term. The optimization continues until either the tolerance is $ | \mathbf{v}^k - \mathbf{v}^{k + 1} | < \tau $, or is achieved the maximum number of iteration.

\item Farneb{\"a}ck

This method proposes to approximate a neighborhood of pixels in an image by local quadratic polynomial expansion such as \cite{FARNEBACK2003},
\begin{align}
    f_1 \left( \mathbf{x} \right)=\mathbf{x}^\top \mathbf{A}_1 \mathbf{x} + \mathbf{b}_1^\top \mathbf{x} + c_1,
\end{align}
so that the displacement in the same neighborhood between two consecutive images is,
\begin{align} 
    f_2 \left( \mathbf{x} \right) = f_1 \left( \mathbf{x} - \mathbf{d} \right) = \mathbf{x}^\top \mathbf{A}_2 \mathbf{x} + \mathbf{b}_2^\top \mathbf{x} + c_2.
\end{align}

After rearranging the local quadratic expansion for the approximation of coefficients, it is obtained that they are given by,
\begin{align} 
    \mathbf{A} \left( \mathbf{x} \right) &= \frac{\mathbf{A}_1 \left( \mathbf{x} \right) - \mathbf{A}_2 \left( \mathbf{x} \right)}{2} \\
    \Delta \mathbf{b} \left( \mathbf{x} \right) &= - \frac{\mathbf{b}_2 \left( \mathbf{x} \right) - \mathbf{b}_1 \left( \mathbf{x} \right)}{2} \\
    \Delta \mathbf{b} \left( \mathbf{x} \right) &=
    \mathbf{A} \left( \mathbf{x} \right) \mathbf{d} \left( \mathbf{x} \right)
\end{align}
where $\mathbf{d} \left( \mathbf{x} \right) $ is the local displacement in the neighborhood.

In order to add robustness to the displacement estimation, the field is configured according to a motion model that follows theses equations,
\begin{align}
    d_x \left( x, y \right) &= a_1 + a_2 x + a_3 y + a_7 x^2 + a_8 x y, \\ 
    d_y \left(x, y \right) &= a_4 + a_5 x + a_-6 y + a_7 x y + a_8 y^{2}.
\end{align}

The coefficients of the weighted least-squares (WLS) problem for this particular motion model are, 
\begin{align}
    \mathbf{d} &= \mathbf{S} \mathbf{p} \\
    \mathbf{S} &= \left[ 
    \begin{array}{ccccccccc}{1} & {x} & {y} & {0} & {0} & {0} & {0} & {x^2} & {x y} \\
    {0} & {0} & {0} & {1} & {x} & {y} & {x} & {y} & {y^2}
    \end{array} \right] \\
    \mathbf{p} &= \left[ 
    \begin{array}{cccccccc}{a_1} & {a_2} & {a_3} & {a_4} & {a_5} & {a_6} & {a_7} & {a_8} \end{array} \right]^\top
\end{align}

The WLS solution of the displacement in the same neighborhood of pixel in two consecutive images is such as,
\begin{align}
    \mathbf{p} = \left( \sum^{W^2}_{i = 1} w_i \cdot \mathbf{S}_i^\top \mathbf{A}_i^\top \mathbf{A}_i \mathbf{S}_i \right)^{-1} \sum^{W^2}_{i = 1} w_i \cdot \mathbf{S}_i^\top \mathbf{A}_i^\top \Delta \mathbf{b}_i.
\end{align}
The weights $w_i$ are computed evaluating the pixels on the sliding windows of size $W$ with a Gaussian distribution with standard deviation $\sigma^2$.

This algorithm includes prior information to approximate small displacements, an approach based on multiple scales layers from coarse to finer, that propagates the displacement to obtain better result as the layers scale increases.

\end{itemize}

\subsubsection{Particle Image Velocimetry}

The methods that are predominant in the fluid mechanic literature \cite{ADRIAN2011}, are based on the CC of same region in an image between two consecutive frames. The computation of the CC is expensive, so these methods are implemented as a sparse approach instead of a dense one, which is the case of the methods based on optical flow. Nevertheless, when we apply the cross-correlation theorem, which says that a convolution in the time domain becomes a dot product in the frequency domain,
\begin{align}
     \mathcal{I}_1 \left( x, y \right) \star  \mathcal{I}_2 \left( x, y \right) = \mathcal{F}^{-1} \left\{ \overline{\mathcal{F} \left\{  \mathcal{I}_1 \left( x, y \right) \right\}} \cdot \mathcal{F} \left\{  \mathcal{I}_2 \left( x, y \right) \right\} \right\}, \quad \mathcal{I}_1,\mathcal{I}_2 \in \mathbb{R}^{N \times D}
\end{align}
where $\star$ denotes a convolution. The CC computation indeed requires less time in the frequency domain. Although the solution of this problem still requires a sparse approach or a GPU in case of large resolution data. 

The sliding windows that goes over the images in the frequency domain is,
\begin{align}
    \mathbf{W}_1 = \mathcal{F} \left\{ \mathcal{I}_1 \left( x, y \right)  \right\}, \ \mathrm{and} \
    \mathbf{W}_2 = \mathcal{F} \left\{ \mathcal{I}_2 \left( x, y \right) \right\}, \quad \mathbf{W}_1,\mathbf{W}_2 \in \mathbb{R}^{W \times W}
\end{align}
where $\mathcal{I}_1$ is the previous image, and $\mathcal{I}_2$ is the current image, $W$ is the sliding window's size.

\begin{itemize}  

\item Spatial Cross-Correlation

The 2-dimensional CC function in the time domain is such as,
\begin{align}
    \mathbf{R}_{x, y} = \sum_{x = k}^{k + W} \sum_{y = m}^{m + W} \mathcal{I}_1 \left( x, y \right) \cdot \mathcal{I}_2 \left( x - k, y - m \right), \\ 
    \quad \quad \forall x = 1, \ldots, N, \ \forall y = 1, \ldots, D,
\end{align}
and when we apply the convolution theorem to the CC function between the two image's region, we obtain that it is equivalent to the following formula in frequency domain,
\begin{align}
    \mathbf{R}_{x, y} = \mathbf{W}_1 \odot \mathbf{W}_2^*, \quad \forall x = 1, \ldots, N, \ \forall y = 1, \ldots, D,
\end{align}
where $\odot$ represents element-wise multiplication.

\item Normalize Cross-Correlation

An alternative to the CC is to compute the Normalized Cross-Correlation (NCC). The function of NCC for a 2-dimensional convolution has following expression,
\begin{align}
    \mathbf{R}_{x, y} = \frac{ \sum_{x = k}^{k + W} \sum_{y = m}^{m + W} \left( \mathcal{I}_1 \left( x, y \right) - \mu_1 \right) \cdot \left( \mathcal{I}_2 \left(x - m, y - k \right) - \mu_2 \right) }{ \sqrt{ \sigma_1 \left( x, y \right) \cdot \sigma_2 \left(x - m, y - k \right) }}, \\
    \forall x = 1, \ldots, N, \ \forall y = 1, \ldots, D,
\end{align}
where $\mu_{x, y}$ represents the sample mean of the pixels in the sliding window, 
\begin{align}
    \mu_{x, y} = \frac{1}{W^2} \sum_{x = k}^{k + W} \sum_{y = m}^{m + W} \mathcal{I} \left(x, y\right),
\end{align}

and $\sigma_{x, y}$ is the sample variance of the pixels in the same window,
\begin{align}
    \sigma_{x, y} = \frac{1}{W^2} \sum_{x = k}^{k + W} \sum_{y = m}^{m + W} \left( \mathcal{I} \left( x, y \right) - \mu\right)^2,
\end{align}
where $W$ are the pixels in the window. When we apply the CC theorem to the NCC function, the equivalent formula obtained in the frequency domain is such as,
\begin{align}
    \mathbf{R}_{x, y} = \frac{ \mathbf{W}_1 \odot \mathbf{W}_2^*}{\left| \mathbf{W}_1 \odot \mathbf{W}_2^* \right|}, \quad \forall x = 1, \ldots, N, \ \forall y = 1, \ldots, D,
\end{align}

The results of computing the CC or NCC can be transform back to the time domain,
\begin{align}
    \Gamma_{x, y} = \mathcal{F}^{-1} \left\{ \mathbf{R}_{x, y} \right\}, \quad  \Gamma_{x, y}\in \mathbb{R^{W \times W}}
\end{align}
$\Gamma_{x, y}$ is matrix where the entry that has higher CC corresponds to the translation occur in the window. 

\end{itemize}

However, if the sliding window has low resolution, it will not possible to determine the exact maximum. We propose to fit a polynomial expansion to the window's pixels CC and implement an optimization approach via numerical gradient to find the maximum of this smooth function that is,
\begin{align}
    \Delta x_{x, y}, \Delta y_{x, y} = \operatorname{argmin} \ - \mathcal{P} \left( \Gamma_{x, y} \right)^d, \quad \Delta \mathbf{X} , \Delta \mathbf{Y} \in \mathbb{R}^{N \times D},
\end{align}
where $\Delta x_{x, y}$ and $\Delta y_{x, y}$ is the displacement for each pixel in region of interest, depending on whether it is a dense or sparse implementation, and $\Delta \mathbf{X}$ and $\Delta \mathbf{Y}$ are the velocity vectors.

\subsection{Bayesian optimization}

The aim is to maximize an acquisition function. There exists multiple acquisition functions that can served to this porpoise \cite{WILSON2018, SHAHRIARI2016}. In spite of that, all of these functions take advantage of the predictive distributions obtained from a Gaussian Process (GP). 

\subsubsection{Gaussian Process Regression}

A regression problem with noise is formulated such as,
\begin{align}
    y^* = f \left( \mathbf{x}^* \right) + \epsilon.
\end{align}

In a GP, the predictive distribution of $y^*$ for a new observation $\mathbf{x}^*$ given $\mathcal{D} = \{y_m, \mathbf{x}_m\}^{M}_{m = 1}$ is
\begin{align}
    y^* \sim \mathcal{N} (\mu, \Sigma),
\end{align}
where $\mu = \mathbf{k}^{*\top}_\theta \mathbf{K}^{-1}_\theta \mathbf{y}$, and $\Sigma = k^{**}_\theta - \mathbf{k}_\theta^{*\top} \mathbf{K}^{-1}_\theta \mathbf{k}_\theta^*$. The matrices,
\begin{align}
    \mathbf{K}_\theta = \mathrm{Cov} \left( \mathbf{x}_n, \mathbf{x}_m \right), \quad
    \mathbf{k}^*_\theta = \mathrm{Cov} \left(\mathbf{x}_n, \mathbf{x}^* \right), \quad
    k^{**}_\theta = \mathrm{Cov} \left( \mathbf{x}^*, \mathbf{x}^* \right),
\end{align}
are obtained evaluating a covariance function \cite{RASMUSSEN2005}. For instance, a common covariance function is the Mat\'ern,
\begin{align}
    \mathrm{Cov} \left( \mathbf{x}_n, \mathbf{x}_m \right) = \frac{2^{1 - \nu}}{ \Gamma \left( \nu \right)} \left( \frac{\sqrt{2\nu r }}{\ell} \right)^{\nu} \mathcal{K}_\nu \left( r \right) \left( \frac{\sqrt{2\nu r}}{\ell} \right),
\end{align}
where $\nu$ and $\ell$ are the kernel hyperparameters, $r = \| \mathbf{x}_n - \mathbf{x}_m \|$ is the Euclidean distance, $\Gamma \left( \nu \right)$ is the gamma function, and $\mathcal{K}_\nu \left( r \right)$ is a modified Bessel function of the second kind, where $\nu$ is the order. For further details in the gamma function, and Bessel functions see Chapter III. The general formula of $\mathcal{K}_\nu \left( r \right)$ is,
\begin{align}
    \mathcal{K}_\nu \left( r \right) &= \sqrt{\frac{\pi}{2}} \cdot \frac{\exp\left( -r\right)}{\sqrt{r}} \sum_{i = 0}^{p} \frac{\left( p + i \right)!}{i! \left( p - i  \right)!} \left( 2r \right)^{-i}, \quad p = | \nu | - \frac{1}{2}.
\end{align}
Nevertheless, the cases more interesting of the Mat\'ern covariance functions are,
\begin{align}
    \mathrm{Cov}_{\nu = 3/2} \left( \mathbf{x}_n, \mathbf{x}_m \right) &= \left( 1 + \frac{\sqrt{3}r}{\ell} \right) \cdot \exp \left( - \frac{\sqrt{3}r}{\ell} \right), \\
    \mathrm{Cov}_{\nu = 5/2} \left( \mathbf{x}_n, \mathbf{x}_m \right) &= \left( 1 + \frac{\sqrt{5}r}{\ell} + \frac{\sqrt{5}r^2}{3\ell^2} \right) \cdot \exp \left( - \frac{\sqrt{5}r}{\ell} \right).
\end{align}

The optimal set of hyperparameters of a covariance function can be obtained via minimizing the negative marginal log-likelihood, or also known as evidence,
\begin{align}
    \log p \left( \mathbf{y} \mid \mathbf{X}, \theta \right) = -\frac{N}{2} \log \left( 2 \pi \right) - \frac{1}{2} \log \mid \mathbf{K}_\theta \mid  - \frac{1}{2} \mathbf{y}^\top \mathbf{K}_\theta \mathbf{y},
\end{align}
where $|\cdot|$ denotes the determinate of a matrix.

\subsubsection{Acquisition Function}
    
We propose the implementation of Expected Improvement (EI). This acquisition function quantifies the amount of improvement calculating the expectation of an improvement function that is,
\begin{align}
    \mathcal{A}_{EI} \left( \mathbf{x} ; \theta, \mathcal{D} \right) = \int_y \max \left( 0, y_{best} - y \right) \cdot p \left( y \mid \mathbf{x}, \theta, \mathcal{D} \right) dy,
\end{align}
we can compute analytically the integral with this formula,
\begin{align}
    \mathcal{A}_{EI} \left( \mathbf{x} ; \theta, \mathcal{D} \right) = \sigma \left( \mathbf{x}; \theta, \mathcal{D} \right) \left( \gamma \left( \mathbf{x} \right) \cdot \varphi \left( \gamma \left( \mathbf{x} \right) \right) \right) + \mathcal{N} \left(\gamma \left( \mathbf{x} \right); 0, 1 \right)
\end{align}
where $\varphi (\cdot)$ represents the standard normal cumulative distribution function, and $\gamma (x)$ is
\begin{align}
    \gamma \left( \mathbf{x} \right) = \frac{f \left( \mathbf{x}_{best} \right) - \mu \left( \mathbf{x} ; \theta, \mathcal{D} \right) + \xi }{\sigma \left( \mathbf{x} ; \theta, \mathcal{D} \right)}
\end{align}
the parameter $\xi$ controls the trade off between exploration and exploitation during the optimization.

\subsubsection{Acquisition Function Optimization}
The following optimization problem has to be solved to find the point maximizes the acquisition function \cite{SNOEK2012},
\begin{align}
    \mathbf{x}^* = \underset{\mathbf{x} \in \mathcal{X}}{\operatorname{argmax}} \ \mathcal{A} \left( \mathbf{x} \right).
\end{align}
In order do that, it is possible to implement a fully-Bayesian treatment of the hyperparameters, thus it is necessary to integrate over the hyperparamters to marginalize acquisition function,
\begin{align}
    \mathcal{\hat{A}} \left( \mathbf{x}; \{y_m, \mathbf{x}_m\} \right) = \int \mathcal{A} \left( \mathbf{x} ; \{y_m, \mathbf{x}_m\}, \theta \right) p \left( \theta \mid \{ y_m, \mathbf{x}_m \}^M_{m = 1} \right) d\theta.
\end{align}
However, the minimum of the acquisition function can be also found via gradient-based optimization methods such as steepest descent,
\begin{align}
    \mathbf{x}_{k + 1} = \mathbf{x}_k - \eta \cdot \nabla \mathcal{A} \left( \mathbf{x}_k \right), \quad \forall k = 1, \ldots, L.
\end{align}
The analytical form of the gradient available is not available, but it can be approximated via numerical differentiation evaluating the acquisition function,
\begin{align}
    \mathcal{A}^{\prime} \left( \mathbf{x}_k \right) = \lim_{\varepsilon \to 0} \frac{ \mathcal{A} \left( \mathbf{x}_k + \varepsilon \right) - \mathcal{A} \left( \mathbf{x}_k \right)}{\varepsilon},
\end{align}
and as $\varepsilon \to 0$, we obtain that the forward finite difference equation is,
\begin{align}
    \Delta_{\varepsilon} \mathcal{A} \left( \mathbf{x}_k \right) = \mathcal{A} \left( \mathbf{x}_k + \varepsilon \right) - \mathcal{A} \left( \mathbf{x}_k \right).
\end{align}

\subsection{Parameters Cross-Validation}

The methodology propose to validate the optimal set of parameters for each one of the velocity vectors techniques is to utilize them approximate a known wind velocity field on an sequence of images. The wind velocity field is generated defining beforehand the streamlines and potential lines. This simulates either linear or a non-linear flow.
\begin{figure}[!htbp]
    \begin{subfigure}{0.245\textwidth}
        \centering
        \includegraphics[scale = 0.15]{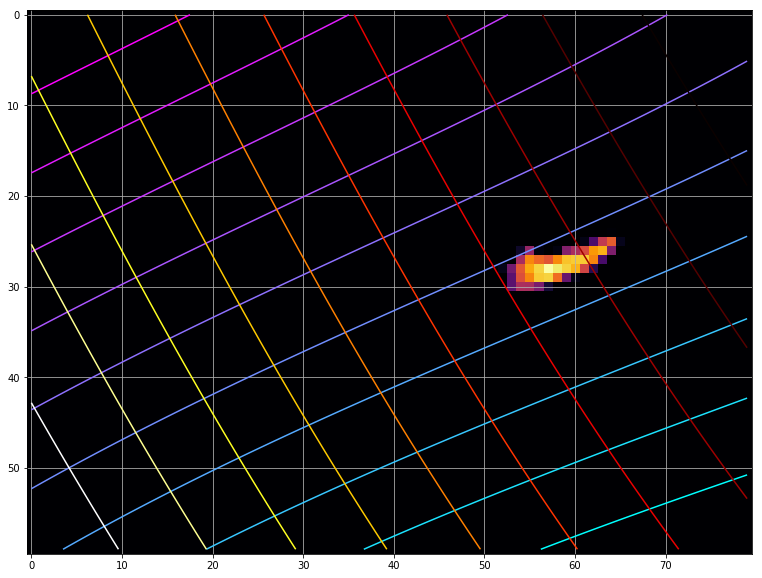}
    \end{subfigure}
    \begin{subfigure}{0.245\textwidth}
        \centering
        \includegraphics[scale = 0.15]{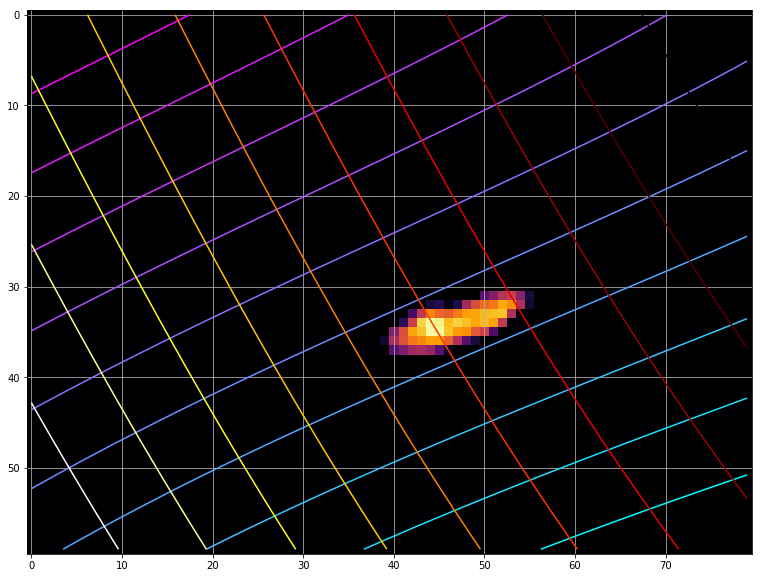}
    \end{subfigure}
    \begin{subfigure}{0.245\textwidth}
        \centering
        \includegraphics[scale = 0.15]{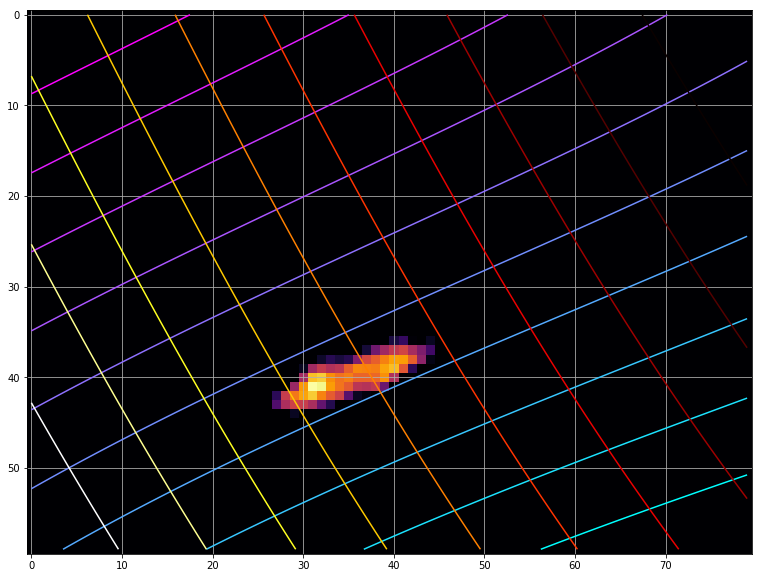}
    \end{subfigure}
    \begin{subfigure}{0.245\textwidth}
        \centering
        \includegraphics[scale = 0.15]{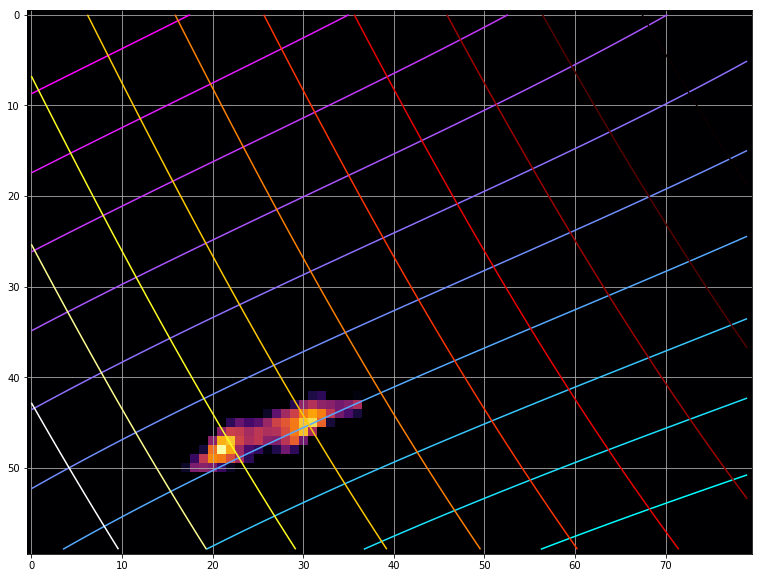}
    \end{subfigure}
\caption{One of the four sequences of images generated to cross-validate the optimal parameters for each method. In the cool colors are displayed the potential lines, in warm colors are displayed the streamlines. The generated flow displayed in the image is non-linear.}
\label{fig:cloud_in_flow}
\end{figure}

\begin{table}[!htb]
    \centering
    \small
    \begin{tabular}{lcccccccc}
        \toprule
        \textbf{Flow} & \textbf{Pyr. Scale} & \textbf{No. Pyrs.} & $\boldsymbol{\mathcal{W}_{p\times p}}$ & \textbf{No. Iters.} & $\boldsymbol{\mathcal{P}^n}$ & \textbf{Filter} $\boldsymbol{\sigma}^2 $ & \textbf{t [s]} & \textbf{RMSE} \\
        \midrule
        Linear & 0.6529 & 7 & 2 & 11 & 5 & 1.3398 & 0.0120 & 0.1804 \\
        Non-Linear & 0.2607 & 9 & 2 & 78 & 1 & 0.6368 & 0.0163 & 0.2620 \\
        \bottomrule
    \end{tabular}
    \caption{This table shows the results of the Farneb{\"a}ck method. The parameters validated were the number of pyramids and their scale, the size of the sliding window, the number of interactions, and the order of the polynomial expansion and the variance of the Gaussian filter.}
    \label{tab:farnerback}
\end{table}

\begin{figure}[!htbp]
    \centering
    \includegraphics[scale = 0.375]{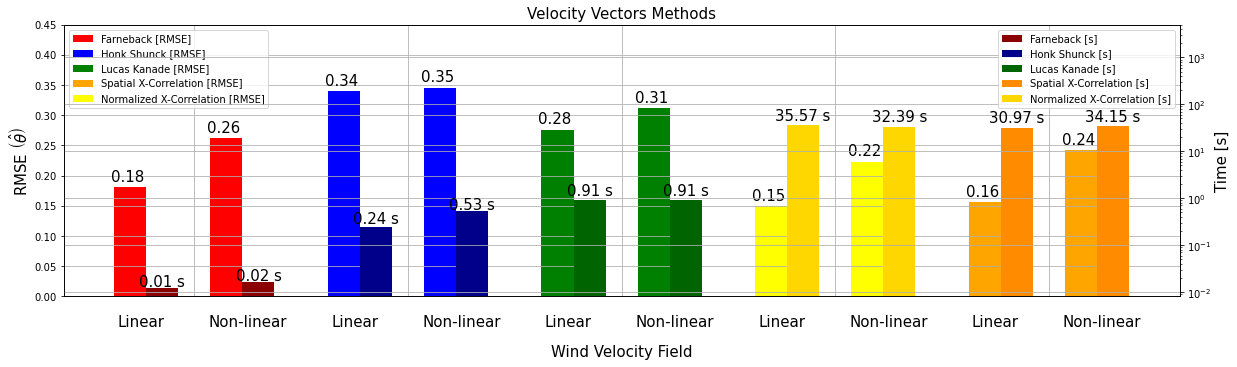}
\caption{This graphs shows the comparison of each method performances in RMSE and computing time for a linear flow, and a non-linear flow which has a small vorticity and divergence. The best trade-off between computing time and error is obtained by Farneb{\"a}ck.}
\label{fig:vel_chart}
\end{figure}

An actual IR image sequence of a cloud with non-simulated dynamics is cropped out of its original frames. The position of reference is the cloud's mass center in the frame. The cropped image of the cloud, which is centered in the cloud's mass center, and placed in a new frame with the simulated flow. The update process to generate a trajectory is: in frame (1) the cloud is displaced to new position in frame (2), according to the simulated velocity vector increments at the cloud's mass center coordinates in frame (1). This is update rule is repeated along the whole sequence of images. Therefore, the trajectory of the cloud is known, it can be compared with velocity vector computed using each method. This process is repeated for 4 different simulated flows, an example is on figure \ref{fig:cloud_in_flow}.

As the partial derivatives in our problem are not available, the optimal parameters for each technique are found by means of BO. This optimization method explores the most likely set of parameters to produce an improvement on an acquisition function at each iteration. In particular, the acquisition functions aims to infer the function of the errors computing the velocity vectors by each one of methods. The function of the errors is estimated via a GP for regression. The optimal set of parameters for each method, are displayed in tables \ref{tab:honk_Schunck}, \ref{tab:lucas_kanade}, \ref{tab:farnerback}, \ref{tab:spatial_xcorr}, and \ref{tab:norm_xcorr}.

\begin{table}[!htb]
    \centering
    \small
    \begin{tabular}{lcccc}
        \toprule
        \textbf{Flow} & $\boldsymbol{\mathcal{W}_{p \times p}}$ & \textbf{Eig. Thrs.} & \textbf{t [s]} & \textbf{RMSE} \\
        \midrule
        Linear &  11 & $1 \times 10^{-8}$ & 0.9128 & 0.2759 \\
        Non-Linear & 11 & $1 \times 10^{-8}$ & 0.9114 & 0.3115 \\
        \bottomrule
    \end{tabular}
    \caption{This table shows the results of the Lucas-Kanade method. The parameters validate where the size of the sliding window, and the noise threshold in the eigenvalues of the LS solution.}
    \label{tab:lucas_kanade}
\end{table}

\begin{table}[!htb]
    \centering
    \small
    \begin{tabular}{lcccccc}
        \toprule
        \textbf{Flow} & \textbf{Inc. Size Opt.} & \textbf{Diff. Kernel Amp.} & \textbf{Tol.} & \textbf{Max. Iters.} & \textbf{t [s]} & \textbf{RMSE} \\
        \midrule
        Linear & 13.6420 & 739.9022 &  $1 \times 10^{-5}$ & 1000 & 0.2444 & 0.3398 \\
        Non-Linear & 3.3449 & 244.2650 & $1 \times 10^{-5}$ & 1000 & 0.5328 & 0.3452 \\
        \bottomrule
    \end{tabular}
    \caption{The table shows the results obtained implementing the Honk-Schunck method with constant tolerance and a maximum number of iteration in the optimization, The parameters validated were the increments multiplayer in the optimization, and the amplitude of the differentiation kernel.}
    \label{tab:honk_Schunck}
\end{table}

\begin{table}[!htb]
    \centering
    \small
    \begin{tabular}{lcccc}
        \toprule
        \textbf{Flow} & $\boldsymbol{\mathcal{W}_{p \times p}}$ & $\boldsymbol{\mathcal{P}^n}$ & \textbf{t [s]} & \textbf{RMSE} \\
        \midrule
        Linear & 23 & 14 & 35.5664 & 0.1495  \\
        Non-Linear & 20 & 17 & 32.3930 &  0.2231 \\
        \bottomrule
    \end{tabular}
    \caption{This table shows the results of the Normalized Cross-Correlation method. The parameters validated where the size of the sliding window, and the order of the polynomial expansion implemented to approximate the normalized cross-correlation.}
    \label{tab:norm_xcorr}
\end{table}

\begin{table}[!htb]
    \centering
    \small
    \begin{tabular}{lcccc}
        \toprule
        \textbf{Flow} & $\boldsymbol{\mathcal{W}_{p \times p}}$ & $\boldsymbol{\mathcal{P}^n}$ & \textbf{t [s]} & \textbf{RMSE} \\
        \midrule
        Linear & 22 & 11 & 30.9664 & 0.1557 \\
        Non-Linear & 23 & 19 & 34.1549 &  0.2422 \\
        \bottomrule
    \end{tabular}
    \caption{This table shows the results of the Spatial Cross-Correlation method. The parameters validated where the size of the sliding window, and the order of the polynomial expansion implemented to approximate the cross-correlation function.}
    \label{tab:spatial_xcorr}
\end{table}

\subsection{Velocity Vectors Regularization}

The dense approximation of velocity vectors between two consecutive have outlier vectors. This is because of the assumption of constant intensity, and small time increments, that these methods require in their formulation, and that is sometimes violated. Therefore, we propose to regularized the velocity vectors to remove the information provided by the outliers at each computation. The cloud velocity vectors in the image $k$ is,
\begin{align}
    \mathbf{v}_{i,j} = \left\{ \left(u_{i,j}, v_{i,j} \right) \mid \mathbf{v}_{i,j} \in \mathbb{R}^2, \ \forall i = 1, \ldots, M, \ \forall j = 1, \ldots, N \right\}.
\end{align}

A lower and upper threshold is applied to find the vectors, which magnitude is so low enough that they are likely noise, or so high that are unfeasible in the real world,
\begin{align}
    \mathbf{v}_{i,j} = 
    \begin{cases}
    \mathbf{v}_{i,j} & \ \tau_{lower} < \| \mathbf{v}_{i,j} \| \leq \tau_{upper} \\
    \mathbf{0} & \mathrm{Otherwise}
    \end{cases}
    \quad \forall m,n = -w, \ldots w,
\end{align}
where $\tau_{lower} = 0.1$ and $\tau_{upper} = 10$. The information lost by deleting these velocity vectors, can be partially recovered inferring it from neighbor vectors, For that, a median filter is applied on the velocity vectors after the regularization, 
\begin{align}
    \mathbf{v}_{i,j} = 
    \begin{cases}
    \mathbf{v}_{i,j} & \ \tau_{lower} < \| \mathbf{v}_{i,j} \| \leq \tau_{upper} \\
    \mathrm{Median} \left( \mathrm{v}_{i + m,j + n} \right) & \mathrm{Otherwise} \\
    \end{cases}
    \ \forall m,n = -w, \ldots w
\end{align}
The filtered velocity vectors are recovered, only in those places where were regularized to zero, in order to replace them.

\begin{figure}[!htbp]
    \centering
    \includegraphics[scale = 0.355]{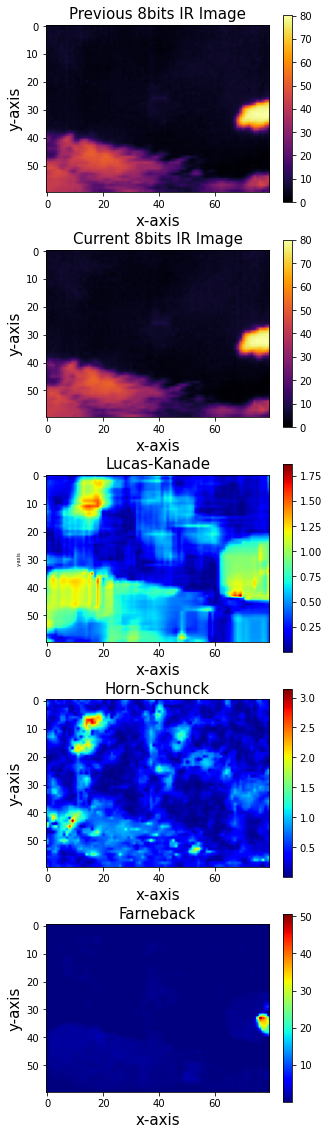}
    \includegraphics[scale = 0.355]{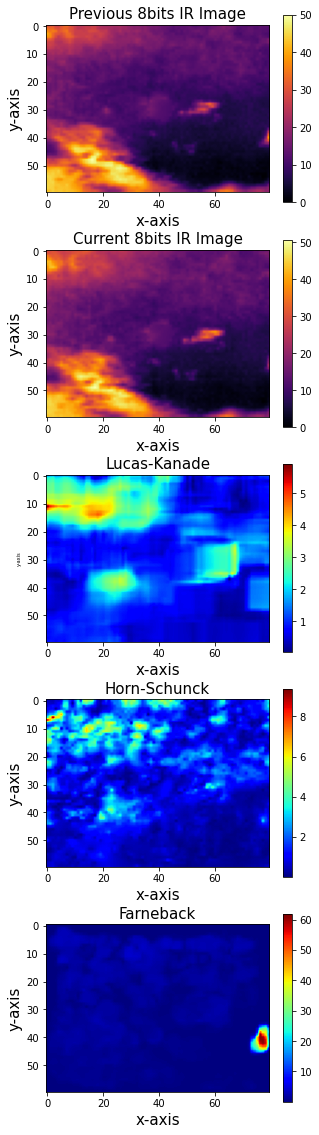}
    \includegraphics[scale = 0.355]{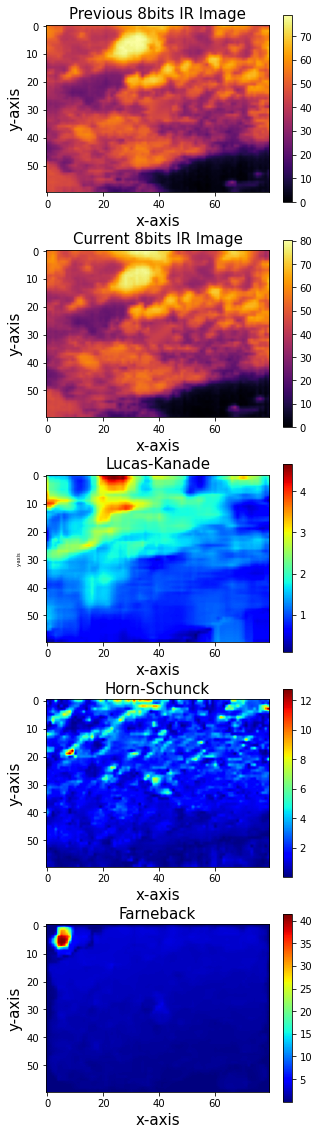}
    \includegraphics[scale = 0.355]{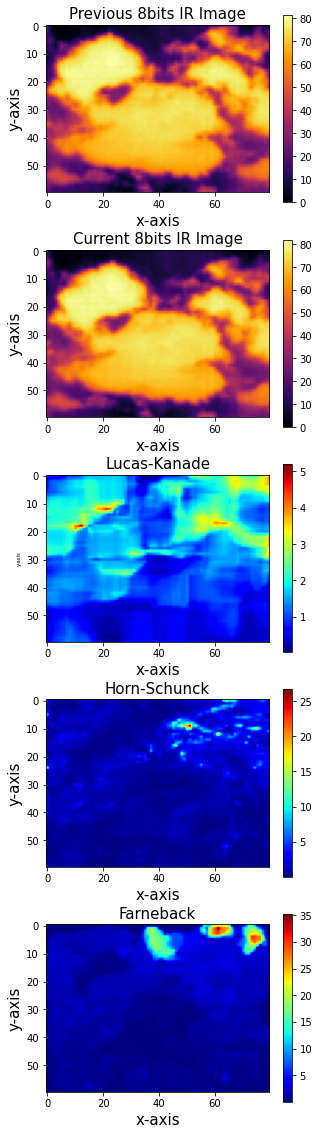}
\caption{The shown velocity field was computed from the images in the first row, and the images in the second row, which are organized as different sequences from left to right. The magnitude of the velocity field computed by Lucas-Kanade, Horn-Schunck, and Farneb{\"a}ck, are shown in the third, fourth and fifth row respectively. In spite of having the lowest error in the validation, the Farneb{\"a}ck method is very sensitive to outliers during the implementation.}
\label{fig:vel_errors}
\end{figure}

\subsection{Implementation}

After the cross-validation of the parameters for each method, these were implemented in the features extraction algorithm, see figure \ref{fig:vel_errors}. Despite of the fact that the Farneb{\"a}ck achieves the lowest computing time with a relative small error, it was found that produces outlier during the implementation. In fact, the region of the outliers in the image is so large that information cannot be recovered from the regularized vectors. As these outliers sometimes occur during consecutive frames, information can affect to the forecast performances. Therefore, the most suitable method is concluded to be the Lucas-Kanade after having in consideration all facts in this analysis.

\section{Acknowledgments}

This work has been supported by NSF EPSCoR grant number OIA-1757207 and the King Felipe VI endowed Chair. Authors would like to thank the UNM Center for Advanced Research Computing, supported in part by the National Science Foundation, for providing the high performance computing and large-scale storage resources used in this work.

\bibliographystyle{unsrt}  
\bibliography{mybibfile}

\end{document}